\documentclass[12pt]{article}
\pdfoutput=1
\usepackage{amsmath, amsthm,comment}
\usepackage{amsfonts}
\usepackage{amssymb,graphics,psfrag}
\usepackage{array,epsfig,stmaryrd,graphicx}
\usepackage{cite}
\usepackage{graphicx}
\usepackage{makeidx}
\usepackage{multicol}
\usepackage{geometry}
\usepackage{amsfonts}
\usepackage{mathrsfs}
\usepackage{amssymb}
\usepackage{amsmath}
\usepackage{slashed}
\usepackage{heck}%
\providecommand{\U}[1]{\protect\rule{.1in}{.1in}}
\newcommand{\beq}{\begin{equation}}
\newcommand{\eeq}{\end{equation}}
\newcommand{\be}{\begin{equation}}
\newcommand{\ee}{\end{equation}}
\newcommand{\bea}{\begin{eqnarray}}
\newcommand{\eea}{\end{eqnarray}}
\newcommand{\ben}{\begin{eqnarray*}}
\newcommand{\een}{\end{eqnarray*}}
\newcommand{\ba}{\begin{aligned}}
\newcommand{\ea}{\end{aligned}}
\newcommand{\bt}{\begin{tabular}}
\newcommand{\et}{\end{tabular}}
\newcommand{\bc}{\begin{center}}
\newcommand{\ec}{\end{center}}

\newcommand{\cref}{{\bf [check ref]}}

\newcommand{\bs}{\begin{subarray}{c}}
\newcommand{\es}{\end{subarray}}

\def\IZ{{\mathbb Z}}

\newcommand{\rd}{{\rm d}}

\numberwithin{equation}{section}
\begin{document}

\date{April, 2009}
\title{F-theory and Neutrinos:\\ Kaluza-Klein Dilution of Flavor Hierarchy}

\preprint{arXiv:0904.1419}

\institution{HarvardU}{\centerline{Jefferson Physical Laboratory, Harvard University, Cambridge,
MA 02138, USA}}

\authors{Vincent Bouchard\footnote{e-mail: {\tt
bouchard@physics.harvard.edu}},
Jonathan J. Heckman\footnote{e-mail: {\tt
jheckman@fas.harvard.edu}}, \\[2mm]
Jihye Seo\footnote{e-mail: {\tt
jihyeseo@fas.harvard.edu}}, and Cumrun
Vafa\footnote{e-mail: {\tt
vafa@physics.harvard.edu}}}

\abstract{We study minimal implementations of Majorana and Dirac neutrino scenarios in F-theory GUT
models. In both cases the mass scale of the neutrinos $m_{\nu} \sim M^{2}_{\text{weak}}/\Lambda_{\text{UV}}$
arises from integrating out Kaluza-Klein modes, where $\Lambda_{\text{UV}}$ is
close to the GUT scale. The participation of non-holomorphic Kaluza-Klein mode wave functions dilutes the mass hierarchy in comparison
to the quark and charged lepton sectors, in agreement with experimentally measured
mass splittings. The neutrinos are predicted to exhibit a ``normal" mass hierarchy, with masses
$(m_{3},m_{2},m_{1}) \sim .05 \times(1,\alpha^{1/2}_{GUT},\alpha_{GUT})$ eV. When the interactions of the neutrino and charged
lepton sectors geometrically unify, the neutrino mixing matrix exhibits a mild hierarchical
structure such that the mixing angles $\theta_{23}$ and $\theta_{12}$ are large and comparable,
while $\theta_{13}$ is expected to be smaller and close to the Cabibbo angle: $\theta_{13} \sim \theta_{C} \sim \alpha^{1/2}_{GUT}	\sim 0.2$.
This suggests that $\theta_{13}$ should be near the current experimental upper bound.}%

%TCIMACRO{\TeXButton{Maketitle}{\maketitle}}%
%BeginExpansion
\maketitle
%EndExpansion
\tableofcontents

\section{Introduction}

The observation of neutrino oscillations \cite{homestake98,superkamiokande98}
has revealed that neutrinos have small non-zero masses. However, non-zero neutrino masses cannot be accommodated in the Standard Model without introducing extra ingredients. As such, neutrino physics offers a concrete and exciting window
into physics beyond the Standard Model.

The seesaw mechanism is perhaps the simplest theoretical model which describes
small neutrino masses. By introducing very heavy right-handed Majorana
neutrinos, the seesaw mechanism produces an effective light Majorana mass for
the left-handed neutrinos. For the masses of the left-handed
neutrinos to be consistent with experimental bounds, the right-handed
neutrinos must have Majorana masses around the scale $\Lambda_{\text{UV}}\sim10^{14}-10^{15}$
GeV, which is close to the GUT scale. Hence, the seesaw mechanism
suggests that neutrino physics should be somehow related to the dynamics of
GUT theories.

However, in four-dimensional GUT models additional ingredients must be added just to
accommodate the seesaw mechanism. For instance, in $SO(10)$ GUTs, this necessitates
additional fields transforming in higher dimensional representations developing suitably large
vevs, or higher dimension operators (see for example \cite{Mohapatra:1986uf}
for a review of such mechanisms in the context of four-dimensional GUTs). Therefore, it
is worth asking whether string theory may offer new insights into neutrino physics.

In recent work on GUTs realized in F-theory (F-theory GUTs) the observation
that $M_{GUT}/M_{pl}\sim10^{-3}$ is a small number has been promoted in
\cite{BHVI,BHVII} to the vacuum selection criterion that there exists a limit
in the compactification where it is in principle possible to decouple the
effects of gravity by taking $M_{pl}\rightarrow\infty$, with $M_{GUT}$ kept
finite. See
\cite{DonagiWijnholt,WatariTATARHETF,DonagiWijnholtBreak,HVGMSB,HVLHC,Font:2008id,HVCKM,Blumenhagen:2008zz,Blumenhagen:2008aw,FGUTSCosmo,Bourjaily:2009vf,Hayashi:2009ge,Andreas:2009uf,Chen:2009me,HKSV,DonagiWijnholtIII}
for some other recent work on F-theory GUTs. Aspects of flavor physics in
F-theory GUTs have been studied in \cite{HVCKM}, where it was shown that
with the minimal number of geometric ingredients necessary for achieving one
heavy generation, the resulting flavor hierarchies in the
quark and charged lepton sectors are in accord with observation. The aim
of this paper is to extend this minimal framework to include a neutrino
sector with viable flavor physics\footnote{See \cite{RandallSimmonsDuffin} for other forthcoming work on flavor
physics in the context of F-theory GUT models.}.

We study both Majorana and Dirac neutrinos in minimal $SU(5)$ F-theory GUTs, finding scenarios which lead to phenomenologically consistent models of neutrino flavor. In both cases, integrating out massive Kaluza-Klein modes generates higher dimension operators which lead to viable neutrino masses. The neutrino mass scale $m_{\nu}$ is roughly related to the weak scale and a scale close to $M_{GUT}$ through the numerology of the seesaw mechanism:
\begin{equation}
m_{\nu} \sim \frac{M^{2}_{\text{weak}}}{\Lambda_{\text{UV}}}.
\end{equation}
In the Majorana scenario, an infinite tower of massive modes trapped on a Riemann surface
play the role of right-handed neutrinos, and generate the F-term
\begin{equation}
\lambda_{ij}^{\text{Maj}} \int \rd^{2}\theta \frac{(H_{u}L^i) (H_u L^j)}{\Lambda_{\text{UV}}}
\label{e:MAJ}
\end{equation}
through an effective Kaluza-Klein seesaw mechanism.  When $H_u$ develops a vev $\langle H_u\rangle \sim M_{\text{weak}}$ this induces
a Majorana mass. In the Dirac scenario, the D-term
\begin{equation}
\lambda_{ij}^{\text{Dirac}} \int \rd^{4}\theta \frac{H_{d}^{\dag}L^i N_{R}^j }{\Lambda_{\text{UV}}}
\label{e:DIR}
\end{equation}
is generated by integrating out massive modes on the Higgs curve. Supersymmetry breaking leads to an F-term for $H_{d}^{\dag}$ of
order ${F_{H_d}}\sim \mu H_u \sim M^2_{\text{weak}}$
which induces a Dirac mass.
We show that the participation of an infinite tower of massive states can boost the overall scale of the neutrino masses. This is welcome, since the two higher dimension operators \eqref{e:MAJ} and \eqref{e:DIR} with scale
$\Lambda_{\text{UV}} = M_{GUT}$ would produce light neutrino masses which are slightly too low.

Owing to the rigid structure present in F-theory GUTs, it is perhaps not surprising that the supersymmetry breaking sector of \cite{HVGMSB} naturally enters the discussion of neutrino physics. In \cite{HVGMSB}, the absence of a bare $\mu$ term in the low energy theory was ascribed to the presence of a $U(1)$ Peccei-Quinn symmetry, derived from an underlying $E_{6}$ GUT structure. This choice of $U(1)_{PQ}$ charges turns out
to also exclude the higher dimension operator \eqref{e:MAJ} appearing in the Majorana scenario. Interestingly, we find
a unique alternative choice of $U(1)$ charge assignments which is simultaneously compatible with a higher unification structure and the operator \eqref{e:MAJ}.\footnote{Even though this new $U(1)_{PQ}$ does not change the general scenario of F-theory GUTs, it does change some of the detailed numerical estimates of the \textquotedblleft PQ deformation\textquotedblright\ away from minimal gauge
mediation studied in \cite{HVGMSB,HKSV}. It would be worth investigating this
further.}

Estimating the form of the Yukawa matrices for the two operators \eqref{e:MAJ} and \eqref{e:DIR}, we find that in both scenarios the neutrinos exhibit a ``normal" hierarchy, where the two lightest neutrinos are close in mass. The participation of Kaluza-Klein modes dilutes the mass hierarchy in comparison to the quark and charged lepton sectors. More precisely, the resulting neutrino mass hierarchy is roughly:
\begin{equation}
m_{1}:m_{2}:m_{3} \sim \alpha_{GUT}:\alpha^{1/2}_{GUT}:1
\end{equation}
which is in reasonable accord with the observed neutrino mass splittings.

The structure of the neutrino mixing matrix depends on whether the neutrino
and lepton interactions localize near each other, or are far apart. When these interactions are geometrically unified at a single point, the mixing matrix displays a mild hierarchical structure. The two mixing angles $\theta_{12}$ and $\theta_{23}$ are found to be comparable, and in rough agreement with experiments. The mixing angle $\theta_{13}$, which measures mixing between the heaviest and lightest neutrino (in our normal hierarchy), is predicted to be roughly given
(in radians) by:%
\begin{equation}
\theta_{13}\sim\theta_{C}\sim\alpha_{GUT}^{1/2}\sim0.2\text{,}%
\end{equation}
where $\theta_{C}$ denotes the Cabibbo angle. These results, in conjunction with the analysis of \cite{HVCKM}, points towards the possibility of a higher unification structure. Along these lines, in both the Majorana and Dirac scenarios we present models where \emph{all} of the interactions of the MSSM unify at a single $E_8$ interaction point in the geometry.

We also study geometries where the neutrino and lepton interaction terms do not unify. In this case, the neutrino mixing matrix is a generic unitary matrix with no particular structure. As a result, large mixing angles are expected, and in particular the angle $\theta_{13}$ should be close to the current experimental upper bound. Assuming that the neutrino mixing matrix is given by a random unitary matrix, we explain how randomness suggests that $\theta_{12}$ and $\theta_{23}$ should be comparable, while $\theta_{13}$ should be slightly smaller, which is in qualitative agreements with neutrino oscillation experiments.

The organization of the rest of the paper is as follows. In section \ref{NEUTREV} we
review the main features of neutrino physics. Section \ref{FREV} provides
a short review of those aspects of F-theory GUTs which are of relevance to
neutrino physics. We present a minimal implementation of the
Majorana scenario in section \ref{KKMAJ}. In this same section, we study the
presence of monodromies in seven-brane configurations, and explain the crucial
role this geometric ingredient plays in the Kaluza-Klein seesaw. In section
\ref{NEUTMASSHIER} we estimate the Majorana scenario Yukawas. Next, in section \ref{sec:DIRAC} we discuss a minimal Dirac mass
scenario, which surprisingly exhibits similar numerology to that of the Majorana scenario.
Our results for the neutrino masses and mixing angles are compared with
experiments in section \ref{s:obs}. Section \ref{CONC} contains our conclusions.
Appendices A, B and C discusses other aspects of F-theory neutrinos, and Appendix D
contains a discussion of probability measures for random unitary matrices.

\section{Review of Neutrino Physics}

\label{NEUTREV}

In this section we review the main features of neutrino physics. We first describe background material on the
masses and mixing angles of the neutrino sector in subsection \ref{BKG}, and then review current observational constraints in subsection \ref{EXPCONST}. This is followed in subsection \ref{NeutrinoANDUV} by a brief
discussion of the suggestive appearance of UV physics in the neutrino sector and potential sources of tension with string based models.

\subsection{Neutrino Masses and Mixing Angles}

\label{BKG}

In this subsection we define the neutrino masses and mixing angles. In order
to maintain continuity with the superfield notation employed later, we let $L$
denote the lepton $SU(2)$ doublet superfield of the MSSM, and $N_{L}$ the
left-handed neutrino component of this doublet. We shall also denote by
$E_{L}$ the charged lepton component of the doublet $L$, and by $E_{R}$ the
right-handed charged lepton superfields. We emphasize that this notation is
adopted for notational expediency. Indeed, at the energy scales where the
neutrinos develop masses, supersymmetry has already been broken.

Neutrino mass can in principle originate from one of two possible effective
chiral couplings, which below the electroweak symmetry breaking scale can be
written as:
\begin{align}
W_{\text{Majorana}}  &  \supset m_{ij}^{\text{Maj}}\cdot N_{L}^{i}N_{L}%
^{j}\label{MajMASS}\\
W_{\text{Dirac}}  &  \supset m_{ij}^{\text{Dirac}}\cdot N_{L}^{i}N_{R}%
^{j}\text{,} \label{DirMASS}%
\end{align}
where in the second case, $N_{R}$ denotes a right-handed neutrino, and
$i,j=e,\mu,\tau$ index the three generations of left-handed
neutrinos. These mass terms correspond respectively to Majorana and
Dirac mass terms. The full lepton sector of the theory can then be written as:%
\begin{equation}
W_{\text{Lepton}}\supset m_{ij}^{(\nu)}\cdot N_{L}^{i}N^{j}+m_{ij}^{(l)}\cdot
E_{L}^{i}E_{R}^{j}\text{,}%
\end{equation}
where the first term corresponds to either of the two mass terms given in lines \eqref{MajMASS} and \eqref{DirMASS}.

As usual, we introduce matrices $U_{L}^{(\nu)}$ and $U_{R}^{(\nu)}$, and
matrices $U_{L}^{(l)}$ and $U_{R}^{(l)}$, diagonalizing the mass matrices in
the lepton sector:%
\begin{align}
U_{L}^{(\nu)}m^{(\nu)}\left(  U_{R}^{(\nu)}\right)  ^{\dag}  &  =\text{diag}%
(m_{1},m_{2},m_{3})\\
U_{L}^{(l)}m^{(l)}\left(  U_{R}^{(l)}\right)  ^{\dag}  &  =\text{diag}%
(m_{e},m_{\mu},m_{\tau})\text{.}%
\end{align}
Using the fact that $N_{L}^{i}$ and $E_{L}^{i}$ transform as $SU(2)$ doublets, we can define a mixing matrix, as in the quark sector. The neutrino mixing
matrix is given by\cite{Pontecorvo:1957cp,Maki:1962mu}:%
\begin{equation}
U_{PMNS}=U_{L}^{(l)}\left(  U_{L}^{(\nu)}\right)  ^{\dag} = \left(
\begin{array}{ccc}
U_{e1} & U_{e2} & U_{e3} \\
U_{\mu 1} & U_{\mu 2} & U_{\mu 3} \\
U_{\tau 1} & U_{\tau 2} & U_{\tau 3}%
\end{array}%
\right)  \text{.}
\label{e:mixing}%
\end{equation}

Introducing the parametrization of the unitary matrix in terms of the mixing
angles $0 \leq\theta_{ij} \leq90^{\circ}$, we can write:%
\begin{equation}
\label{anglePARAM}U_{PMNS}=\left(
\begin{array}
[c]{ccc}%
c_{12}c_{13} & s_{12}c_{13} & s_{13}e^{-i\delta}\\
-s_{12}c_{23}-c_{12}s_{23}s_{13}e^{i\delta} & c_{12}c_{23}-s_{12}s_{23}%
s_{13}e^{i\delta} & s_{23}c_{13}\\
s_{12}s_{23}-c_{12}c_{23}s_{13}e^{i\delta} & -c_{12}s_{23}-s_{12}c_{23}%
s_{13}e^{i\delta} & c_{23}c_{13}%
\end{array}
\right)  \cdot D_{\alpha}\text{,}%
\end{equation}
where $D_{\alpha}=$ diag$(e^{i\alpha_{1}/2},e^{i\alpha_{2}/2},1)$,
$c_{ij}=\cos\theta_{ij}$ and $s_{ij}=\sin\theta_{ij}$. Here, $\delta$, $\alpha_1$ and $\alpha_2$ are
CP violating phases. In the Dirac scenario, only $\delta$ corresponds to a physical phase, whereas
in the Majorana scenario all three angles are physical.

\subsection{Experimental Constraints}

\label{EXPCONST}

Neutrino oscillation experiments have established that neutrinos are indeed
massive \cite{homestake98,superkamiokande98}. While we do not know the
absolute mass eigenvalues $m_{1}$, $m_{2}$ and $m_{3}$, experiments have
measured small mass splittings.  It is important to note that neither the relative spacing between the three
neutrino masses, nor the lower bound on the neutrino masses has been
established. There are three relative mass spacings which are in principle
possible, corresponding to $m_{1}\sim m_{2}\sim m_{3}$, $m_{1}<m_{2}\ll m_{3}$
and $m_{3}\ll m_{1}<m_{2}$, which are respectively known as
\textit{degenerate/democratic, normal hierarchy} and \textit{inverted
hierarchy} mass spectra.
As reviewed in \cite{GonzalezGarcia:2007ib,GonzalezGarcia:2009ij},
solar and atmospheric measurements of neutrino oscillation lead to the mass splittings:%
\begin{align}
\Delta m_{21}^{2}  &  =m_{2}^{2}-m_{1}^{2}=\left(  7.06-8.34\right)
\times10^{-5}\text{ eV}^{2},\nonumber\\
|\Delta m_{31}^{2}|  &  =\left\vert m_{3}^{2}-m_{1}^{2}\right\vert =\left(
2.13-2.88\right)  \times10^{-3}\text{ eV}^{2}\text{.} \label{twothrsplit}%
\end{align}

The ambiguity in determining the type of neutrino hierarchy is in part due to
the large amount of mixing in the neutrino sector. As reviewed for example in
\cite{GonzalezGarcia:2007ib,GonzalezGarcia:2009ij}, at the $3\sigma$ level of observation, the
magnitude of the entries of the neutrino mixing matrix \eqref{e:mixing} are:
\begin{equation}
\left\vert U_{PMNS}^{3\sigma}\right\vert \sim\left(
\begin{array}
[c]{ccc}%
0.77-0.86 & 0.50-0.63 & 0.00-0.22\\
0.22-0.56 & 0.44-0.73 & 0.57-0.80\\
0.21-0.55 & 0.40-0.71 & 0.59-0.82
\end{array}
\right)  \text{.}%
\end{equation}
Aside from the upper right-hand entry, the content of this mixing matrix has a
very different structure from the CKM matrix in the quark sector:%
\begin{equation}
\left\vert V_{CKM}\right\vert \sim\left(
\begin{array}
[c]{ccc}%
0.97 & 0.23 & 0.004\\
0.23 & 0.97 & 0.04\\
0.008 & 0.04 & 0.99
\end{array}
\right)  \text{.}%
\end{equation}

Returning to the parametrization of the mixing matrix given in equation
(\ref{anglePARAM}), the current lack of distinguishability between Majorana
and Dirac masses implies that there is at present no conclusive observational
data on the CP violating phases $\delta, \alpha_{1}$ and $\alpha_{2}$.
The
experimental values for the mixing angles have been extracted in \cite{GonzalezGarcia:2007ib}, and at the $3\sigma$ level are given by:
\begin{align}
\theta_{12}  &  \sim 30.5^{\circ}-39.3^{\circ}\\
\theta_{23}  &  \sim 34.6^{\circ}-53.6^{\circ}\\
\theta_{13}  &  \sim 0^{\circ}-12.9^{\circ}\text{.}%
\end{align}
Current bounds on $\theta_{13}$ from the CHOOZ collaboration \cite{Chooz} are
expected to be improved by MINOS \cite{MINOS0901}.

At the $1\sigma$ level, global fits to solar and atmospheric oscillation data
obtained by KAMLAND and SNO suggest a non-zero value for $\theta_{13}$
\cite{Fogli:2008jx}. In fact, a non-zero value for $\theta_{13}$ near the
current upper bound has recently been announced by MINOS \cite{MINOS}%
.\footnote{We thank G. Feldman for bringing this result to our attention,
which we learned of after the results of this paper had already been
obtained.}

\subsection{Neutrinos and UV\ Physics\label{NeutrinoANDUV}}

Having described the main experimental constraints, we now review some of the
primary features of Dirac and Majorana mass terms in the context of the MSSM.\footnote{
We refer the interested reader to the review article \cite{Mohapatra:2005wg} for further discussion.}
After this, we review the fact that in spite of the suggestive link
between neutrinos and high energy physics, there is a certain amount of
tension in string based models which aim to incorporate neutrinos.

At a theoretical level, there are two features of the neutrino sector which
are quite distinct from the Standard Model. First, the overall mass scale of
the neutrino sector is far below the scale of electroweak symmetry breaking, but retains a suggestive link to the GUT scale, in that roughly speaking:
\begin{equation}\label{massrel}
m_{\nu} \sim \frac{M_{\text{weak}}^{2}}{\Lambda_{\text{UV}}},
\end{equation}
where $\Lambda_{\text{UV}} \sim 10^{14} - 10^{15}$ GeV is close to the GUT scale.
Second, the mixing angles are far larger than their counterparts in the
CKM\ matrix. These observations suggest that neutrino Yukawas may have
a very different origin from the other couplings of the Standard Model.

Let us first consider the case of Dirac neutrinos. Simply mimicking the mass
terms of the Standard Model, the Dirac type interaction:%
\begin{equation}
W\supset\lambda_{ij}^{(\nu)}H_{u}L^{i}N_{R}^{j}%
\end{equation}
would then generate a mass term for the neutrinos far above $0.05$ eV,
unless the entries of the corresponding Yukawa matrix are quite small, on the
order of $10^{-13}$. This however is rather fine-tuned, and it then becomes
necessary to explain why all of the other matter fields of the MSSM\ have
order one Yukawas, whereas the neutrino sector happens to have such small
couplings. We will find in section \ref{sec:DIRAC} that the relation of equation
(\ref{massrel}) can actually be accommodated quite naturally through the presence
of a higher dimension operator in the MSSM.

Leaving aside Dirac neutrinos for the moment, next consider Majorana neutrinos. Although a Majorana mass term as in
\eqref{MajMASS} is incompatible with the gauge symmetries of the Standard
Model, an effective mass term correlated with the vev of $H_{u}$ can be
introduced through the higher dimension operator:%
\begin{equation}
W_{eff}\supset\lambda_{ij}^{(\nu)}\frac{\left(  H_{u}L^{i}\right)  (H_{u}%
L^{j})}{\Lambda_{\text{UV}}}\text{,} \label{WeffMajorana}%
\end{equation}
where $\Lambda_{\text{UV}}$ is an energy scale far above the scale of
electroweak symmetry breaking. Once $H_{u}$ develops a vev on the order of the
weak scale, this will induce a Majorana mass term of the type given by
(\ref{MajMASS}). This operator breaks the accidental global $U(1)$ lepton
number symmetry of the Standard Model. Assuming that at least one of the
eigenvalues of $\lambda_{ij}^{(\nu)}$ is an order one number, this will
induce the neutrino mass scale of equation (\ref{massrel}).

The higher dimension operator of \eqref{WeffMajorana} can be generated in
seesaw models with heavy right-handed neutrinos. For example, in the type
I seesaw model (considering for simplicity the case of a single generation),
the superpotential term%
\begin{equation}
W\supset\lambda H_{u}LN_{R}+M_{\text{maj}}N_{R}N_{R}\text{,}
\label{righthandedupstairs}%
\end{equation}
will induce the requisite effective operator once the heavy $N_{R}$ field has
been integrated out. This can be generalized to all three generations of
leptons, and to an arbitrary number of $n$ right-handed neutrinos ($i,j=1,2,3$
and $I,J=1,2,\cdots,n)$:
\begin{equation}
W\supset\lambda_{iJ}H_{u}L^{i}N_{R}^{J}+M_{IJ}N_{R}^{I}N_{R}^{J}\text{.}%
\end{equation}

While any number of right-handed neutrinos are in principle allowed, in the
context of four-dimensional $SO(10)$ GUTs the appearance of three copies of
$N_{R}$ is especially natural. This is because in addition to the chiral
matter of the Standard Model, each spinor $\mathbf{16}$ of $SO(10)$ contains
an additional singlet $N_{R}$ state. Indeed, the presence of three
right-handed neutrino states renders the $U(1)_{B-L}$ symmetry non-anomalous.
However, we note here that in the context of string theory, anomalous $U(1)$
symmetries are quite common, and so the motivation for precisely three $N_{R}%
$'s is perhaps less obvious.

While the appearance of a scale close to $M_{GUT}$ is quite suggestive, the
bare matter content necessary to accommodate the Standard Model and
right-handed neutrinos is typically insufficient to generate a realistic
neutrino sector. For example, although it is a very non-trivial and elegant
fact that three copies of the spinor $\mathbf{16}$ in four-dimensional
$SO(10)$ GUTs contain just the chiral matter of the Standard Model, as well as
the right-handed neutrinos, this by itself is not sufficient for generating a
Majorana mass term for the right-handed neutrinos. Indeed, $\mathbf{16}
\times\mathbf{16}$ is not a gauge invariant operator.

In four-dimensional $SO(10)$ GUT\ models, it is therefore common to
incorporate additional degrees of freedom which can generate an appropriate
Majorana mass term for the right-handed neutrinos. These extra degrees of
freedom can either correspond to additional vector-like pairs in the
$\mathbf{16}_{\text{extra}}\oplus\mathbf{\overline{16}}_{\text{extra}}$, or to
higher dimensional representations such as the $\overline{\mathbf{126}}_{\text{extra}}$
of $SO(10)$.\footnote{We recall that the $\overline{\mathbf{126}}$ corresponds to the
five-index anti-self-dual anti-symmetric tensor of $SO(10)$.} The corresponding
operators:%
\begin{align}
W_{\mathbf{16~16~\overline{16}~\overline{16}}}  &  =\frac{\mathbf{16}%
_{M}\times\mathbf{16}_{M}\times\mathbf{\overline{16}}_{\text{extra}}%
\times\mathbf{\overline{16}}_{\text{extra}}}{M_{\text{UV}}},\\
W_{\mathbf{16~16~\overline{126}}}  &  =\mathbf{16}_{M}\times\mathbf{16}_{M}%
\times\overline{\mathbf{126}}_{\text{extra}},
\end{align}
can then generate Majorana mass terms for the neutrino component of the spinor
once either the $\mathbf{16}_{\text{extra}} \oplus\overline{\mathbf{16}%
}_{\text{extra}}$ or the $\overline{\mathbf{126}}_{\text{extra}}$ develops a vev. In the
above, $M_{\text{UV}}$ denotes a suppression scale which could either
correspond to the string or Planck scale. The second possibility is quite
problematic in the context of string based constructions, since typically, the
massless mode content will only contain matter in the $\mathbf{10}$,
$\mathbf{16}$, $\overline{\mathbf{16}}$ or $\mathbf{45}$ of $SO(10)$. However,
the first possibility, involving the presence of higher dimension operators,
is compatible with string considerations, and has figured prominently in many
string based constructions. Note that this type of interaction term will also
be present in $SU(5)$ GUT\ models once suitable GUT\ group singlets are
included. For a recent example of this type where a suitable combination of
singlet fields develop vevs, see \cite{Buchmuller:2007zd}.

Even in the context of $SU(5)$ GUT\ models, selection rules in the effective
field theory can be quite problematic. For example, in intersecting D-brane
configurations, the right-handed neutrinos will typically correspond to
bifundamentals between two D-brane gauge group factors. In such cases, the
gauge symmetries of the D-brane configuration forbid the coupling $N_{R}N_{R}%
$. As noted in \cite{BlumenhagenWeigandINST,IbanezUrangaMajorana,Cvetic:2007ku}, the additional gauge symmetries of the D-branes are often anomalous and so can be violated by stringy instanton effects. Because the characteristic size of this instanton is \emph{a priori} uncorrelated with the size of instantons in the GUT brane, an appropriate instanton effect might generate a Majorana mass term in the requisite range of $10^{12}-10^{15}$ GeV.
Nonetheless, achieving precisely the correct Majorana mass scale requires a
certain amount of tuning, because the magnitude of the instanton effect is
quite sensitive to the volume of the cycle which is wrapped by the D-brane
instanton. Worldsheet instanton effects in compactifications of the heterotic
string can also potentially generate a suitable Majorana mass term for
right-handed neutrinos.

It is also in principle possible to associate right-handed neutrinos with
other GUT group singlets, such as moduli fields. In this case, the primary
challenge is to obtain a Majorana mass which is near the GUT scale. Indeed,
moduli stabilization typically will lead either to very heavy masses for such
fields, or potentially, much lighter masses when one loop factors from
instanton effects stabilize a given modulus. This is a possibility which does
not appear to have received much attention in the literature, perhaps because
concrete realizations of the Standard Model with stabilized moduli are not yet available.

Even once the correct Majorana mass term has been generated, there is still
the further issue of addressing more refined features of the neutrino sector,
such as mass splittings, and the overall structure, or lack
thereof, in the neutrino mixing matrix.  While it indeed appears possible to
engineer detailed models of flavor utilizing large discrete symmetries, it is
not completely clear whether all such features can be incorporated
consistently within string based constructions. One of the aims of this paper
is to show that in a very minimal fashion, F-theory GUTs can accommodate mild
mass hierarchies and large mixing angles.

\section{Minimal F-theory GUTs\label{FREV}}

In this section we briefly review the main features of minimal F-theory GUTs,
focusing on those aspects of particular relevance for neutrino physics.\ For
further background and discussion, see for instance
\cite{BHVI,BHVII,HVGMSB,HVLHC,HVCKM,FGUTSCosmo,HKSV}, as well as
\cite{DonagiWijnholt,WatariTATARHETF,DonagiWijnholtBreak,Font:2008id,Blumenhagen:2008zz,Blumenhagen:2008aw,Bourjaily:2009vf,Hayashi:2009ge,Andreas:2009uf,Chen:2009me,DonagiWijnholtIII}%
. We also discuss in greater detail the role of the anomalous
global $U(1)$ Peccei-Quinn symmetry in the supersymmetry breaking sector of the low energy theory, and its interplay with the
neutrino sector.

\subsection{Primary Ingredients}

F-theory is defined as a strongly coupled formulation of IIB string theory in
which the profile of the axio-dilaton $\tau_{IIB}$ is allowed to vary over the
ten-dimensional spacetime. Interpreting $\tau_{IIB}$ as the complex structure
modulus of an elliptic curve, the vacua of F-theory can then be formulated in
terms of a twelve-dimensional geometry. Preserving four-dimensional
$\mathcal{N}=1$ supersymmetry then corresponds to compactifying F-theory on an
elliptically fibered Calabi-Yau fourfold with a section. In this case, the
base of the elliptic fibration corresponds to a complex threefold $B_{3}$.
Within this framework, the primary ingredients correspond to seven-branes
wrapping complex surfaces in $B_{3}$.

In F-theory GUTs, the gauge degrees of freedom of the GUT\ group propagate in
the bulk of the seven-brane wrapping a complex surface $S$, which is defined as a component of the discriminant locus of the elliptic fibration. Depending
on the type of singular fibers over $S$, the GUT group can correspond to
$SU(5)$, or some higher rank GUT\ group. In this paper we shall focus on the
minimal case with GUT group $SU(5)$.

The chiral matter and Higgs fields of the MSSM\ localize on Riemann surfaces (complex curves) in $S$. The massless modes
of the theory are given by the zero modes of these six-dimensional fields in
the presence of a non-trivial background gauge field configuration derived
from fluxes on the worldvolumes of the various seven-branes. The Yukawa
couplings of the model localize near points of the geometry where at least
three such matter curves meet.\footnote{As we will explain in subsequent
sections, this is only true in the cover theory, before we quotient by the
geometric action of the Weyl group defined by the geometric singularity. In
other words, some of the curves may be identified by monodromies, in which
case Yukawa couplings can arise at points where only two curves meet. See
\cite{Hayashi:2009ge} for a recent analysis of such configurations.}

An intriguing feature of F-theory GUTs is that imposing the condition that
gravity can in principle decouple from the GUT\ theory imposes severe
restrictions on the class of vacua suitable for particle physics
considerations. This endows the models with a considerable amount of
predictive power. For example, the existence of a decoupling limit requires
that the GUT\ seven-brane must wrap a del Pezzo surface. In particular, the
zero mode content of the resulting theory does not contain any adjoint-valued
chiral superfields, so that for example, embeddings of standard four-dimensional GUTs in F-theory \textit{cannot}
be decoupled from gravity. Breaking the GUT group requires introducing a
non-trivial flux in the $U(1)_{Y}$ hypercharge direction of the GUT group
\cite{BHVII,DonagiWijnholtBreak}. The resulting unbroken gauge group in four
dimensions is then given by $SU(3)_{C}\times SU(2)_{L}\times U(1)_{Y}$.

The ubiquitous presence of this flux has important ramifications elsewhere in
the model. For example, doublet triplet splitting in the Higgs sector can be
achieved by requiring that this flux pierces the Higgs up and Higgs down
curves. In fact, the requirement that the low energy should not contain any
chiral or even vector-like pairs of exotics also severely limits the class of
admissible fluxes.

This rigid structure also extends to the supersymmetry breaking sector.
Generating an appropriate value for the $\mu$ term in F-theory GUTs requires a
specific scale of supersymmetry breaking $\sqrt{F}\sim10^{8}-10^{9}$ GeV,
which is incompatible with gravity mediated supersymmetry breaking. Instead,
F-theory GUTs appear to more naturally accommodate minimal gauge mediated
supersymmetry breaking scenarios. In fact, the scalar component of the same
chiral superfield responsible for supersymmetry breaking also develops a vev,
breaking a global $U(1)$ Peccei-Quinn symmetry at a scale $f_{a}\sim10^{12}$
GeV. The associated Goldstone mode then corresponds to the QCD\ axion. In addition,
some of the common problems in gravitino cosmology are naturally evaded in F-theory GUTs.\footnote{In \cite{FGUTSCosmo}, a
scenario of leptogenesis in F-theory GUTs based on a non-minimal neutrino sector with Majorana masses in the
range of $10^{12}$ GeV was studied. We will see later that in minimal implementations of
F-theory neutrinos, the natural mass scale of neutrinos is somewhat higher. It would
be interesting to study the associated leptogenesis scenario.}

As the above discussion should make clear, the framework of F-theory GUTs is
surprisingly rigid. Nevertheless, it is in principle possible to introduce
matter content and fields in F-theory models to engineer ever more elaborate
extensions of the MSSM. Given this range of possibilities, we shall focus our
attention on vacua with a \textit{minimal} number of additional geometric and
field theoretic ingredients required to obtain phenomenologically viable low
energy physics.

It turns out that these minimal ingredients are frequently sufficient for
reproducing more detailed features of the MSSM. For example, as shown in
\cite{HVCKM}, minimal realizations of $SU(5)$ F-theory GUTs\ --- with the
minimal number of curves and interaction points necessary for compatibility
with the interactions of the MSSM\ --- automatically contain rank one Yukawa
matrices which receive small corrections due to the presence of the ubiquitous
background hyperflux. More precisely, the hierarchical structure of the
CKM\ matrix further requires the interaction points for the $\mathbf{5}%
_{H}\times\mathbf{10}_{M}\times\mathbf{10}_{M}$ and $\mathbf{\overline{5}}%
_{H}\times\mathbf{\overline{5}}_{M}\times\mathbf{10}_{M}$ couplings to be
nearby, suggestive of a higher unification structure. We will revisit this point
later when we present models with a single $E_{8}$ point of enhancement which
geometrically unifies \textit{all} of the interactions of the MSSM.

But as noted in \cite{BHVII,HVCKM}, there are strong reasons to suspect that
the neutrino sector of F-theory GUTs is qualitatively different. Identifying
the right-handed neutrinos in terms of modes localized on matter curves, the
fact that the right-handed neutrino is a singlet of $SU(5)$ implies that the
corresponding curve only touches the GUT seven-brane at a few distinct points.
In \cite{BHVII}, it was shown that Dirac neutrinos could be accommodated from
an exponential wave function repulsion due to the local curvature of the
GUT\ seven-brane. Moreover, it was also shown in \cite{BHVII} that by
including additional GUT\ group singlets which develop a suitable vev, it is
also possible to accommodate Majorana masses. On the other hand, both of these
scenarios are somewhat non-minimal in that they require the presence of an
additional physical input, such as a particular exponential hierarchy in the
Dirac case, or a new GUT group singlet with a suitable vev in the Majorana
case. In this paper we show that even without introducing a new scale, or a
new set of fields which develop a suitable vev, the geometry of F-theory GUTs
already naturally contains a phenomenologically viable neutrino sector.

\subsubsection{Local Models and Normal Curves}

One of the important advantages of local F-theory GUT models is that some
features pertaining to Planck scale physics can be deferred to a later stage
of analysis. Indeed, this is possible precisely because the dynamics of the
theory localizes near the subspace wrapped by the GUT seven-brane. On the other
hand, by including fields such as right-handed neutrinos which localize on
curves normal to the GUT seven-brane, it may at first appear that such modes
cannot be treated consistently in the context of a local model. As we now
explain, such normal curves can indeed form part of a well-defined local
model. As such, they can be consistently decoupled from Planck scale physics.

Rather than present a general analysis, we discuss an illustrative example.
Consider a local model of F-theory where the threefold base $B_{3}$ is given
as an ALE space fibered over a base $\mathbb{P}_{b}^{1}$. Although the ALE
space is non-compact, it contains a number of homologically distinct fiber
$\mathbb{P}^{1}$'s, which we label as $\mathbb{P}_{(1)}^{1},...,\mathbb{P}_{(n)}^{1}$.
$B_{3}$ defines a local model with compact surfaces defined by the $\mathbb{P}_{(i)}^{1}$'s fibered over the base
$\mathbb{P}_{b}^{1}$. The pairwise intersection of two such surfaces will
occur at a point in the ALE space which is fibered over $\mathbb{P}_{b}^{1}$.
Identifying one such surface as the one wrapped by the GUT
seven-brane, it follows that in this local model, there are compact curves inside the GUT
seven-brane given by a point in the ALE space fibered over $\mathbb{P}_{b}^{1}$. The model
also contains compact normal curves corresponding to fiber $\mathbb{P}_{(i)}^{1}$'s which
intersect the GUT seven-brane at a point. Hence, modes localized on such normal curves
can be consistently defined while remaining decoupled from Planck
scale physics. Although we do not do so here, it would be interesting to study
this more general class of local models by extending the analysis presented in
\cite{BHVI}.

\subsection{$U(1)_{PQ}$ and Neutrinos \label{U1PQNEUT}}

Selection rules in string based
constructions can sometimes forbid interaction terms in the low energy theory.
In the specific context of F-theory GUTs, the $U(1)_{PQ}$ symmetry plays an
especially prominent role in that it forbids a bare $\mu$ and $B\mu$ term in
the low energy theory. Indeed, $U(1)_{PQ}$ symmetry breaking and supersymmetry
breaking are tightly correlated in the deformation away from gauge mediation
found in \cite{HVGMSB}. However, as we now explain, the presence of this
symmetry can also forbid necessary interaction terms in the neutrino sector.
After presenting this obstruction, we show that there is in fact a unique
alternative $U(1)_{PQ}$ compatible with a Majorana scenario.

\subsubsection{Review of $E_{6}$ and $U(1)_{PQ}$}

\label{e6PQ}

An interesting feature of GUTs is the presence of higher rank symmetries.
Indeed, these symmetries can forbid otherwise problematic interaction terms.
For example, in the context of the MSSM, it is quite natural to posit the
existence of a global $U(1)_{PQ}$ symmetry under which the Higgs up and Higgs
down have respective $U(1)_{PQ}$ charges $q_{H_{u}}$ and $q_{H_{d}}$. Provided
that $q_{H_{u}}+q_{H_{d}}\neq0$, this forbids the bare $\mu$-term:
\begin{equation}
\mu H_{u}H_{d}\text{,}%
\end{equation}
thus providing a partial explanation for why $\mu$ can be far smaller than the
GUT scale. Since the Higgs fields interact with the MSSM\ superfields, the
presence of this symmetry then requires that all of the fields of the MSSM are
charged under this symmetry.

In the context of F-theory GUTs, correlating the value of the $\mu$ term with
supersymmetry breaking is achieved through the presence of the higher
dimension operator:%
\begin{equation}
L_{eff}\supset\int \rd^{4}\theta\frac{X^{\dag}H_{u}H_{d}}{\Lambda_{\text{UV}}}\text{,}%
\end{equation}
where in the above, $X$ is a chiral superfield which localizes on a matter
curve normal to the GUT seven-brane. Here, the $X$, $H_{u}$ and $H_{d}$ curves
form a triple intersection and the above operator originates from integrating
out Kaluza-Klein modes on the curve where $X$ localizes. When $X$ develops a
supersymmetry breaking vev:%
\begin{equation}
\left\langle X\right\rangle =x+\theta^{2}F_{X}\text{,}%
\end{equation}
this induces an effective $\mu$ term of order:%
\begin{equation}
\mu\sim\frac{\overline{F_{X}}}{\Lambda_{\text{UV}}}\text{.}%
\end{equation}
As estimated in \cite{HVGMSB}, using the fact that $\Lambda_{\text{UV}}\lsim M_{GUT}$, generating a value for the $\mu$ term near the scale of electroweak symmetry breaking
requires $\sqrt{F_{X}}\sim10^{8}-10^{9}$ GeV \cite{HVGMSB}. In this context,
the $U(1)_{PQ}$ symmetry can be identified with a linear combination of the
$U(1)$ symmetries present on the seven-branes which intersect the GUT
seven-brane. This necessarily requires that $X$ be charged under $U(1)_{PQ}$
with charge:%
\begin{equation}
q_{X}=q_{H_{u}}+q_{H_{d}}\text{.}%
\end{equation}

As explained in \cite{HVGMSB}, this type of structure is quite natural in the
context of F-theory GUTs and is in fact compatible with an underlying $E_{6}$
structure. Indeed, decomposing the $\mathbf{27}$ and $\mathbf{\overline{27}}$
of $E_{6}$ into irreducible representations of $SO(10)\times U(1)_{PQ}$
yields:%
\begin{align}
E_{6}  &  \supset SO(10)\times U(1)_{PQ}\\
\mathbf{27}  &  \rightarrow\mathbf{1}_{4}+ \mathbf{10}_{-2}+ \mathbf{16}_{1}\\
\mathbf{\overline{27}}  &  \rightarrow\mathbf{1}_{-4}+ \mathbf{10}_{2}+
\mathbf{\overline{16}}_{-1}\text{.}%
\end{align}
The MSSM\ chiral matter transform in the $\mathbf{16}_{1}$, while the Higgs
fields transform in the $\mathbf{10}_{-2}$. In addition, $X$ transforms in the
$\mathbf{1}_{-4}$. This structure is also compatible with gauge mediated
supersymmetry breaking, with the messenger fields transforming in the
$\mathbf{10}_{2}$. In this context, the $U(1)_{PQ}$ charges of the various
fields are:%
\begin{equation}%
\begin{tabular}
[c]{|c|c|c|c|c|c|c|c|}\hline
& $X$ & $Y$ & $Y^{\prime}$ & $H_{u}$ & $H_{d}$ & $\mathbf{10}_{M}$ &
$\mathbf{\overline{5}}_{M}$\\\hline
$U(1)_{PQ}$ & $-4$ & $+2$ & $+2$ & $-2$ & $-2$ & $+1$ & $+1$\\\hline
\end{tabular}
\label{U1orig}%
\end{equation}
where in the above, $Y$ and $Y^{\prime}$ denote the messenger fields of the
gauge mediation sector. In addition to forbidding a bare $\mu$ term, a
$\mathbb{Z}_{2}$ subgroup of $U(1)_{PQ}$ can naturally be identified with
matter parity of the MSSM. Indeed, by inspection of the above charges, note
that the charges of the MSSM\ chiral matter are all odd, while the Higgs
fields are even.

The choice of charge assignments obtained by embedding all matter fields in
representations of $E_{6}$ is problematic for neutrino models with a Majorana
mass term which is induced by the F-term $(H_{u}L)^{2}/\Lambda_{\text{UV}}$. Indeed, under the charge assignments of line
(\ref{U1orig}), this operator has charge $-2$. While it is tempting to argue
that a suitable vev for the $X$ field could generate such a term from a higher
dimension operator, note that a non-zero vev for $X$ will simply break
$U(1)_{PQ}$ to the discrete subgroup $\mathbb{Z}_{4}$. Since the operator
$(H_{u}L)^{2}/\Lambda_{\text{UV}}$ is not invariant under this discrete subgroup, we conclude that
compatibility with a Majorana mass term scenario requires incorporating
another GUT\ group singlet with charge $\pm2$. Once this singlet develops a
vev, it is possible to consider models which include this higher dimension
operator. This is somewhat non-minimal, however, so in keeping with the
general philosophy espoused in this paper, we shall seek an alternative
scenario which does not require the presence of an additional symmetry
breaking sector, the sole purpose of which is to solve a single problem.

\subsubsection{Generalizing $U(1)_{PQ}$}

\label{PQalt}

In a broader context, it is possible to consider more general $U(1)_{PQ}$
charge assignments. We now show that compatibility with $SU(5)$ GUT structures imposes strong
restrictions on possible charge assigments. We find that there is essentially a unique
alternative $U(1)_{PQ}$ given by the Abelian factor of $SU(5)\times
U(1)\subset SO(10)$ which is compatible with the requirements of both
supersymmetry breaking and the existence of the operator $(H_{u}L)^{2}/\Lambda_{\text{UV}}$.

To establish this result, we begin by asking more generally what $U(1)_{PQ}$
charge assignments are compatible with the interaction terms of the MSSM.
Assuming that all fields in the $\mathbf{10}_{M}$ and $\mathbf{\overline{5}%
}_{M}$ have respective $U(1)_{PQ}$ charges $q_{10}$ and $q_{\overline{5}}$,
the interaction terms ${\mathbf{5}_{H}\times\mathbf{10}_{M}\times
\mathbf{10}_{M}}$ and $\mathbf{\overline{5}}_{H} \times\mathbf{\overline{5}%
}_{M}\times\mathbf{10}_{M}$ are compatible with $U(1)_{PQ}$ provided:%
\begin{align}
q_{10}  &  =-\frac{1}{2}q_{H_{u}}\\
q_{\overline{5}}  &  =-q_{H_{d}}-q_{10}=-q_{H_{d}}+\frac{1}{2}q_{H_{u}%
}\text{.}%
\end{align}
If we now demand that the operator $(H_{u}L)^{2}/\Lambda_{\text{UV}}$ is invariant under
$U(1)_{PQ}$, we also find:%
\begin{equation}
q_{\overline{5}}+q_{H_{u}}=0\text{.}%
\end{equation}
Solving for all PQ\ charge assignments yields:%
\begin{equation}%
\begin{tabular}
[c]{|c|c|c|c|c|c|c|c|}\hline
& $X$ & $Y$ & $Y^{\prime}$ & $H_{u}$ & $H_{d}$ & $\mathbf{10}_{M}$ &
$\mathbf{\overline{5}}_{M}$\\\hline
$U(1)_{PQ}^{\prime}$ & $+5$ & $-2$ & $-3$ & $+2$ & $+3$ & $-1$ & $-2$\\\hline
\end{tabular}
\label{Uprime}%
\end{equation}
up to an overall common rescaling of all charges.

It is quite remarkable that this structure is in fact compatible with the
representation theory of $SO(10)$. Indeed, decomposing the $\mathbf{16}$,
$\mathbf{\overline{16}}$ and $\mathbf{10}$ of $SO(10)$ into irreducible
representations of $SU(5)\times U(1)$ yields:%
\begin{align}
SO(10)  &  \supset SU(5)\times U(1)\\
\mathbf{16}  &  \rightarrow\mathbf{1}_{-5}+\mathbf{\overline{5}}%
_{+3}+\mathbf{10}_{-1}\\
\mathbf{\overline{16}}  &  \rightarrow\mathbf{1}_{+5}+\mathbf{5}_{-3}+
\mathbf{\overline{10}}_{+1}\\
\mathbf{10}  &  \rightarrow\mathbf{5}_{+2}+ \mathbf{\overline{5}}_{-2}\text{.}%
\end{align}
Note that in contrast to the conventional assignments within the $\mathbf{16}$
of $SO(10)$, now, $H_{d}\in\mathbf{16}$ and $\mathbf{\overline{5}}_{M}%
\in\mathbf{10}$. While this may seem anti-thetical from the perspective of
grand unification, one of the important features of F-theory GUTs is that
locally, the chiral matter can organize into the $\mathbf{16}$, although in
the global geometry, this identification is ambiguous. Indeed, the
localization of interaction terms at points of the geometry can naturally
accommodate both the presence of higher unification structures, as well as the
identification of this new $U(1)$ symmetry.

In addition to global symmetries, it is also important to check that matter
parity remains intact. In fact, with respect to these new charge assignments,
note that nothing forbids the interaction term $H_{u}L$. Indeed, because the
global $U(1)_{PQ}$ symmetry is compatible with $(H_{u}L)^{2}/\Lambda_{\text{UV}}$, it cannot
forbid $H_{u}L$. However, additional discrete symmetries of the geometry can
in principle forbid such interaction terms, and as proposed in \cite{BHVII},
could potentially be identified with matter parity. We will return to this
point in the context of the Kaluza-Klein seesaw in section \ref{KKMAJ}.

\subsubsection{F-theory Neutrinos and the LHC}

One of the distinctive features of F-theory GUTs is that integrating out the
gauge boson of the anomalous $U(1)_{PQ}$ gauge theory in general shifts the
soft mass terms of the scalars away from the value predicted in minimal gauge
mediated supersymmetry breaking so that the soft scalar mass squared of an
MSSM\ superfield $\Phi$ obeys the messenger scale relation:%
\begin{equation}
m_{\Phi}^{2}=\widehat{m}_{\Phi}^{2}+4\pi\alpha_{PQ}e_{X}e_{\Phi}\left\vert
\frac{F_{X}}{M_{U(1)_{PQ}}}\right\vert ^{2}\text{,}%
\end{equation}
where $\widehat{m}_{\Phi}$ denotes the soft mass in minimal gauge mediation,
$M_{U(1)_{PQ}}$ denotes the mass of the anomalous $U(1)_{PQ}$ gauge boson and
$\alpha_{PQ}$ the associated fine structure constant of the gauge theory. In
\cite{HVGMSB}, a particular choice of $U(1)_{PQ}$ charges compatible with an
$E_{6}$ unification structure was considered. This choice leads to a
predictive deformation away from gauge mediation, with potentially measurable
consequences at the LHC \cite{HVGMSB,HVLHC,HKSV}. Here, we see that
considerations from neutrino physics can prefer a different choice of charge
assignments inducing a different shift in the soft masses. Thus, determining
the form of the mass shift constrains the form of the neutrino sector, and the
converse holds as well! More generally, note that we have identified the two
Abelian factors in the breaking pattern:%
\begin{equation}
E_{6}\supset SO(10)\times U(1)_{b}\supset SU(5)\times U(1)_{a}\times U(1)_{b}%
\end{equation}
as potential $U(1)_{PQ}$ symmetries. However, Majorana neutrino masses
single out $U(1)_{a}$ as the PQ symmetry.

Having shown that there is in principle no obstruction to accommodating
neutrino physics in F-theory GUT scenarios, we now proceed to study the
geometry of such configurations. The suggestive link between the suppression
scale $\Lambda_{\text{UV}}$ and the higher dimension operator $(H_{u}%
L)^{2}/\Lambda_{\text{UV}}$ indicates the presence of GUT scale physics, and so
we now turn to Majorana neutrinos in F-theory GUTs.

\section{Majorana Neutrinos and the Kaluza-Klein Seesaw \label{KKMAJ}}

In this section we study minimal implementations of the Majorana scenario in
F-theory GUTs. This amounts to determining geometries which contain the
terms:
\begin{equation}
W_{eff}\supset\lambda_{ij}^{(\nu)}\frac{\left(  H_{u}L^{i}\right)  \left(
H_{u}L^{j}\right)  }{\Lambda_{\text{UV}}}\text{.} \label{Majorana}%
\end{equation}
As explained in subsection \ref{NeutrinoANDUV}, this type of operator can
naturally originate from a type I\ seesaw mechanism with a superpotential term
of the form:%
\begin{equation}
W\supset y_{iJ}^{(\nu)}\cdot H_{u}L^{i}N_{R}^{J}+M_{IJ}\cdot N_{R}^{I}%
N_{R}^{J}%
\end{equation}
for some number of right-handed neutrinos $N_{R}^{I}$ labeled by the index
$I$. Here, $M_{IJ}$ denotes the entries of a Majorana mass matrix. In matrix
notation, the coupling $\lambda_{ij}^{(\nu)}/\Lambda_{\text{UV}}$ is then
given by:%
\begin{equation}
\frac{\lambda_{(\nu)}}{\Lambda_{\text{UV}}}=y_{(\nu)}\cdot \frac{1}{M} \cdot
y_{(\nu)}^{T}\text{.}%
\end{equation}
For simplicity, in this section we exclusively consider scenarios where the
right-handed neutrinos localize on curves. Indeed, the implementation of bulk
mode right-handed neutrinos appears less straightforward in the context of
F-theory GUTs, although we shall briefly comment on this possibility later in the context of Dirac neutrino models.

Neutrino physics in F-theory GUTs has been discussed previously in
\cite{BHVII}. In that context, the Yukawa coupling $y$ turned out to be
somewhat smaller than an order one number. Moreover, upon estimating the
expected vev of GUT singlets to be $10^{12}$ GeV, it was argued that an
appropriate GUT\ group singlet $P$ could generate the requisite Majorana mass
$M_{\text{maj}}\sim\left\langle P\right\rangle $. Similar seesaw mechanisms
induced by higher dimension operators of flipped $SU(5)$ F-theory GUTs have
also been studied in \cite{BHVII,Jiang:2008yf}.

In a certain sense, however, such scenarios must be viewed as incomplete until
we specify how $P$ develops a suitable vev. While suitable brane constructions
are likely available to achieve this goal, in this section we instead
investigate minimal constructions which do not require additional low energy
field theory dynamics. To this end, we demonstrate that in F-theory GUTs, it
is also quite natural to treat right-handed neutrinos as Kaluza-Klein modes.
Thus, rather than specify a means by which such fields develop a mass, the
fact that they are massive modes is already present, by construction. See for example, \cite{ConlonNeut} for other scenarios which attempt to realize a seesaw mechanism using heavy modes of the compactification.

In subsection \ref{RRN} we analyze the effective field theory of the Kaluza-Klein
seesaw. Since the mass term for the Kaluza-Klein modes pairs $N_{R}$ with
$N_{R}^{c}$, while only $N_{R}$ directly couples to the MSSM, we explain why
the Kaluza-Klein seesaw requires an identification between $N_{R}$ and
$N_{R}^{c}$. We then turn to explicit realizations of the Kaluza-Klein seesaw
in F-theory. We first give in \ref{TOYGEO} a simple realization where the
identification comes from the geometry itself. We then provide more natural
F-theory realizations where the identification is provided by monodromy in
seven-brane configurations. The main properties of such monodromies are
reviewed in subsection \ref{MONOD}. In subsection \ref{TOY} we present a toy model based on an $SU(7)$ interaction
point which implements the Kaluza-Klein seesaw. This example turns out to be
only semi-realistic in that it requires the Higgs up and lepton doublets to
localize on the same matter curve. In subsection \ref{BIGBOY} we present a
more refined example based on an $E_{8}$ interaction point which accommodates
a richer class of interaction terms. We also provide a complete model where
all interactions of the MSSM unify in $E_{8}$.

\subsection{Right-Handed Neutrinos as Kaluza-Klein Modes\label{RRN}}

At a conceptual level, it is somewhat ambiguous to interpret the right-handed
neutrinos of a GUT\ scale seesaw as zero modes. Indeed, assuming that
$\lambda_{ij}^{(\nu)}$ has an order one eigenvalue, the resulting Majorana
mass scale is quite close to the Kaluza-Klein scale. This observation opens up
the possibility that \emph{right-handed neutrinos may in fact be Kaluza-Klein
modes}.\footnote{We note that the idea of using Kaluza-Klein modes as
right-handed neutrinos has appeared for instance in \cite{Antoniadis:2002qm}
(see also references therein), albeit in a different context.} From this
perspective, it becomes unclear whether any right-handed neutrino zero modes
are in fact necessary.

Although seemingly quite simple, there is one immediate objection to such a
proposal. Indeed, when right-handed neutrinos localize on matter curves, the
natural expectation is that the six-dimensional field transforms as a
bifundamental under the gauge groups of two distinct seven-branes. The mass
term for the Kaluza-Klein right-handed neutrinos pairs $N_{R}$ and $N_{R}^{c}$
so that the actual interaction term is of the schematic form:
\begin{equation}
W\supset y\cdot H_{u}LN_{R}+M_{N}^{KK}\cdot N_{R}N_{R}^{c}\text{,}%
\end{equation}
where for simplicity we have included the contribution from a single
generation of lepton doublets. In this subsection we will suppress all such
generational indices to avoid cluttering the discussion. To obtain the higher
dimension operator of \eqref{Majorana} after integrating out the
Kaluza-Klein modes, we would also need a coupling of the form $H_{u}LN_{R}%
^{c}$. Note, however, that the gauge symmetries of the other seven-branes will
forbid such a term!

\begin{figure}[ptb]
\begin{center}
\includegraphics[
height=2.7735in,
width=5.3566in
]%
{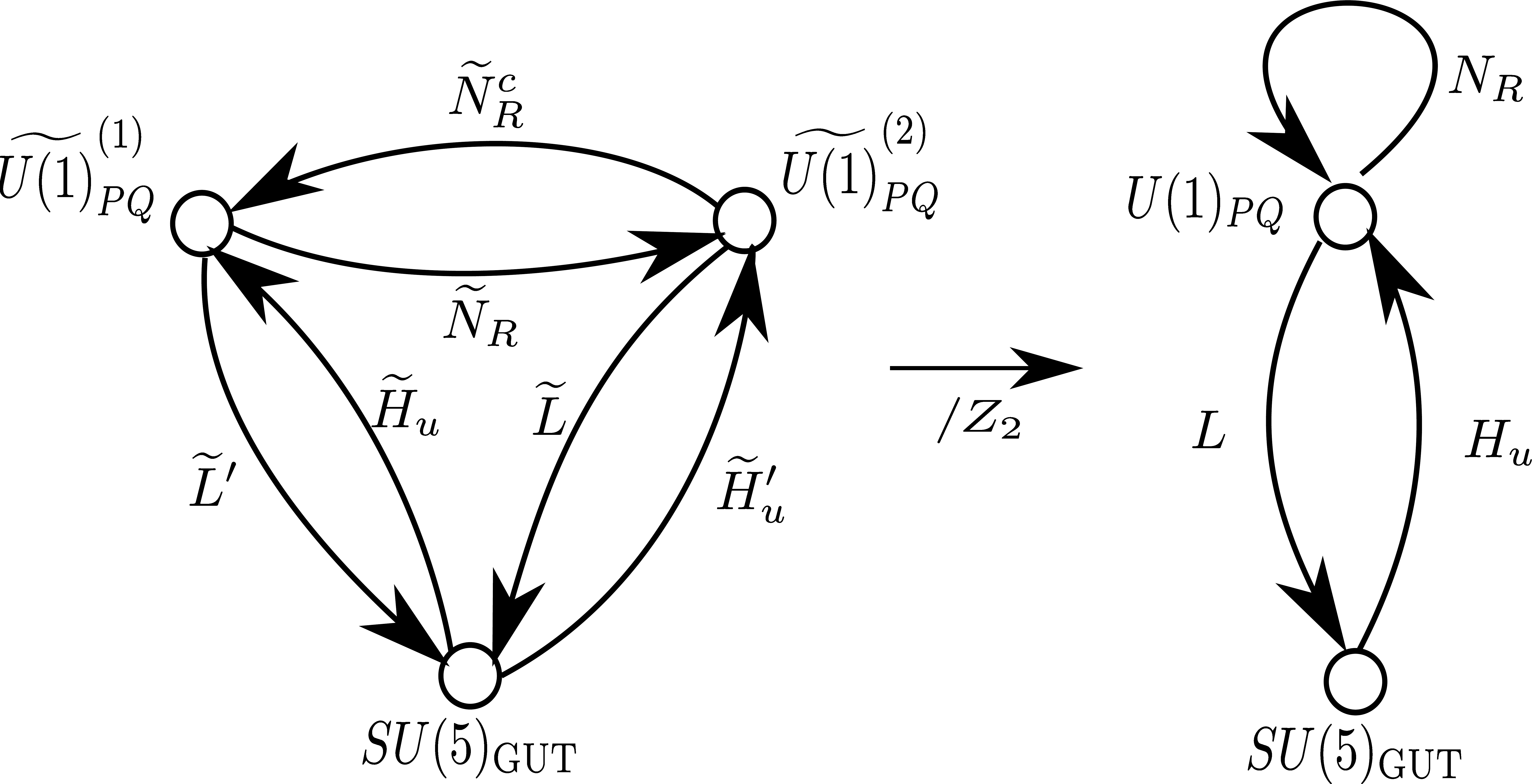}%
\caption{Quiver diagram of the field theory associated to the Kaluza-Klein seesaw in the covering theory (left) and the quotient theory (right).}
\label{z2quiver}
\end{center}
\end{figure}

The problematic nature of this coupling stems from the fact that the
right-handed neutrino transforms in a complex representation of the
seven-branes. There is, however, a more general possibility in F-theory due to
the interplay between geometric and field theoretic degrees of freedom. In
particular, an appropriate discrete group action can identify the resulting
seven-branes, so that $N_{R}$ and $N_{R}^{c}$ transform in a real
representation. This effect can be analyzed purely in field theoretical terms
by passing to a covering theory with additional fields which are to be
identified in a suitably defined quotient theory. In the simplest example, we
shall be interested in a covering theory with matter fields $\widetilde{L}$,
$\widetilde{L}^{\prime}$, $\widetilde{H}_{u}$, $\widetilde{H}_{u}^{\prime}$,
$\widetilde{N}_{R}$ and $\widetilde{N}_{R}^{c}$. The quotient theory is
defined by quotienting by the $\mathbb{Z}_{2}$ identification:%
\begin{equation}
\widetilde{L}\leftrightarrow\widetilde{L}^{\prime}\text{, }\widetilde{H}%
_{u}\leftrightarrow\widetilde{H}_{u}^{\prime}\text{, }\widetilde{N}%
_{R}\leftrightarrow\widetilde{N}_{R}^{c}\text{.}\label{e:iden}%
\end{equation}
Physically, the $\mathbb{Z}_{2}$ group action corresponds to an interchange of
the seven-branes under which the various bifundamentals are charged. A variant
of the seesaw mechanism is present in the covering theory provided that
$\widetilde{N}_{R}$ couples to $\widetilde{H}_{u}$ and $\widetilde{L}$, with a
similar coupling between $\widetilde{N}_{R}^{c}$ and $\widetilde{H}%
_{u}^{\prime}$ and $\widetilde{L}_{u}^{\prime}$ so that:%
\begin{equation}
\widetilde{W}\supset\widetilde{y}\cdot\widetilde{H}_{u}\widetilde{L}%
\widetilde{N}_{R}+\widetilde{y}^{\prime}\widetilde{H}_{u}^{\prime}%
\widetilde{L}^{\prime}\widetilde{N}_{R}^{c}+\widetilde{M}_{N}^{KK}%
\cdot\widetilde{N}_{R}\widetilde{N}_{R}^{c}\label{CoverPot}%
\end{equation}
Integrating out $\widetilde{N}_{R}$ and $\widetilde{N}_{R}^{c}$ then generates
an effective superpotential term:%
\begin{equation}
\widetilde{W}_{eff}\supset\widetilde{y}\widetilde{y}^{\prime}\cdot
\frac{(\widetilde{H}_{u}\widetilde{L})(\widetilde{H}_{u}^{\prime}\widetilde
{L}^{\prime})}{\widetilde{M}_{N}^{KK}}\text{.}%
\end{equation}
Note that in order for $\widetilde{W}_{eff}$ to be invariant under the
exchange symmetry, we must have $\widetilde{y}=\widetilde{y}^{\prime}$.
Descending to the quotient theory, the corresponding field theory will then
contain the effective term:%
\begin{equation}
W_{eff}\supset y^{2}\cdot\frac{(H_{u}L)(H_{u}L)}{M_{N}^{KK}}\text{,}%
\label{Weffquot}%
\end{equation}
in the obvious notation. See figure \ref{z2quiver} for a depiction of the
quiver theory associated with these interaction terms in both the covering and
quotient theory. This type of mechanism clearly extends to multiple
generations, and will therefore induce masses for the remaining neutrinos.

It is in fact possible to generalize the field theory construction above to
Kaluza-Klein seesaw models where we quotient by the action of a more general
finite group $\mathfrak{S}$, such that $N_{R}$ and $N_{R}^{c}$ are identified. In the covering theory, the action of the finite group $\mathfrak{S}$ will map the matter fields to one another. Let us group the matter fields in the covering theory in terms of orbits under the action of $\mathfrak{S}$. Consider a covering theory with matter content:
\begin{align}
Orb(\widetilde{H}_{u})  & \equiv\left\{  \sigma(\widetilde{H}_{u})|\sigma
\in\mathfrak{S}\right\} , \\
Orb(\widetilde{L})  & \equiv\left\{  \sigma(\widetilde{L})|\sigma
\in\mathfrak{S}\right\} , \\
Orb(\widetilde{N}_{R})  & \equiv\left\{  \sigma(\widetilde{N}_{R})|\sigma
\in\mathfrak{S}\right\} , \\
Orb(\widetilde{N}_{R}^{c})  & \equiv\left\{  \sigma(\widetilde{N}_{R}%
^{c})|\sigma\in\mathfrak{S}\right\}  \text{.}%
\end{align}
In the quotient theory, all fields belonging to the same orbit will be
identified. In particular, since we want $N_R$ and $N_R^c$ to be identified by the action of $\mathfrak{S}$, we must require that $Orb(\widetilde{N}_{R})=Orb(\widetilde{N}_{R}^{c})$. To realize the Kaluza-Klein seesaw, the superpotential of the covering theory must contain terms
of the form:%
\begin{equation}
\widetilde{W}\supset\widetilde{y}_{ijk}\cdot\widetilde{H}_{u}^{i}\widetilde
{L}^{j}\widetilde{N}_{R}^{k}+\widetilde{y}_{i^{\prime}j^{\prime}k}%
\widetilde{H}_{u}^{i^{\prime}}\widetilde{L}^{j^{\prime}}(\widetilde{N}_{R}%
^{c})^{k}+\widetilde{M}_{k}\cdot\widetilde{N}_{R}^{k}(\widetilde
{N}_{R}^{c})^{k},
\label{e:couplings}
\end{equation}
where the indices $i,j,i',j',k$ label elements in the group orbits such that the associated terms form gauge invariant combinations.

In the context of F-theory, the finite group $\mathfrak{S}$ will descend from
a geometrical symmetry of the compactification. Since the matter fields
localize on curves, this geometric identification will also identify
curves in the covering theory. Thus, distinct matter curves in the quotient theory must come from curves in the covering theory lying in distinct orbits of $\mathfrak{S}$. For example, compatibility with doublet triplet splitting
requires the Higgs up and lepton doublet to localize on distinct curves. Hence,
these fields must lie in \textit{distinct} orbits in the covering theory.

We now turn to explicit realizations of the Kaluza-Klein seesaw in F-theory.%

%TCIMACRO{\FRAME{ftbpF}{5.3566in}{2.7735in}{0pt}{}{}{z2quiver.pdf}%
%width 5.3566in;  height 2.7735in;  depth 0pt;  original-width 30.8444in;
%original-height 15.9298in;  cropleft "0";  croptop "1";  cropright "1";
%cropbottom "0";  filename 'z2quiver.pdf';file-properties "XNPEU";}} }%
%BeginExpansion
%EndExpansion%

\subsection{A Geometric Realization of the Kaluza-Klein Seesaw\label{TOYGEO}}

In this subsection we present a first realization of the Kaluza-Klein seesaw
by directly interpreting the ingredients of figure \ref{z2quiver} as an
intersecting seven-brane configuration which admits a ${\mathbb{Z}}_{2}$ group
action. Let us first study the covering theory. In terms of the local
geometry, this can be modelled in terms of two interaction points where
$SU(5)$ enhances to $SU(7)$ so that the $\widetilde{H}_{u}\widetilde
{L}\widetilde{N}_{R}$ localizes at a point $P$, while $\widetilde{H}%
_{u}^{\prime}\widetilde{L}^{\prime}\widetilde{N}_{R}^{c}$ localizes at a point
$P^{\prime}$. Geometrically, the required quotienting procedure amounts to the
following identification of curves:%
\begin{equation}
\Sigma_{\widetilde{L}}\leftrightarrow\Sigma_{\widetilde{L}^{\prime}}%
,\Sigma_{\widetilde{H}_{u}}\leftrightarrow\Sigma_{\widetilde{H}_{u}^{\prime}%
},\Sigma_{\widetilde{N}_{R}}\leftrightarrow\Sigma_{\widetilde{N}_{R}}\text{,}%
\end{equation}
so that the Higgs and leptons of the covering theory are correctly identified,
while the neutrino curve maps to itself. Note that the $%
%TCIMACRO{\U{2124} }%
%BeginExpansion
\mathbb{Z}
%EndExpansion
_{2}$ group action will in general \textit{not }leave the curve $\Sigma
_{\widetilde{N}_{R}}$ fixed pointwise, even though it is mapped to itself. The
intersection points $P$ and $P^{\prime}$ are mapped to each other.

\begin{figure}
[ptb]
\begin{center}
\includegraphics[
width=6.4in
]%
{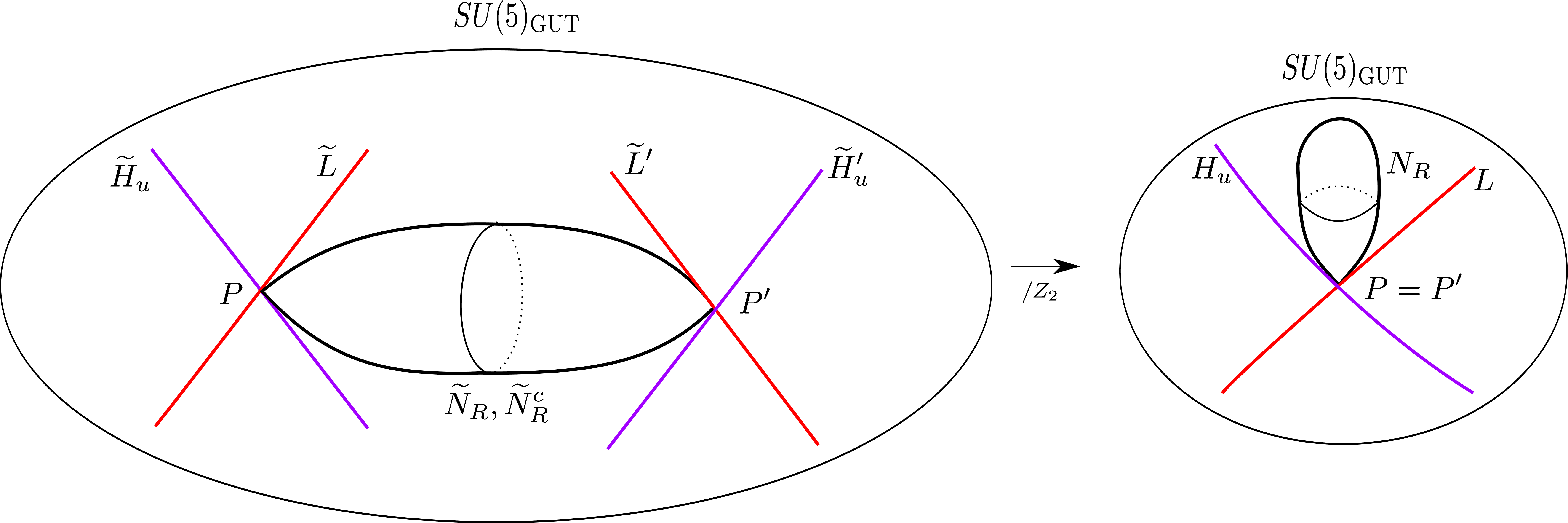}%
\caption{Depiction of a minimal implementation of the Kaluza-Klein seesaw in
which the fields of the covering theory are identified in the quotient theory.
This geometrical action also identifies the two interaction points of the covering theory.}%
\label{z2P}%
\end{center}
\end{figure}

The covering quiver of figure \ref{z2quiver} contains three distinct gauge
group factors, which we identify with three seven-branes wrapping divisors in
the threefold base. Labelling these divisors as $\Gamma_{GUT}$, $\Gamma_{+}$,
and $\Gamma_{-}$, the corresponding matter curves are contained in the
pairwise intersections:%
\begin{align}
\Sigma_{\widetilde{H}_{u}},\Sigma_{\widetilde{L}^{\prime}}  &  \subset
\Gamma_{+}\cap\Gamma_{GUT}\\
\Sigma_{\widetilde{H}_{u}^{\prime}},\Sigma_{\widetilde{L}},  &  \subset
\Gamma_{-}\cap\Gamma_{GUT}\\
\Sigma_{\widetilde{N}}  &  \subset\Gamma_{+}\cap\Gamma_{-}\text{.}%
\end{align}
Note that while the divisors wrapped by the seven-branes are irreducible, the
intersection curves may be reducible. Therefore, in this context the lepton
and Higgs can in principle live on distinct curves, and there is \emph{a
priori} no obstruction to realizing doublet triplet splitting, although this
must be checked explicitly in a given geometric model. The exchange symmetry
then identifies the divisors $\Gamma_{+}$ and $\Gamma_{-}$, while
$\Gamma_{GUT}$ remains invariant:%
\begin{equation}
\Gamma_{+}\leftrightarrow\Gamma_{-},\Gamma_{GUT}\leftrightarrow\Gamma
_{GUT}\text{.}%
\end{equation}
See figure \ref{z2P} for a depiction of the covering and quotient theory associated with this realization of the Kaluza-Klein seesaw mechanism.
%TCIMACRO{\FRAME{ftbpFU}{6.6763in}{2.2269in}{0pt}{\Qcb{CAPTION}}{\Qlb{z2P}%
%}{z2p.pdf}{\special{ language "Scientific Word";  type "GRAPHIC";
%width 6.6763in;  height 2.2269in;  depth 0pt;  original-width 54.8974in;
%original-height 18.2233in;  cropleft "0";  croptop "1";  cropright "1";
%cropbottom "0";  filename 'z2P.pdf';file-properties "XNPEU";}} }%
%BeginExpansion
%EndExpansion

As an example which realizes this type of configuration, we consider a
configuration of colliding $A$-type singularities. To this end, let $u,v,z$
denote three local coordinates of the threefold base such that $z=0$ is the
location of the GUT\ seven-brane. The $z$ coordinate labels the direction
normal to the seven-brane, and so the right-handed neutrino curve will be
parameterized by this coordinate. In this language, the $\widetilde{H}%
_{u}\widetilde{L}\widetilde{N}_{R}$ interaction point $P$ descends from the
codimension three enhancement in the singularity type:%
\begin{equation}
\text{near }P:y^{2}=x^{2}+z^{5}(u-v-a)(u+v-a)\text{.}%
\end{equation}
This corresponds to an $SU(5)$ GUT\ seven-brane at $z=0$, and locally defined
$U(1)$ factors at $u-v-a=0$ and $u+v-a=0$. Labelling a patch containing $P$ of
a given subvariety $V$ as $\mathcal{U}_{P}(V)$, the local profile of the
divisors, matter curves, and local interaction point are:%
\begin{align}
\mathcal{U}_{P}\left(  \Gamma_{GUT}\right)   &  =(z=0),\nonumber\\
\mathcal{U}_{P}\left(  \Gamma_{\pm}\right)   &  =(u\pm v-a=0)\text{,}%
\nonumber\\
\mathcal{U}_{P}\left(  \Sigma_{\widetilde{H}_{u}}\right)   &  =(z=0)\cap
(u+v-a=0),\nonumber\\
\mathcal{U}_{P}\left(  \Sigma_{\widetilde{L}}\right)   &  =(z=0)\cap
(u-v-a=0),\nonumber\\
\mathcal{U}_{P}\left(  \Sigma_{\widetilde{N}}\right)   &  =(u+v-a=0)\cap
(u-v-a=0)=(u=a)\cap(v=0),\nonumber\\
\mathcal{U}_{P}(P) &  =(z=0)\cap(v=0)\cap(u=a)\text{.}%
\end{align}
Similarly, the $\widetilde{H}_{u}^{\prime}\widetilde{L}^{\prime}\widetilde
{N}_{R}^{c}$ interaction point descends from the codimension three enhancement
in the singularity type:%
\begin{equation}
\text{near }P^{\prime}:y^{2}=x^{2}+z^{5}(u-v-b)(u+v-b)\text{,}%
\end{equation}
where now we can model the local profile of the divisors, matter curves, and
interaction point as:%
\begin{align}
\mathcal{U}_{P^{\prime}}\left(  \Gamma_{GUT}\right)   &  =(z=0),\nonumber\\
\mathcal{U}_{P^{\prime}}\left(  \Gamma_{\pm}\right)   &  =(u\pm v-b=0)\text{,}%
\nonumber\\
\mathcal{U}_{P^{\prime}}\left(  \Sigma_{\widetilde{L}^{\prime}}\right)   &
=(z=0)\cap(u+v-b=0),\nonumber\\
\mathcal{U}_{P^{\prime}}\left(  \Sigma_{\widetilde{H}_{u}^{\prime}}\right)
&  =(z=0)\cap(u-v-b=0),\nonumber\\
\mathcal{U}_{P^{\prime}}\left(  \Sigma_{\widetilde{N}}\right)   &
=(u+v-b=0)\cap(u-v-b=0)=(u=b)\cap(v=0),\nonumber\\
\mathcal{U}_{P^{\prime}}\left(  P^{\prime}\right)   &  =(z=0)\cap
(v=0)\cap(u=b)\text{.}%
\end{align}
Comparing the local data defined by these two patches, we conclude that the ${\mathbb{Z}}_{2}$ group action which interchanges $P$ and $P^{\prime}$ is
given by a reflection in the $v$ coordinate and an interchange of the
parameters $a$ and $b$:%
\begin{align}
z &  \mapsto z\text{, }u\mapsto u\\
v &  \mapsto-v\text{, }a\leftrightarrow b\text{.}%
\end{align}

While this provides a simple realization of the Kaluza-Klein seesaw, it is
somewhat unsatisfactory, in the sense that it requires an additional geometric
ingredient to be added by hand. Even so, geometries with an appropriate $%
%TCIMACRO{\U{2124} }%
%BeginExpansion
\mathbb{Z}
%EndExpansion
_{2}$ can in principle be manufactured, providing a straightforward realization of the
Kaluza-Klein seesaw. Nevertheless, as we now explain, there are other geometric
realizations of the Kaluza-Klein seesaw where the finite group action comes
from monodromy around codimension three singularities. Such monodromies occur
quite generically in compactifications of F-theory, and therefore provide
another means by which to realize the Kaluza-Klein seesaw.

\subsection{Weyl Groups and Monodromies\label{MONOD}}

In F-theory compactifications there is a natural set of discrete group actions
which are especially prevalent, corresponding to monodromies around
codimension three singularities. In subsequent subsections we will present
explicit realizations of the Kaluza-Klein seesaw which use the presence of
this natural identification. In this context, both the quotient and the
covering theories will have only one interaction point, in contrast to the
geometric example of the previous subsection.

Let us start by reviewing the appearance of monodromies in F-theory seven-brane
configurations. We refer the reader to \cite{Hayashi:2009ge}
for a recent study of monodromies in compactifications of F-theory with
codimension three singularities.

To illustrate the main feature of monodromies, we consider a stack of $N$
D7-branes wrapping a hypersurface defined by the local equation $z=0$. This
corresponds to a local $A_{N-1}$ singularity:%
\begin{equation}
y^{2}=x^{2}+z^{N}\text{.}%
\end{equation}
Assuming that the geometry admits a suitable deformation to lower degree
terms, the $A_{N-1}$ can break to $A_{N-3}$ as:%
\begin{equation}
y^{2}=x^{2}+z^{N-2}(z-t_{1})(z-t_{2}),\label{specialdeform}%
\end{equation}
where the $t_{i}$ may be viewed as non-trivial polynomials in the coordinates
defined on the threefold base $B_{3}$. Equation (\ref{specialdeform}) defines
a configuration of $\left(  N-2\right)  $ D7-branes wrapping $z=0$, and a
single D7-brane wrapping each hypersurface $z=t_{i}$. In the original $SU(N)$
gauge theory, this corresponds to allowing an adjoint-valued chiral superfield
develop vevs in the Cartan of $SU(N)$ such that $SU(N)$ breaks to
$SU(N-2)\times U(1)\times U(1)\subset SU(N)$. Thus, suitable vevs in the
Cartan of the gauge group translate into deformations of the corresponding geometry.

Expanding out equation (\ref{specialdeform}), note that we can also write this
singularity as:%
\begin{equation}
y^{2}=x^{2}+z^{N-2}(z^{2}+az+b)\text{,}%
\end{equation}
with $a=-t_{1}-t_{2}$ and $b=t_{1}t_{2}$. Geometrically, however, it is now
immediate that there is a broader class of geometries where $a$ and $b$ do not
necessarily decompose in terms of polynomial $t_{i}$'s. Indeed, formally
solving for the $t_{i}$ in terms of $a$ and $b$ yields:%
\begin{align}
t_{1} &  =-\frac{a+\sqrt{a^{2}-4b}}{2}\\
t_{2} &  =-\frac{a-\sqrt{a^{2}-4b}}{2}\text{.}%
\end{align}
The presence of the branch cut structure indicates that in this more general
case, monodromy around the brane configuration will now interchange the
location of the D7-branes wrapping $z=t_{1}$ and $z=t_{2}$. In other words,
quotienting by the $%
%TCIMACRO{\U{2124} }%
%BeginExpansion
\mathbb{Z}
%EndExpansion
_{2}$ symmetry which interchanges the two branches, we obtain a single smooth
irreducible surface wrapped by a D7-brane. This more general breaking pattern
corresponds to the decomposition $SU(N-2)\times SU(2)\times U(1)\subset
SU(N)$. The branch cut structure reflects a deformation by the Cartan
subalgebra of $SU(2)\times U(1)$ modulo the Weyl group. This entire system can
be studied in terms of a covering theory with local coordinates $t_{1}$ and
$t_{2}$ subject to an overall quotient by this Weyl group. Note that we can
always study the covering theory, and then perform a suitable quotient.

We now formalize the above procedure of parameterizing deformations in terms
of directions in the Cartan modulo the Weyl group. Starting from a seven-brane
with gauge group $G_{GUT}$, we consider an intersection point where the gauge
group is enhanced by rank $r\geq2$ to $G$. Let $G_{GUT}\times G_{\bot}\subset
G$ denote a maximal subgroup. The local field theory near this intersection
point is determined by a theory of deformations of $G$ preserving $G_{GUT}$.
Such deformations are parameterized by the Cartan subalgebra of $G_{\bot}$,
which we denote by $g_{\bot}$, modulo the Weyl group of $G_{\bot}$, which we
denote by $W(G_{\bot})$ \cite{KatzMorrison}.

To analyze the action of the Weyl group, we first describe the covering theory
where deformations are parameterized by $g_{\bot}$, and then quotient by
$W(G_{\bot})$. Let $\{t_{1},\ldots,t_{r}\}$ parameterize directions in
$g_{\bot}$. Suitable vanishing loci for the $t_{i}$'s then define the
locations of enhancements in the singularity type of the F-theory
compactification. As in \cite{KatzVafa,BHVI}, to work out the matter content
of the covering theory, we proceed as follows. First, we write down the
decomposition of the adjoint of $G$ under the maximal subgroup $G_{GUT}\times
G_{\bot}\subset G$:
\begin{align}
G &  \supset G_{GUT}\times G_{\bot}\nonumber\\
\text{adjoint}(G) &  \rightarrow\bigoplus_{i}(R_{i},R_{i}^{\prime}),
\end{align}
where $R_{i}$ and $R_{i}^{\prime}$ are irreducible representations of
$G_{GUT}$ and $G_{\bot}$ respectively. Next, to each representation
$R_{i}^{\prime}$ is associated a set of weights, which are points in the dual
space $g_{\bot}^{\ast}$ to the Cartan subalgebra. The weights of the
representations $R_{i}^{\prime}$ give the $U(1)^{r}$ charges of the
decomposition of $R_{i}^{\prime}$ under the branching $G_{\bot}\supset
U(1)^{r}$ to the Cartan subgroup. We have thus obtained the matter content in
the covering theory. By duality, the weights also give linear combinations of
the Cartan parameters $\{t_{1},\ldots,t_{r}\}$, whose vanishing loci define
the matter curves where matter in the representation $R_{i}$ of $G_{GUT}$
localizes. The interactions arising at such intersection points in the
covering theory can be found by writing down gauge invariant combinations of
the matter content.

We now study the action of the Weyl group, in order to understand the quotient
theory. The Weyl group acts on the Cartan subalgebra parameterized by the
$t_{i}$'s. In terms of the geometry, this corresponds to a group action on the
vanishing loci in the geometry, so that the Weyl group $W(G_{\bot})$ will in
general identify some of the matter curves of the covering theory. What is
particularly interesting is that, as we have just seen, in studying a more
generic class of deformations of the geometry, such identifications by the
Weyl group occur \textit{generically} in compactifications of F-theory!

In fact, for a generic choice of complex structure all curves of a given Weyl
group orbit will be identified. In particular, all six-dimensional fields with
a given representation under $SU(5)_{GUT}$ would then be forced to localize on
the same matter curve in the quotient theory. This turns out to be too
constraining for us since, as we noted earlier, the Higgs and leptons (which
both descend from six-dimensional fields in the $\mathbf{5} \oplus
\overline{\mathbf{5}}$ of $SU(5)_{GUT}$) must localize on different curves in
the quotient theory. However, at the expense of losing a bit of generality, it
is also possible to consider geometries where only a subgroup of the full Weyl
group acts to produce the quotient theory. This requires a somewhat
more specific choice of complex structure in the geometry, since those
deformations of the singularity are not fully generic.

Our discussion of matter curves has been at the level of a quotienting procedure.
This raises the interesting question to what extent localized modes of
the covering theory descend to localized modes of the quotient theory. To a certain extent,
the notion of localized modes depends on the profile of the K\"ahler metric in both the
cover and quotient. As an example, consider local coordinates of a covering theory $\widetilde{x}$
and $\widetilde{y}$ with K\"ahler form given by:
\begin{equation}
\widetilde{\omega}
= i(4 |\widetilde{x}|^2 \cdot d\widetilde{x} \wedge d\overline{\widetilde{x}} + d\widetilde{y} \wedge d\overline{\widetilde{y}}).
\end{equation}
Assuming that the quotient acts by sending $\widetilde{x} \rightarrow - \widetilde{x}$ with
$\widetilde{y}$ invariant, we now make the identifications $\widetilde{x}^{2} = x$ and $\widetilde{y} = y$. The K\"ahler form of the quotient theory is then
of canonical form:
\begin{equation}
\omega
= i(dx \wedge d\overline{x} + dy \wedge d\overline{y}).
\end{equation}
The precise form of the K\"ahler form in the covering and quotient theories differ, and so will affect the extent to which the corresponding modes satisfying the Dirac equation are indeed localized along specific loci. Strictly speaking,
however, it is not necessary to specify the global profile of the K\"ahler form. Indeed, we shall often be interested in only the local profile of modes
near a given interaction point. Such effects are controlled by the local
curvature of the metric and gauge fields, and so we shall typically assume
that an appropriate notion of localization is available in such cases. Indeed, in the specific context of Majorana neutrino scenarios where we shall consider massive mode excitations anyway, the notion of localization on a matter curve is itself less well-defined. The important point, however, is that an appropriate notion of massive modes with non-vanishing profile near an interaction point is still available, and so we shall sometimes abuse terminology and refer to ``matter curves'' in such instances as well.

Having presented a general discussion of the potential applications of such
monodromies in seven-brane configurations, we now restrict our attention to
some geometric examples. As a first toy model, we consider the interaction
between $H_{u}$, $L$ and $N_{R}$ derived from an $SU(7)$ enhancement point.
Some deficiencies in this example will then be rectified when we present a
neutrino sector derived from an $E_{8}$ enhancement point.

\subsection{$SU(7)$ Toy Model\label{TOY}}

As a toy model of the Kaluza-Klein seesaw, we first consider an interaction
between the Higgs up, lepton doublet and right-handed neutrino curve which
originates from a point of enhancement to $SU(7)$ in the $SU(5)$ bulk
worldvolume theory. Here, the Higgs up and lepton doublet localize on two
curves where $SU(5)$ enhances to $SU(6)$. The right-handed neutrino localizes
on a curve which only touches the GUT seven-brane at the $SU(7)$ point of enhancement.

It turns out to be easier to consider instead the parent $U(7)$ theory,
locally Higgsed down to $U(6)$ on the matter curves, and $U(5)$ in the bulk of
the GUT\ seven-brane. The analysis is equivalent, but it maintains contact
with the perturbative IIB\ description.

In the absence of monodromies, the breaking pattern $U(7) \supset U(5) \times
U(1)_{1} \times U(1)_{2}$ determines three D7-branes with gauge groups $U(5)$,
$U(1)_{1}$ and $U(1)_{2}$ wrapping distinct complex surfaces in the threefold
base $B_{3}$. A six-dimensional bifundamental localizes at each pairwise
intersection of the seven-branes. Two of these bifundamentals localize on
curves inside of the GUT seven-brane, and may therefore be identified with
$H_{u}$ and $L$. The final bifundamental is neutral under the $U(5)$ factor
and as a GUT group singlet localizes on a curve normal to the seven-brane.

To incorporate the effects of seven-brane monodromies, we now pass to a
description in terms of deformations by the Cartan, modulo the Weyl subgroup.
The maximal subgroup of $U(7)$ containing $U(5)$ is $U(5) \times U(2) \subset
U(7)$. Generic deformations of $U(7)$ preserving $U(5)$ are parameterized by
the Cartan subalgebra $g$ of $U(2)$, modulo the Weyl group $W(U(2))$. Letting
$\{e_{1},e_{2}\}$ denote an orthonormal basis, the Cartan subalgebra $g$ is
given by the vector space $\{ t_{1} e_{1} + t_{2} e_{2} \}$. Thus, the Cartan
parameters are $\{t_{1}, t_{2}\}$. The Weyl group $W(U(2))$ is the permutation
group $S_{2} = {\mathbb{Z}}_{2}$ acting on $\{t_{1},t_{2}\}$.

In this language, the matter content in the covering theory is given as
follows. Under the breaking pattern $U(7)\supset U(5)\times U(2)$, the adjoint
of $U(7)$ decomposes as:
\begin{align}
U(7)  &  \supset U(5)\times U(2)\nonumber\\
\mathbf{49}\rightarrow &  (\mathbf{25},\mathbf{1})+(\mathbf{1},\mathbf{4}%
)+(\mathbf{5},\overline{\mathbf{2}})+(\mathbf{\overline{5}},\mathbf{2})\text{.}%
\end{align}
The weights of the vector representations $\mathbf{2}$ of $U(2)$ are simply
$e_{1}^{\ast}$ and $e_{2}^{\ast}$ in the dual space $g^{\ast}$, and the
weights of the $\overline{\mathbf{2}}$ are just minus the weights of the
$\mathbf{2}$. Thus, by duality we obtain two matter curves in the covering
theory where the $\mathbf{5} \oplus \overline{\mathbf{5}}$ localize, namely
$t_{1}=0$ and $t_{2}=0$. The weights of the adjoint $\mathbf{4}$ are
$\pm(e_{1}^{\ast}-e_{2}^{\ast})$ and twice the zero weight. There is finally one
curve where the singlet $\mathbf{1}$ lives, which is defined by $t_{1}=t_{2}$.

Translating into charges under the branching $U(7)\supset U(5)\times U(1)_{1}
\times U(1)_{2} $, we obtain the decomposition
\begin{align}
U(7)  &  \supset U(5)\times U(1)_{1} \times U(1)_{2}\nonumber\\
\mathbf{49}\rightarrow &  \mathbf{25}_{0,0}+\mathbf{1}_{-1,+1}+\mathbf{1}%
_{+1,-1}+\mathbf{1}_{0,0}+\mathbf{1}_{0,0}+\mathbf{5}_{-1,0}+\mathbf{\overline
{5}}_{0,+1}+\mathbf{5}_{0,-1}+\mathbf{\overline{5}}_{+1,0}\text{,}%
\end{align}
where the subscripts denote the respective $U(1)$ charges. This recovers the
perturbative description of three six-dimensional bifundamentals
$\mathbf{5}_{-1,0}\oplus\mathbf{\overline{5}}_{+1,0}$, $\mathbf{5}%
_{0,-1}\oplus\mathbf{\overline{5}}_{0,+1}$ and $\mathbf{1}_{+1,-1}%
\oplus\mathbf{1}_{-1,+1}$.

At the $U(7)$ enhancement point, we find the interaction terms:%
\begin{equation}
W\supset\mathbf{5}_{-1,0}\times\mathbf{\overline{5}}_{0,+1}\times
\mathbf{1}_{+1,-1}+\mathbf{5}_{0,-1}\times\mathbf{\overline{5}}_{+1,0}%
\times\mathbf{1}_{-1,+1}+\widetilde{M}_{N}^{KK}\cdot\mathbf{1}_{+1,-1}%
\times\mathbf{1}_{-1,+1}\text{,}%
\end{equation}
where in addition to the cubic interaction terms derived from the $U(7)$
interaction point, we have also included the Kaluza-Klein mass associated with
the $U(5)$ singlets. In other words, we assume that there are no zero modes
transforming in the singlets, but only massive Kaluza-Klein modes. Upon making
the assignments:
\begin{align}
\widetilde{H}_{u}  &  \in\mathbf{5}_{-1,0}\text{, }\widetilde{L}%
\in\mathbf{\overline{5}}_{0,+1}\text{, }\widetilde{N}_{R}\in\mathbf{1}%
_{+1,-1}\text{, }\nonumber\\
\widetilde{H}_{u}^{\prime}  &  \in\mathbf{5}_{0,-1}\text{, }\widetilde
{L}^{\prime}\in\mathbf{\overline{5}}_{+1,0}\text{, }\widetilde{N}_{R}^{c}%
\in\mathbf{1}_{-1,+1}\text{,} \label{e:identif}%
\end{align}
the covering theory superpotential contains the terms:%
\begin{equation}
\widetilde{W}\supset\widetilde{H}_{u}\widetilde{L}\widetilde{N}_{R}%
+\widetilde{H}_{u}^{\prime}\widetilde{L}^{\prime}\widetilde{N}_{R}%
^{c}+\widetilde{M}_{N}^{KK}\cdot\widetilde{N}_{R}\widetilde{N}_{R}^{c}\text{.}%
\end{equation}

Let us now see whether the Kaluza-Klein seesaw mechanism can be implemented in
this geometric model. To go to the quotient theory, we must quotient by the
action of the Weyl group. The ${\mathbb{Z}}_{2}$ Weyl group acts by permuting
$t_{1}$ and $t_{2}$. In terms of $U(1)$ charges, it thus permutes the
$U(1)_{1}$ and $U(1)_{2}$ factors. Therefore, quotienting by the Weyl group
corresponds to the identification:
\begin{equation}
\mathbf{5}_{-1,0}\leftrightarrow\mathbf{5}_{0,-1}\text{, }\mathbf{\overline
{5}}_{0,+1}\leftrightarrow\mathbf{\overline{5}}_{+1,0}\text{, }\mathbf{1}%
_{+1,-1}\leftrightarrow\mathbf{1}_{-1,+1}\text{.} \label{Ztwotoy}%
\end{equation}
Using \eqref{e:identif}, this indeed becomes the required identification
\begin{equation}
\widetilde{H}_{u}\leftrightarrow\widetilde{H}_{u}^{\prime}\text{, }%
\widetilde{L}\leftrightarrow\widetilde{L}^{\prime}\text{, }\widetilde{N}%
_{R}\leftrightarrow\widetilde{N}_{R}^{c}\text{.}%
\end{equation}
It follows that the Kaluza-Klein seesaw will then generate the desired
dimension five operator in the quotient theory.

\begin{figure}[ptb]
\begin{center}
\includegraphics[
height=2.3601in,
width=6.0528in
]{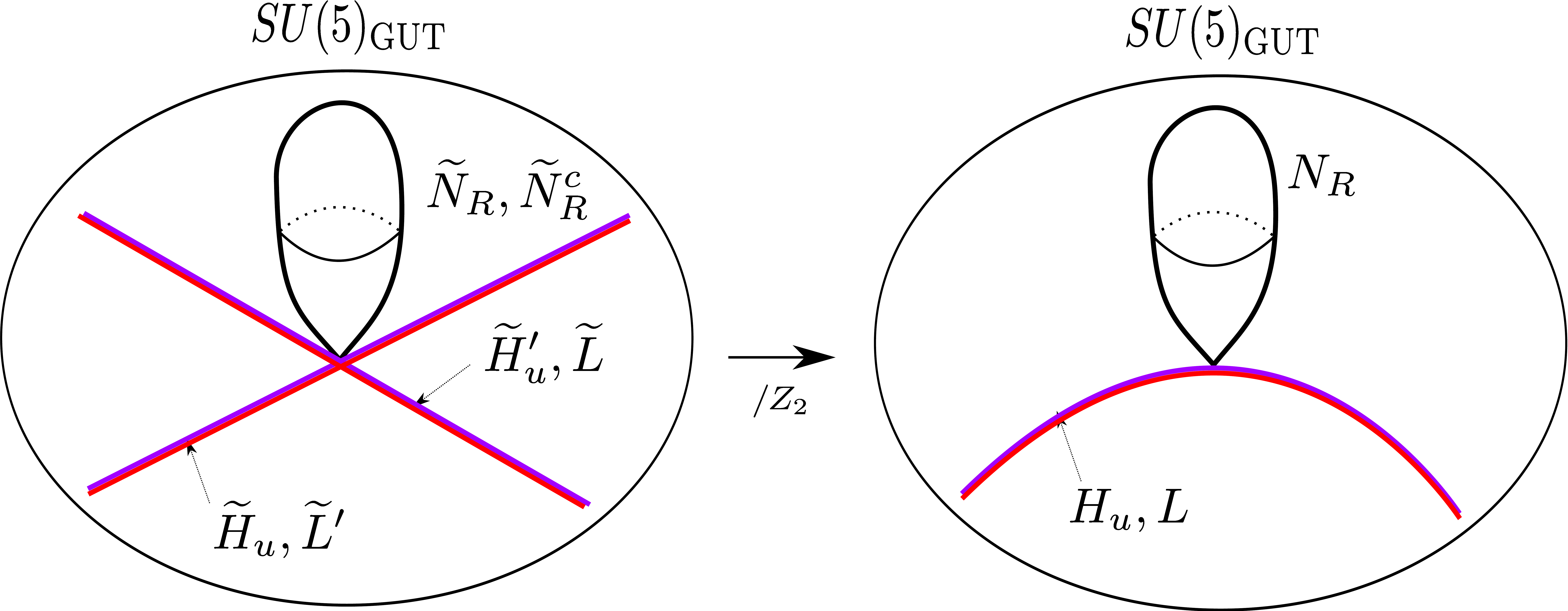}
\end{center}
\caption{Depiction of the $SU(7)$ toy model described in subsection \ref{TOY}.
In the covering theory (left) $\widetilde{H}_{u}$ and $\widetilde{L}^{\prime}$ localize
on the same curve, and the same is true for $\widetilde{H}_{u}^{\prime}$ and $\widetilde{L}$. As a
consequence, in the quotient theory (right) $H_{u}$ and $L$ localize on the
same matter curve.}%
\label{su7maj}%
\end{figure}

Unfortunately, this toy model is difficult to merge with the other requisite
elements of F-theory GUTs. The essential problem is that in the covering
theory, the fields $\widetilde{H}_{u}$ and $\widetilde{L}^{\prime}$ correspond
to conjugate representations which localize on the \emph{same} matter curve,
and similarly for $\widetilde{H}_{u}^{\prime}$ and $\widetilde{L}$ (see figure
\ref{su7maj}). As a consequence, in the quotient theory, $H_{u}$ and $L$ also
localize on the same matter curve. This was of course to be expected, since in
this example we quotiented by the whole Weyl group $S_{2}={\mathbb{Z}}_{2}$ so
that all the curves in the covering theory corresponding to the $\mathbf{5}%
\oplus\overline{\mathbf{5}}$ of $SU(5)_{GUT}$ are identified by the Weyl
group. Doublet triplet splitting of the $\mathbf{5}_{H}$ requires
a non-trivial hyperflux to pierce the Higgs curve. This is incompatible
with the requirement that the entire $\mathbf{\overline
{5}}_{M}$ $SU(5)$ GUT\ multiplet is a zero mode on the lepton curve. Moreover,
if $\widetilde{H}_{u}$ and $\widetilde{L}^{\prime}$ correspond to conjugate
representations, there is no reason for the bare coupling $\widetilde{H}%
_{u}\widetilde{L}^{\prime}$ to be prevented in the covering theory, since it
is gauge invariant. This coupling descends to the bare coupling $H_{u}L$ in
the quotient theory. Therefore, to prevent this bare coupling from appearing
in the quotient theory, we must again require that $H_{u}$ and $L$ live on
different curves. This geometric sequestering effectively plays the role of
matter parity. We now turn to an explicit realization of the Kaluza-Klein
seesaw which incorporates these elements.

\subsection{$E_{8}$ Kaluza-Klein Seesaw\label{BIGBOY}}

In the previous subsection we observed that although generic monodromies
in a configuration of seven-branes would indeed generate a Kaluza-Klein seesaw
mechanism, the limited number of matter curves in the covering theory always
identified the lepton doublet and Higgs curve. To overcome this obstruction,
we now proceed to configurations with additional matter curves in the covering
theory. The most straightforward way to implement this additional condition is
to require that the $SU(5)$ F-theory GUT\ contain an enhancement to a higher
rank singularity of type $G$ such that the corresponding gauge group $G\supset
SU(5)\times U(1)^{4}$ and $\widetilde{H}_{u}$, $\widetilde{H}_{u}^{\prime}$,
$\widetilde{L}$ and $\widetilde{L}^{\prime}$ localize on four distinct matter curves.

Adhering to the general requirement that all of the interactions of interest
embed consistently within $E_{8}$ gauge theory structures, it is therefore
most natural to consider the rank eight singularity $G=E_{8}$. Let us now
analyze the configuration of curves meeting at this $E_{8}$ intersection point.

The $SU(5)$ GUT group embeds in $E_{8}$ in the maximal subgroup $SU(5)_{GUT}%
\times SU(5)_{\bot}\subset E_{8}$. Generic deformations of $E_{8}$ preserving
$SU(5)_{GUT}$ are parameterized by the Cartan subalgebra $g_{\bot}$ of
$SU(5)_{\bot}$, modulo the Weyl group $W(SU(5)_{\bot})$. Let $\{e_{1},\ldots
e_{5}\}$ be an orthonormal basis. The Cartan subalgebra $g_{\bot}$ of
$SU(5)_{\bot}$ is given by the vector space $\{t_{1}e_{1}+\ldots+t_{5}e_{5}%
\}$, subject to the tracelessness condition $\sum_{i=1}^{5}t_{i}=0$. These
$t_{i}$'s define the Cartan parameters. The Weyl group $W(SU(5)_{\bot})$ is
isomorphic to the symmetric group $S_{5}$ which acts by
permutations of the $t_{i}$'s.

The decomposition of the adjoint representation is then given by:%
\begin{align}
E_{8}\supset &  SU(5)_{GUT}\times SU(5)_{\bot}\nonumber\\
\mathbf{248}\rightarrow &  (\mathbf{1},\mathbf{24})+(\mathbf{24}%
,\mathbf{1})+(\mathbf{5},\mathbf{\overline{10}})+(\mathbf{\overline{5}%
},\mathbf{10})+(\mathbf{10},\mathbf{5})+(\mathbf{\overline{10}}%
,\mathbf{\overline{5}})\text{.}%
\end{align}
The Higgs up, lepton doublets and right-handed neutrinos respectively
transform in the $\mathbf{5}_{H}$, $\overline{\mathbf{5}}_{M}$ and
$\mathbf{1}_{N}$ of $SU(5)_{GUT}$.  Hence, they must descend from the
irreducible representations of $SU(5)_{GUT}\times SU(5)_{\bot}$ given as:%
\begin{equation}
H_{u}\in(\mathbf{5},\mathbf{\overline{10}})\text{, }\qquad L\in
(\mathbf{\overline{5}},\mathbf{10})\text{, }\qquad N_{R}\in(\mathbf{1}%
,\mathbf{24})\text{.}%
\end{equation}

Since we will ultimately need to describe the action of the Weyl group of
$SU(5)_{\bot}$ on the matter curves of the covering theory, we now identify
the corresponding weights of $SU(5)_{\bot}$ associated with each matter curve.
Consider the matter curves on which six-dimensional $\mathbf{10}_{GUT}%
\oplus\overline{\mathbf{10}}_{GUT}$'s of $SU(5)_{GUT}$ localize. These fields
transform in the $\mathbf{5}_{\bot}\oplus\overline{\mathbf{5}}_{\bot}$ of
$SU(5)_{\bot}$. Since the weights of the $\mathbf{5}_{\bot}$ of $SU(5)_{\bot}$
are given by $e_{1}^{\ast}$, $e_{2}^{\ast}$, $e_{3}^{\ast}$, $e_{4}^{\ast}$
and $e_{5}^{\ast}$ (with opposite signs for the weights of the $\overline
{\mathbf{5}}_{\bot}$), it follows that in the covering theory, there are five
curves where a six-dimensional field in the $\overline{\mathbf{10}}_{GUT} \oplus \mathbf{10}_{GUT}$
of $SU(5)_{GUT}$ localizes, specified by the five distinct vanishing loci of the $t_{i}$'s. Note, however, that an
appropriate choice of fluxes can avoid the presence of any four-dimensional
zero modes from such matter curves.

Next consider matter curves where a six-dimensional field in the
$\mathbf{5}_{GUT}\oplus\overline{\mathbf{5}}_{GUT}$ of $SU(5)_{GUT}$ localize.
Such curves will support the Higgs up and lepton fields. These matter fields
transform in the $\overline{\mathbf{10}}_{\bot} \oplus \mathbf{10}_{\bot}$ of
$SU(5)_{\bot}$. Since the weights of the $\mathbf{10}_{\bot}$ are $e_{i}%
^{\ast}+e_{j}^{\ast}$, for $i,j=1,\ldots,5$, $i\neq j$, there are ten distinct
curves with matter content in the $\mathbf{5}_{GUT}\oplus\overline{\mathbf{5}%
}_{GUT}$ of $SU(5)_{GUT}$.

Finally, we consider the right-handed neutrinos. These fields correspond to
singlets under $SU(5)_{GUT}$, and transform as an adjoint of $SU(5)_{\bot}$.
The weights of the adjoint are $e_{i}^{*} - e_{j}^{*}$, $i,j=1,\ldots,5$,
$i\neq j$, and four times the zero weight.

As a result, we obtain that the matter content intersecting at this $E_{8}$
point lies in the following directions of the Cartan subalgebra:
\begin{align}
\mathbf{10}_{GUT}: &  \qquad t_{i},\qquad\ \ \ \ \ \ i=1\ldots5;\nonumber\\
\mathbf{\overline{5}}_{GUT}: &  \qquad t_{i}+t_{j},\qquad i,j=1,\ldots,5,i\neq
j;\nonumber\\
\mathbf{1}_{GUT}: &  \qquad t_{i}-t_{j},\qquad i,j=1,\ldots,5,i\neq
j,\label{e:U1charges}%
\end{align}
with opposite signs for the conjugate representations. Note that these are
subject to the tracelessness condition $\sum_{i=1}^{5}t_{i}=0$. For the
$\mathbf{1}$, we omitted the four singlets uncharged under $U(1)^{4}$.

This gives us the matter content of the covering theory, where deformations
are parameterized by the Cartan subalgebra itself. To each of these directions
there is an associated matter curve, given by the vanishing locus of the
linear combination of the $t_{i}$'s written above. Again, we stress that an
appropriate choice of flux can forbid most of these curves from acquiring a
non-trivial zero mode content. As such, it is enough to focus exclusively on
the geometric arrangement of curves. In the examples of the next subsections
we will specify precisely on which curves we allow zero modes.

The next step is to mod out by the monodromy group to obtain the quotient theory.
Note, however, that quotienting by the entire Weyl group would identify
\emph{all} the matter curves corresponding to a given representation of
$SU(5)_{GUT}$. This is not consistent with our requirement that the Higgs and
lepton localize on distinct curves in the quotient theory. Therefore, we will
only quotient by a subgroup of the Weyl group; we need to identify which
subgroup we will be interested in. We first present a simple example where we
identify a ${\mathbb{Z}}_{2}$ subgroup of the Weyl group realizing the
Kaluza-Klein seesaw with distinct curves in the quotient theory. After this we
present a more involved example in which \textit{all} of the interaction terms
of the MSSM unify at the $E_{8}$ enhancement point.

\subsubsection{A ${\mathbb{Z}}_{2}$ Model}

Let us first present a Kaluza-Klein seesaw where we quotient by a
${\mathbb{Z}}_{2}$ subgroup of the Weyl group. That is, we consider a geometry
where the deformations of the $E_{8}$ singularity are parameterized by the
Cartan subalgebra of $SU(5)_{\bot}$ modulo a ${\mathbb{Z}}_{2}$ subgroup of
the Weyl group $W(SU(5)_{\bot})$.

\begin{figure}[ptb]
\begin{center}
\includegraphics[
height=2.3099in,
width=5.924in
]{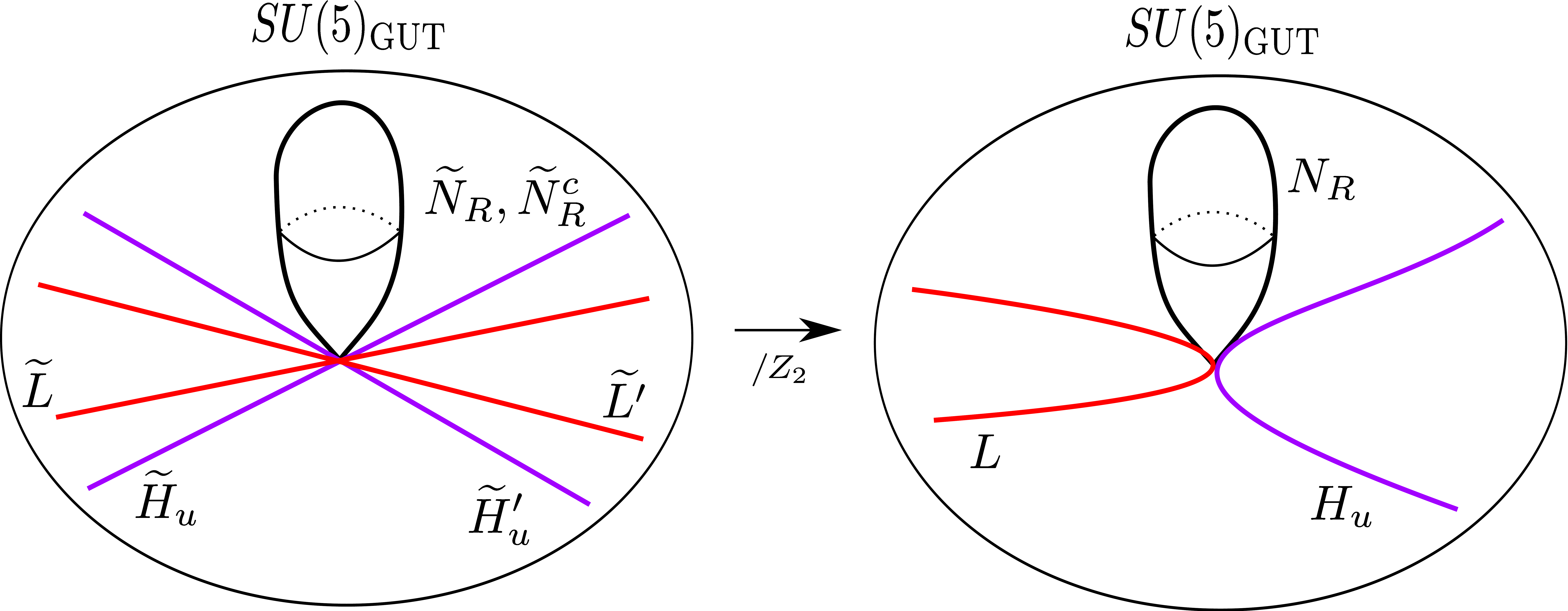}
\end{center}
\caption{Depiction of the matter curves in the Kaluza-Klein seesaw associated
with an $E_{8}$ intersection point. As opposed to the matter curve
configuration of figure \ref{su7maj}, here $H_{u}$ and $L^{\prime}$ localize
on different curves in the covering theory (left). In the quotient theory
(right), $H_{u}$ and $L$ localize on two distinct matter curves.}%
\label{e8maj}%
\end{figure}
Our aim is now to identify a ${\mathbb{Z}}_{2}$ subgroup of the Weyl group $W(SU(5)_{\bot})$ which
generates the Kaluza-Klein seesaw mechanism. We need the following matter fields
in the covering theory: $\widetilde{H}_{u}$, $\widetilde{H}_{u}^{\prime}$ in the
$\mathbf{5}$, and $\widetilde{L}$, $\widetilde{L}^{\prime}$ in the
$\mathbf{\overline{5}}$, and the singlets $\widetilde{N}_{R}$ and
$\widetilde{N}_{R}^{c}$, such that:
\begin{itemize}
\item $\widetilde{H}_{u}$ and $\widetilde{H}_{u}^{\prime}$ lie in a single
orbit of the ${\mathbb{Z}}_{2}$ subgroup; similarly, $\widetilde{L}$ and
$\widetilde{L}^{\prime}$ form a single orbit, as well as $\widetilde{N}_{R}$
and $\widetilde{N}_{R}^{c}$. This ensures that the
${\mathbb{Z}}_{2}$ subgroup provides the required identification of \eqref{e:iden};
\item The two $\IZ_2$ orbits for the Higgs and lepton doublet are \emph{distinct}, so that $H_{u}$ and $L$ localize on distinct matter curves in the quotient theory;
\item $\widetilde{N}_{R}$ and $\widetilde{N}_{R}^{c}$ must have opposite
Cartan directions, since they are conjugate fields;
\item The Cartan directions allow for the gauge invariant operators given in \eqref{CoverPot}.
\end{itemize}

\begin{figure}
[ptb]
\begin{center}
\includegraphics[
height=4.9182in,
width=6.1237in
]%
{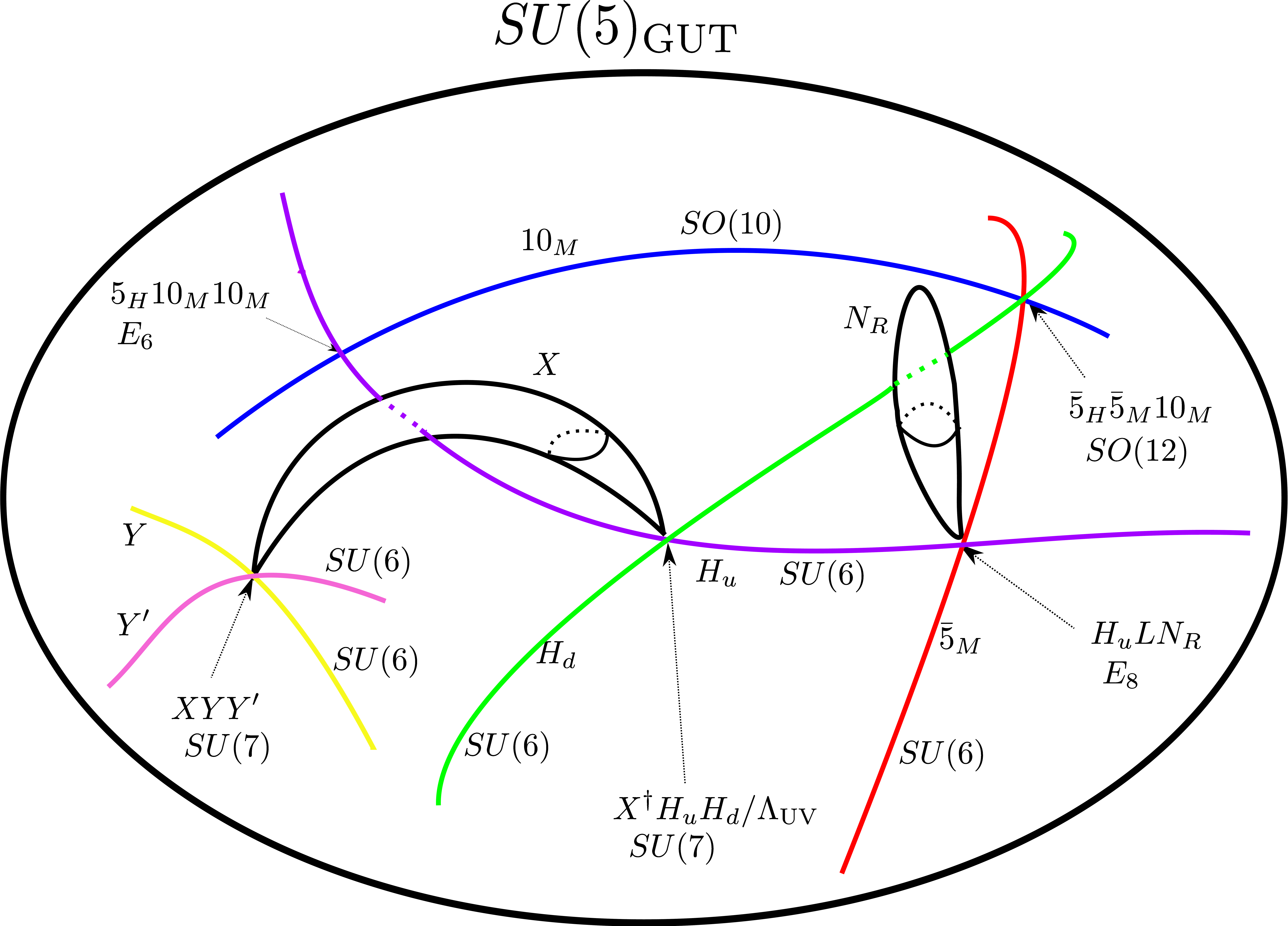}%
\caption{Depiction of a minimal F-theory GUT with a Majorana neutrino sector. In this case,
the Higgs up curve forms a triple intersection with the lepton doublet curve and the
right-handed neutrino curve. Integrating out the massive right-handed neutrino states
generates the quartic operator $(H_{u}L)^{2}/\Lambda_{\text{UV}}$ in the low energy effective
theory.}%
\label{fgutnug}%
\end{center}
\end{figure}

Let us now provide an explicit identification of this matter content and
${\mathbb{Z}}_{2}$ action. We consider the ${\mathbb{Z}}_{2}$ subgroup of the
Weyl group given by the permutation $(12)(34)$, which acts on the Cartan
parameters as:
\begin{equation}
(t_{1},t_{2},t_{3},t_{4},t_{5})\mapsto(t_{2},t_{1},t_{4},t_{3},t_{5}).
\label{e:Z2perm}%
\end{equation}
Using the Cartan directions found in \eqref{e:U1charges}, we make the following matter content assignments in the covering theory:
\begin{align}
\widetilde{H}_{u}:  &  -t_{1}-t_{3}, \qquad\widetilde{L}: t_{2}+t_{3},
\qquad\widetilde{N}_{R}: t_{1}-t_{2},\nonumber\\
\widetilde{H}_{u}^{\prime}:  &  -t_{2} -t_{4}\text{, }\qquad\widetilde
{L}^{\prime}: t_{1}+t_{4}\text{, }\qquad\widetilde{N}_{R}^{c}: -t_{1}%
+t_{2}\text{.} \label{assign}%
\end{align}
Note that by construction $\widetilde{H}_{u}$, $\widetilde{H}_{u}^{\prime}$,
$\widetilde{L}$ and $\widetilde{L}^{\prime}$ all localize on different curves,
and $\widetilde{N}_{R}$ and $\widetilde{N}_{R}^{c}$ are conjugate fields.
These assignments are consistent with the superpotential terms in the covering
theory:%
\begin{equation}
\widetilde{W}\supset\widetilde{H}_{u}\widetilde{L}\widetilde{N}_{R}%
+\widetilde{H}_{u}^{\prime}\widetilde{L}^{\prime}\widetilde{N}_{R}%
^{c}+\widetilde{M}_{N}^{KK}\cdot\widetilde{N}_{R}\widetilde{N}_{R}^{c}.
\end{equation}
The ${\mathbb{Z}}_{2}$ permutation \eqref{e:Z2perm} acts by:
\begin{equation}
\widetilde{H}_{u}\leftrightarrow\widetilde{H}_{u}^{\prime}\text{, }%
\widetilde{L}\leftrightarrow\widetilde{L}^{\prime}\text{, }\widetilde{N}%
_{R}\leftrightarrow\widetilde{N}_{R}^{c}\text{,}%
\end{equation}
as required for the Kaluza-Klein seesaw mechanism. See figure \ref{e8maj} for
a depiction of this interaction structure in the covering and quotient theories, and figure \ref{fgutnug}
for a depiction of how this interaction embeds in a minimal $SU(5)$ F-theory GUT.

%TCIMACRO{\FRAME{ftbpFU}{4.7366in}{3.4091in}{0pt}{\Qcb{CAPTION}}%
%{\Qlb{fgutnugmaje8}}{fgutnugmaje8.pdf}{\special{ language "Scientific Word";
%type "GRAPHIC";  maintain-aspect-ratio TRUE;  display "USEDEF";
%valid_file "F";  width 4.7366in;  height 3.4091in;  depth 0pt;
%original-width 44.3078in;  original-height 31.8303in;  cropleft "0";
%filename 'fgutnug.pdf';file-properties "XNPEU";}} }%
%BeginExpansion

In the above analysis, we have presented one particular choice of
$\mathbb{Z}_{2}$ group action and matter assigment in the covering theory. In
principle, there could be other choices compatible with the Kaluza-Klein seesaw.
In fact, in the next subsection we provide an alternative choice which
realizes the Kaluza-Klein seesaw and geometrically unifies all the MSSM
interactions at the $E_{8}$ interaction point.

\subsubsection{Geometric $E_{8}$ Unification of All MSSM Interactions}

In the previous subsection we showed that the Kaluza-Klein seesaw can indeed
be accomodated by an $E_{8}$ enhancement point. In a certain sense, however,
it is not particularly economical to include such a high rank enhancement
simply to incorporate a neutrino sector. Indeed, the presence of this higher
unification structure suggests that the other interactions of the MSSM\ might
also unify at this same point. In fact, as shown in \cite{HVCKM}, the
hierachical structure of the CKM\ matrix \textit{requires} the $\mathbf{5}%
_{H}\times\mathbf{10}_{M}\times\mathbf{10}_{M}$ and $\mathbf{\overline{5}}%
_{H}\times\mathbf{\overline{5}}_{M}\times\mathbf{10}_{M}$ interaction points
to be close to each other. It is therefore quite natural to consider
geometries where all of the interaction terms geometrically unify.

\begin{figure}
[ptb]
\begin{center}
\includegraphics[
height=4.9182in,
width=6.1237in
]%
{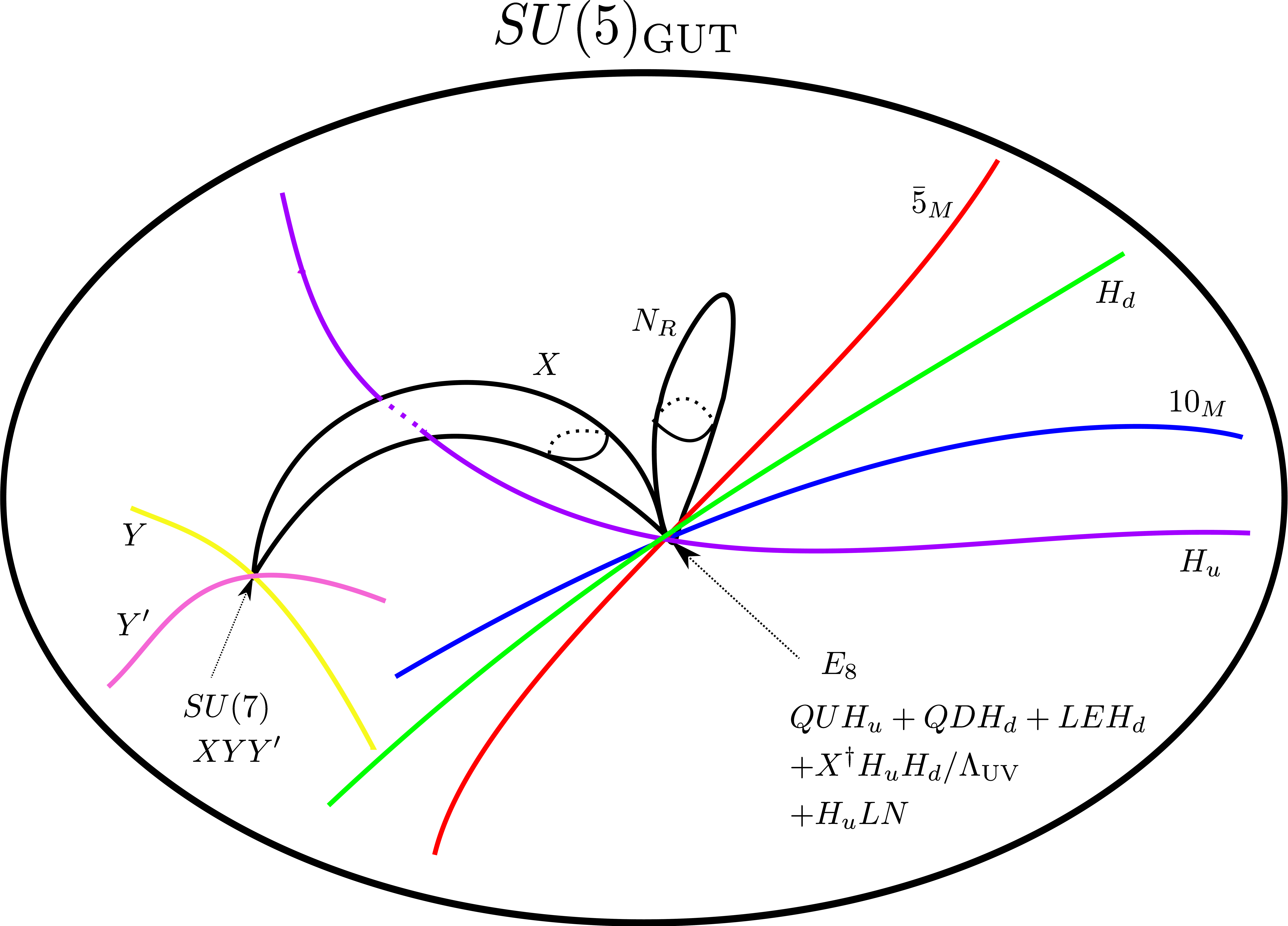}%
\caption{Depiction of a Kaluza-Klein seesaw model in which all of the interaction terms geometrically unify at a single point of $E_{8}$ enhancement.}%
\label{fgutnugmaje8}%
\end{center}
\end{figure}

In this subsection we present a
geometry where \emph{all} MSSM interactions descend from a single $E_{8}$
singularity. In the example we present, only the supersymmetry breaking
messenger sector localizes at a different point of the geometry. Monodromies
play an especially prominent role, both in terms of the Kaluza-Klein seesaw,
and also through the condition that in the quotient theory, the $\mathbf{10}%
_{M}$'s localize on a single curve. This latter condition is important in
ensuring that the up type quarks have one heavy generation (see
\cite{Hayashi:2009ge} for further discussion). To do so, we must however leave
the simple realm of ${\mathbb{Z}}_{2}$ identifications, and consider the
action of a bigger finite subgroup of the Weyl group. See figure \ref{fgutnugmaje8} for a
depiction of this geometry with all interaction terms geometrically unified.

%TCIMACRO{\FRAME{ftbpFU}{4.7366in}{3.4091in}{0pt}{\Qcb{CAPTION}}%
%{\Qlb{fgutnugmaje8}}{fgutnugmaje8.pdf}{\special{ language "Scientific Word";
%type "GRAPHIC";  maintain-aspect-ratio TRUE;  display "USEDEF";
%valid_file "F";  width 4.7366in;  height 3.4091in;  depth 0pt;
%original-width 44.3078in;  original-height 31.8303in;  cropleft "0";
%filename 'fgutnugmajE8.pdf';file-properties "XNPEU";}} }%
%BeginExpansion
We consider the subgroup $\mathfrak{S}\subset W(SU(5)_{\bot})$ generated by
the order $2$ element $g_{1}=(12)(34)$ and the order $4$ element
$g_{2}=(1234)$. These elements act on the Cartan parameters as:
\begin{equation}
g_{1}:(t_{1},t_{2},t_{3},t_{4},t_{5})\mapsto(t_{2},t_{1},t_{4},t_{3}%
,t_{5}),\qquad g_{2}:(t_{1},t_{2},t_{3},t_{4},t_{5})\mapsto(t_{2},t_{3}%
,t_{4},t_{1},t_{5}).\label{finitegroup}%
\end{equation}
Note that $t_{5}$ is invariant under this subgroup.

We make the following matter assignments in the covering theory. We group
the matter fields in terms of orbits under the action of $\mathfrak{S}$;
we indicate next to each type of field its corresponding orbit.
\begin{align}
\widetilde{\mathbf{5}}_{H}: &  \{-t_{1}-t_{3},-t_{2}-t_{4}\},\nonumber\\
\widetilde{\overline{\mathbf{5}}}_{M}: &  \{t_{1}+t_{2},t_{2}+t_{3}%
,t_{3}+t_{4},t_{1}+t_{4}\}\nonumber\\
\widetilde{N}_{R},\widetilde{N}_{R}^{c}: &  \{\pm(t_{1}-t_{2}),\pm(t_{2}%
-t_{3}),\pm(t_{3}-t_{4}),\pm(t_{4}-t_{1})\},\nonumber\\
\widetilde{\mathbf{10}}_{M}: &  \{t_{1},t_{2},t_{3},t_{4}\},\nonumber\\
\widetilde{\overline{\mathbf{5}}}_{H}: &  \{t_{1}+t_{5},t_{2}+t_{5}%
,t_{3}+t_{5},t_{4}+t_{5}\},\nonumber\\
\widetilde{X}: &  \{t_{5}-t_{1},t_{5}-t_{2},t_{5}-t_{3},t_{5}-t_{4}%
\}\text{,}\label{e:matterfull}%
\end{align}
where $t_{5}=-t_{1}-t_{2}-t_{3}-t_{4}$. We include here the singlet $X$ which
is required for supersymmetry breaking, as explained in section \ref{FREV}.

Each line in \eqref{e:matterfull} corresponds to an orbit under the action of
$\mathfrak{S}$. Therefore,
all of the fields in a given line are identified in the quotient theory. By
looking at the Cartan directions for each field, it is easy to work out all
the gauge-invariant interactions in the covering theory. The list is rather
long, so we will not include it here and instead focus on the most salient features.

First, all Kaluza-Klein neutrino mode $\widetilde{N}_{R}^{k}$ and
$(\widetilde{N}_{R}^{c})^{k}$ have interactions of the form given in line
\eqref{e:couplings}. Since by construction the $\widetilde{\mathbf{5}}_{H}^i$, the
$\widetilde{\overline{\mathbf{5}}}_{M}^{j}$ and the neutrinos live in orbits of the finite group, this
is sufficient to realize the Kaluza-Klein seesaw mechanism and generate the
higher-dimension operator \eqref{Weffquot} in the quotient theory.

Second, after quotienting by the finite group, the MSSM interaction terms
$\mathbf{\overline{5}}_{H}\times\mathbf{\overline{5}}_{M}\times\mathbf{10}%
_{M}$ $\ $and $\mathbf{5}_{H}\times\mathbf{10}_{M}\times\mathbf{10}_{M}$ are
both present in the quotient theory. Moreover, all
the $\widetilde{\mathbf{10}}_{M}$ curves in the covering theory are identified
by the finite group. Therefore, there is only one $\mathbf{10}_{M}$ curve in
the quotient theory, as required for one heavy up type quark generation.

Finally, the interaction term $X H_{u}H_{d}$ is \emph{not} gauge invariant, and
so is not present in the quotient theory. However, $X^{\dagger}H_{u}H_{d}/\Lambda_{\text{UV}}$ is
gauge invariant. This operator can be produced by integrating out Kaluza-Klein
modes, as explained in \cite{HVGMSB}. Once $X$ develops a supersymmetry breaking
vev, this generates a $\mu$-term.

As a result, we obtain an F-theory $SU(5)$ GUT where all the MSSM interactions geometrically
unify at a single $E_{8}$ enhancement point. The local effective theory
near this $E_{8}$ interaction point is parameterized by the Cartan of
$SU(5)_{\bot}$, modulo the particular subgroup $\mathfrak{S}$ of the Weyl group generated by \eqref{finitegroup}.

The only remaining interaction concerns the supersymmetry breaking messenger
sector \cite{HVGMSB}, which is given by the superpotential term $XYY^{\prime}$,
where $Y$ and $Y^{\prime}$ are a vector-like pair of messenger fields either
in the $\mathbf{5}\oplus\overline{\mathbf{5}}$ or in the $\mathbf{10}%
\oplus\overline{\mathbf{10}}$. Since this interaction has a very different
origin from the other MSSM interactions, it seems natural not to require that
it unifies inside the same $E_{8}$ interaction point. Note however that it
could be unified inside $E_8$, although either $Y$ or $Y^{\prime}$ (or both)
would then be required to live on the same matter curve as some of the MSSM matter content.

To end this subsection, we comment that we have not done an exhaustive search
for finite subgroups of the Weyl group of $SU(5)_{\bot}$ which allow for the
realization of the Kaluza-Klein seesaw. There could be other choices
compatible with the Kaluza-Klein seesaw. Our main goal here was simply to
demonstrate that such consistent choices exist. It would be interesting to
investigate this issue further.

\subsubsection{$U(1)_{PQ}$ and Matter Parity in the Quotient Theory}

Now that we have realized the Kaluza-Klein seesaw, we can analyze symmetries
of the low energy effective theory directly in the quotient theory. In
particular, we can identify the $U(1)_{PQ}$ gauge symmetry in the quotient
theory, and understand what plays the role of matter parity.

For simplicity, we focus on the two models of the previous
subsection with an $E_{8}$ singularity. Recall that in the covering theory,
the deformation is specified by the Cartan parameters $t_{1},t_{2},t_{3}%
,t_{4}$ generating the Cartan subgroup $U(1)^{4}\subset SU(5)_{\bot}$. In both models, the finite group
that we quotiented the covering theory with left the Cartan parameter $t_{5}$
invariant. By the tracelessness condition, we know that $t_{5}=-t_{1}%
-t_{2}-t_{3}-t_{4}$. Therefore, all fields in the quotient theory will remain
charged under a $U(1)$ subgroup generated by $t_{5}$, which is the diagonal
combination of the four $U(1)$'s in the Cartan subgroup of the covering
theory. It turns out that this invariant $U(1)$ corresponds precisely to the
alternative $U(1)_{PQ}$ presented in subsection \eqref{PQalt}.

Indeed, consider the matter content presented in \eqref{e:matterfull} (the
same analysis holds for the matter content of the ${\mathbb{Z}}_{2}$ model).
The charges of the fields under the diagonal subgroup generated by $t_5$ are:%
\begin{equation}%
\begin{tabular}
[c]{|c|c|c|c|c|c|}\hline
& $X$ & $H_{u}$ & $H_{d}$ & $\mathbf{10}_{M}$ & $\mathbf{5}_{M}$\\\hline
$U(1)_{PQ}$ & $+5$ & $+2$ & $+3$ & $-1$ & $-2$\\\hline
\end{tabular}
\text{ \ ,}%
\end{equation}
which are precisely the charges obtained in section \eqref{PQalt}.

We can also say something about matter parity. More precisely, we want to
understand why the quotient theory admits the quartic superpotential term
$(H_{u}L)^{2}$, while the associated matter parity violating coupling $H_{u}L$
is absent. Note that for the $H_{u}L$ term to be present in the quotient
theory, one would need terms of the form $\widetilde{H}_{u}^{i}\widetilde
{L}^{j}$ for some $i$ and $j$ in the covering theory. For such a term
to be gauge invariant $\widetilde{H}_{u}^{i}$ would need to have opposite
Cartan charges to $\widetilde{L}^{j}$; that is, they would need to be conjugate
fields living on the same matter curve. We may then say that the effective
role of matter parity is played by the requirement that $H_{u}$ and $L$
descend from distinct orbits under the action of the finite group. It is
interesting to note that this requirement is also necessary to implement
doublet-triplet splitting for the Higgs $H_{u}$, which is \emph{a priori}
unrelated to conservation of matter parity.

\section{Yukawas of the Kaluza-Klein Seesaw \label{NEUTMASSHIER}}

In the previous section we showed that the geometry of F-theory
compactifications is flexible enough to accommodate a Kaluza-Klein seesaw
mechanism, whereby an effective Majorana mass for the left-handed neutrinos is
induced through a coupling to Kaluza-Klein right-handed neutrinos. In this
section we estimate the entries of the Yukawa matrix $\lambda_{ij}^{(\nu)}$ of
the higher dimension operator:
\begin{equation}
W_{eff}\supset\lambda_{ij}^{(\nu)}\frac{\left(  H_{u}L^{i}\right)  \left(
H_{u}L^{j}\right)  }{\Lambda_{\text{UV}}}\text{,} \label{quartopagain}%
\end{equation}
obtained through the Kaluza-Klein seesaw mechanism. Here, $i=1,2,3$ is an
index for the three generations of lepton doublets such that $L^{3}$
corresponds to the $\tau$ and $\nu_{\tau}$ doublet.

This type of interaction term originates from integrating out the heavy
right-handed neutrinos. In terms of four-dimensional chiral superfields, the
Kaluza-Klein seesaw is given as:%
\begin{equation}
\widetilde{W}\supset\widetilde{y}_{i,I}\widetilde{H}_{u}\widetilde{L}%
^{i}\widetilde{N}_{I}+\widetilde{y}_{j,J}^{\prime}\widetilde{H}_{u}%
\widetilde{L}^{j}\widetilde{N}_{J}^{c}+\widetilde{M}_{IJ}\widetilde{N}_{I}%
^{c}\widetilde{N}_{J}\text{,} \label{WCOVER}%
\end{equation}
where $I$ and $J$ are indices labelling all of the massive modes of the
compactification. In matrix notation, the Majorana coupling is then given by:%
\begin{equation}
\frac{\lambda^{(\nu)}}{\Lambda_{\text{UV}}}=\widetilde{y}\cdot\frac{1}{M}%
\cdot\widetilde{y}^{T}\text{.}%
\end{equation}

The Yukawas $\widetilde{y}_{i,I}$ of line (\ref{WCOVER}) are given by overlaps
between the Higgs and lepton zero mode wave functions with the massive
right-handed neutrino zero modes:%
\begin{equation}
\widetilde{y}_{i,I}=\underset{\mathcal{U}_{B}}{\int}\widetilde{\Psi}_{H_{u}%
}\widetilde{\Psi}_{L}^{i}\widetilde{\Psi}_{N}^{(I)} \label{ytilde}%
\end{equation}
where $\mathcal{U}_{B}\subset B_{3}$ denotes a neighborhood in $B_{3}$ around
the neutrino interaction point.

The form of the integral in equation (\ref{ytilde}) is to be contrasted with
the Yukawas in the quark and charged lepton sectors which are instead given by
overlap integrals in a two-dimensional neighborhood $\mathcal{U}_{S}\subset S$
which contains the corresponding interaction point \cite{HVCKM}. For example,
the up type quark Yukawa coupling in the interaction term:%
\begin{equation}
W\supset\lambda_{ij}^{(u)}H_{u}Q^{i}U^{j}\text{,}%
\end{equation}
is given by the overlap integral:%
\begin{equation}
\lambda_{ij}^{(u)}=\underset{\mathcal{U}_{S}}{\int}\Psi_{H_{u}}\Psi_{Q}%
^{i}\Psi_{U}^{j} \label{upYUK}%
\end{equation}
where the $\Psi$'s denote the corresponding zero mode wave functions. More
formally, the interaction term of equation (\ref{upYUK}) descends from an
appropriate superpotential coupling in an eight-dimensional quasi-topological
theory. In a perturbative string description, the Yukawa of equation
(\ref{ytilde}) can be interpreted in terms of holomorphic Chern-Simons theory
defined in a patch of the neutrino interaction point. We will return to a more
precise formulation of this overlap integral in subsection
(\ref{MASSIVEOVERLAP}).

We now explain in crude terms our expectation for the form of the Yukawa matrix $\widetilde{y}$. \ The main
point is that whereas zero mode wave functions $\Psi^{(0)}$ satisfy wave
equations of the schematic form%
\begin{equation}
\overline{\partial}\Psi^{(0)}=0,
\end{equation}
Kaluza-Klein mode wave functions $\Psi^{KK}$ are massive modes and as such%
\begin{equation}
\overline{\partial}\Psi^{KK}\neq0\text{.}%
\end{equation}
Thus, whereas there is a notion of holomorphicity for zero mode wave
functions, there is no similar notion for these massive modes. As found in
\cite{HVCKM}, and as we shall review in subsection \ref{YUKREV}, the
holomorphicity of the wave function translates into an approximate set of
$U(1)$ symmetries which are violated by the presence of background fluxes.
These violations then generate subleading corrections to the Yukawa matrices
of the zero modes. \textit{By contrast, because the Kaluza-Klein wave
functions are not holomorphic, these approximate }$U(1)$\textit{ symmetries
will be violated more strongly, leading to milder hierarchies in the neutrino
sector.}

The rest of this section is organized as follows. In subsection \ref{YUKREV}
we review the computation of the Yukawa matrices in the quark and charged
lepton sectors obtained in \cite{HVCKM}. Next, in subsection
\ref{MASSIVEOVERLAP}, we perform the analogous computation in the case of the
Kaluza-Klein neutrino sector. With this result in hand, in subsection
\ref{NEUTYUK}, we compute the form of the neutrino Yukawa coupling
$\lambda_{(\nu)}$ in the low energy effective field theory. Finally, in
subsection \ref{NORMALIZATION} we discuss the overall mass scale expected from
the Kaluza-Klein seesaw, and why the effective seesaw scale can in principle
be lower than the GUT scale.

\subsection{Review of Quark and Charged Lepton Yukawas\label{YUKREV}}

As we will explain in the next subsection, the fact that the right-handed
neutrinos do not correspond to zero modes significantly dilutes the expected
mass hierarchy in the neutrino sector. To see how this comes about, we first
recall the estimate of the quark and charged lepton Yukawa matrices obtained
in \cite{HVCKM}.

For brevity, we focus on the up type quark Yukawa coupling:
\begin{equation}
\lambda_{ij}^{(u)}=\underset{\mathcal{U}_{S}}{\int}\Psi_{H_{u}}\Psi_{Q}%
^{i}\Psi_{U}^{j}. \label{e:upyuk}%
\end{equation}
Although a global description of the wave function profile would be
interesting, it is not necessary to define the requisite wave functions.
Indeed, in a neighborhood of the interaction point, the entire gauge theory on
$S$ can be modelled in terms of a parent gauge theory which is Higgsed down to
the bulk gauge group on $S$ by the vev of a locally defined $(2,0)$ form of
the parent theory. In the presence of suitable background fluxes, the zero modes correspond to solutions to the defect equation of
motion of the eight-dimensional quasi-topological theory found in
\cite{BHVI}:
\begin{align}
\omega_{S}\wedge\partial_{A}\psi+\frac{i}{2}\left[  \overline{\phi}%
,\chi\right]   &  =0 + \cdots ,\label{DEOM}\\
\overline{\partial}_{A}\chi+\frac{1}{2}\left[  \phi,\psi\right]   &
=0 + \cdots \text{,} \label{FEOM}%
\end{align}
where $\omega_{S}$ denotes the K\"{a}hler form of the parent theory, $\phi$
denotes the background value of the $(2,0)$ form of the parent
eight-dimensional topological theory, $\chi$ and $\psi$ respectively
denote $(2,0)$ and $(0,1)$ forms associated with modes localized on curves
where the eigenvalues of the background $\phi$ vanish, and the ``$\cdots$'' correspond to possible higher dimension operator contributions induced by
background fluxes of the compactification. Here it is important to
note that the available background fluxes can correspond to local curvatures of the metric, gauge field strength, and, for example $H$-fluxes of the compactification. The Yukawa coupling matrix is then defined by evaluating the overlap of solutions to the defect equations of motion given above.

Parameterizing the local patch $\mathcal{U}_{S}$ in terms of two coordinates
$z_{Q}$ and $z_{U}$ such that the coordinate of the quark $Q$ curve is $z_{Q}%
$ while that of the $U$ quark curve is $z_{U}$, the local profile of the zero
mode wave functions for $Q$, $U$ and $H_{u}$ are \cite{BHVI,BHVII,HVCKM}:%
\begin{align}
\Psi_{Q}^{i}\sim &  \left(  \frac{z_{Q}}{R_{Q}}\right)  ^{3-i}\exp\left(
-\frac{z_{U}\overline{z}_{U}}{R_{\ast}^{2}}\right)  \cdot\exp\left(
\mathcal{M}_{k\overline{l}}^{(Q)}\cdot z_{k}\overline{z}_{l}\right)
,\nonumber\\
\Psi_{U}^{j}\sim &  \left(  \frac{z_{U}}{R_{U}}\right)  ^{3-j}\exp\left(
-\frac{z_{Q}\overline{z}_{Q}}{R_{\ast}^{2}}\right)  \cdot\exp\left(
\mathcal{M}_{k\overline{l}}^{(U)}\cdot z_{k}\overline{z}_{l}\right)
,\nonumber\\
\Psi_{H_{u}}\sim &  \exp\left(  -\frac{z_{\perp}\overline{z}_{\perp}}{R_{\ast
}^{2}}\right)  \cdot\exp\left(  \mathcal{M}_{k\overline{l}}^{(H_{u})}\cdot
z_{k}\overline{z}_{l}\right)  \text{,}%
\end{align}
where we have organized the zero mode wave functions $\Psi_{Q}^{i}$ and
$\Psi_{U}^{i}$ according to their order of vanishing near the mutual
interaction point $z_{Q}=z_{U}=0$. The coordinate $z_{\perp}$ corresponds to
the coordinate normal to the Higgs curve inside of $S$, and is given by a
linear combination of $z_{Q}$ and $z_{U}$. Finally, $R_{Q}$ and $R_{U}$ denote
the characteristic lengths of the $Q$ and $U$ curves, and $R_{\ast} \sim M^{-1}_{\ast}$ denotes
the characteristic width of localization, where $M^{4}_{GUT}/M^{4}_{\ast} \sim \alpha_{GUT}$.

Once we know the form of the wavefunctions, we can estimate the Yukawa
coupling \eqref{e:upyuk} in the presence of background fluxes. The Yukawa
coupling is then given by the overlap:%
\begin{equation}
\lambda_{ij}^{(u)}=\underset{\mathcal{U}_{S}}{\int}\left(  \frac{z_{Q}}{R_{Q}%
}\right)  ^{3-i}\left(  \frac{z_{U}}{R_{U}}\right)  ^{3-j}\exp\left(
\mathcal{M}_{k\overline{l}}\cdot z_{k}\overline{z}_{l}\right)  \cdot
Gaussian\text{,} \label{upYUKEVAL}%
\end{equation}
where $\mathcal{M}_{i\overline{j}}$ denotes a quadratic form determined by the
background fluxes, and $Gaussian$ corresponds to the contribution from the
Gaussian wave function factors of the form $\exp(- \left\vert z\right\vert ^{2}/R^{2}_{\ast})$
such that each six-dimensional field localizes on the appropriate matter curve.

In the limit where $\mathcal{M}_{i\overline{j}}$ is exactly constant, the
$3\times3$ Yukawa matrix is rank one.\ Indeed, note that in this limit the
local $U(1)\times U(1)$ rephasing of the coordinates:%
\begin{align}
z_{Q}  &  \mapsto\exp(i\alpha_{Q})z_{Q},\\
z_{U}  &  \mapsto\exp(i\alpha_{U})z_{U},%
\end{align}
causes all Yukawas other than the entry $\lambda_{33}^{(u)}$ to vanish. These
$U(1)$'s are broken when $\mathcal{M}_{i\overline{j}}$ has non-trivial
$\overline{z}$ dependence. By considering the Taylor expansion of the
exponential $\exp\left(  \mathcal{M}_{i\overline{j}}\cdot z_{i}\overline
{z}_{j}\right)  $, it was proposed in \cite{HVCKM} that this \textquotedblleft flux distortion\textquotedblright\ of the wave function generates a hierarchical structure in the Yukawa coupling matrix which is characterized by the degree of $U(1)$ charge violation.\footnote{After \cite{HVCKM} and the present paper appeared, much of this hierarchical structure was indeed corroborated in \cite{FGUTSNC}. We shall therefore use the same methodology proposed in \cite{HVCKM} to crudely estimate the structure of
neutrino Yukawas.}

We are now in a position to understand the qualitative difference between the
Yukawas associated with zero modes and massive modes. As we will see, the
internal profile of massive modes will always violate this type of $U(1)$
rephasing symmetry. Indeed, massive modes are characterized by the fact that
$\overline{\partial}\Psi^{KK}\neq0$, so there is no sense in which they will
preserve the rephasing symmetry present in the zero mode sector. This has the
important consequence that in computing the overlap between massive modes and
zero modes, we generically expect to find larger violations of the $U(1)$
rephasing symmetry besides those present due to flux distortion. Indeed, since
it is a subleading correction, we can safely neglect the effects of the flux
distortion in the computation that follows.

\subsection{Hierarchy Dilution from Kaluza-Klein Modes\label{MASSIVEOVERLAP}}

Having reviewed the estimate of the Yukawa matrices present in the quark and
charged lepton sector, we now estimate the covering theory Yukawa matrix
$\widetilde{y}_{i,I}$ of equation (\ref{ytilde}):%
\begin{equation}
\widetilde{y}_{i,I}=\underset{\mathcal{U}_{B}}{\int}\widetilde{\Psi}_{H_{u}%
}\widetilde{\Psi}_{L}^{i}\widetilde{\Psi}_{N}^{(I)}\text{.}%
\end{equation}
To this end, we first discuss the profile of the massive right-handed neutrino
excitations, and then use this behavior to estimate the form of the neutrino
sector Yukawas.

\subsubsection{Massive Mode Wavefunctions\label{MassiveModes}}

Since the Kaluza-Klein seesaw crucially relies on the profile of massive
modes, we now turn to a more explicit description of their internal profiles.
There are two ways in which a given excitation can correspond to a non-zero
mode. The first class of non-zero modes descend from massless
\textit{six-dimensional} fields which localize along a given matter curve.
Upon reducing the profile of these massless six-dimensional field
into harmonics of the curve, we indeed find massive modes with
excitations parallel to the curve. There is, however, another class of modes
corresponding to fields which are already massive in six-dimensions. These
turn out to play an especially prominent role in the context of the neutrino
sector. Since the right-handed neutrino curve is transverse to the GUT
seven-brane, the corresponding wave functions will have a profile in three
directions, corresponding to directions transverse to the neutrino curve which
we parameterize by the coordinates $z_{\bot}$ and $z_{\bot}^{\prime}$, and a
local coordinate $z_{N}$ along the neutrino curve. In this section we show
that the massive six-dimensional fields correspond to harmonic oscillator wave
functions in the $z_{\bot}$ and $z_{\bot}^{\prime}$ coordinates.

In fact, we can analyze these modes in terms of the quasi-topological eight-dimensional
theory studied in \cite{BHVI} by restricting to the two-complex dimensional
patch $\mathcal{U}$ given by $z_{\bot}^{\prime}=0$. The effective action
defined over the patch $\mathbb{R}^{3,1}\times\mathcal{U}$ contains the
terms:
\begin{align}
S_{8d} &  \supset\underset{%
%TCIMACRO{\U{211d} }%
%BeginExpansion
\mathbb{R}
%EndExpansion
^{3,1}\times\mathcal{U}}{\int}Tr\left(  \eta^{(0,0)}\wedge\left(  \omega
_{S}^{(1,1)}\wedge\partial_{A}\psi^{(0,1)}+\frac{i}{2}\left[  \overline{\phi
}^{(0,2)},\chi^{(2,0)}\right]  \right)  \right)  \nonumber\\
&  +\underset{%
%TCIMACRO{\U{211d} }%
%BeginExpansion
\mathbb{R}
%EndExpansion
^{3,1}\times\mathcal{U}}{\int}Tr\left(  \psi^{(0,1)}\wedge\left(
\overline{\partial}_{A}\chi^{(2,0)}+\frac{1}{2}\left[  \phi^{(2,0)}%
,\psi^{(0,1)}\right]  \right)  \right)  \text{,}\label{action8d}%
\end{align}
where we have included the explicit Hodge type of each field, and
$\eta^{(0,0)}$ denotes a zero form of the theory. The presence of the neutrino
curve is given by the condition that $\phi=z_{\bot}t_{1}$, where $t_{1}$
denotes an element of the Cartan. Varying with respect to $\eta^{(0,0)}$ and
$\psi^{(0,1)}$, we obtain the zero mode equations (\ref{DEOM}) and (\ref{FEOM}).

The background $(2,0)$ form $\phi^{(2,0)}$ as well as
the bulk gauge field $A$ both play crucial roles in defining the zero mode
content of the theory. Although the presence of these two contributions at
first may appear to be on different footings, we note that both can be
combined in a generalization of the covariant derivative. Indeed, equations
(\ref{DEOM}) and (\ref{FEOM}) can be written as:%
\begin{equation}
\mathcal{D}_{A+\phi}\Psi=0\text{,}%
\end{equation}
where $\Psi$ is a vector with entries $\eta^{(0,0)}$, $\chi^{(2,0)}$ and
$\psi^{(0,1)}$, and $\mathcal{D}_{A+\phi}$ denotes the implicitly defined
differential operator which depends on the background $A$ and $\phi$. In this
language, the massive modes of the eight-dimensional quasi-topological theory
are eigenmodes of the Hermitian operator:%
\begin{equation}
\Delta_{A+\phi}\equiv\mathcal{D}_{A+\phi}^{\dag}\mathcal{D}_{A+\phi
}+\mathcal{D}_{A+\phi}\mathcal{D}_{A+\phi}^{\dag}\text{.}%
\end{equation}

The formal similarity between $A$ and $\phi$ can be made precise using the
fact that in flat space, eight-dimensional super Yang-Mills theory originates
from the reduction of ten-dimensional super Yang-Mills theory. Thus, in a
suitably local patch, we can similarly view line (\ref{action8d}) as a
reduction of a ten-dimensional theory where $\phi$ simply corresponds to a
component of the ten-dimensional gauge field.

Using this observation, we can
now deduce the profile of the massive modes in the presence of a background
$A$ and $\phi$. Since $\phi$ is linear in the coordinate $z_{\bot}$, the
corresponding bulk gauge field in ten-dimensions defines a constant background
flux. This leads to a multi-dimensional version of Landau's wave function, and
as such, the massless and massive modes correspond to the ground state, and
excited states of a harmonic oscillator in the $z_{\bot}$ direction. Although
this corresponds to two real directions, for notational expediency we will
label the modes $\Psi^{(I_{\bot})}$ in terms of a single integer
index $I_{\bot}\geq0$. Suppressing all dependence on
the coordinate $z_{N}$, the ground state wave function of the harmonic
oscillator is a Gaussian:%
\begin{equation}
\Psi^{(0)}=\exp\left(  -M_{\ast}^{2}z_{\bot}\overline{z_{\bot}}\right)
\text{,}%
\end{equation}
which corresponds to a massless six-dimensional field. The $I_{\bot}$-th
massive mode is similar, and corresponds to exciting the ground state wave
function:%
\begin{equation}
\Psi^{(I_{\bot})}=f^{(I_{\bot})}(z_{\bot},\overline{z_{\bot}})\exp\left(
-M_{\ast}^{2}z_{\bot}\overline{z_{\bot}}\right)  \text{,}\label{mypsi}%
\end{equation}
where $f^{(I_{\bot})}$ denotes a degree $I_{\bot}$ polynomial in $z_{\bot}$
and $\overline{z_{\bot}}$. This corresponds to a massive six-dimensional
field, with mass set by the characteristic oscillation frequency of the
harmonic oscillator, so that for $I_{\bot}\neq0$:%
\begin{equation}
M_{I_{\bot}\neq0}\sim M_{\ast}\text{.}%
\end{equation}
Note that because $\Psi^{(I_{\bot})}$ contains contributions from two
one-dimensional harmonic oscillators in the $\operatorname{Re}z_{\bot}$ and
$\operatorname{Im}z_{\bot}$ directions, $f^{(I_{\bot})}(z_{\bot}%
,\overline{z_{\bot}})$ will generically contain contributions of all lower
degrees as well.

Returning to the actual case of interest defined by the Kaluza-Klein seesaw,
the right-handed neutrinos are defined by the vanishing locus $z_{\bot
}=z_{\bot}^{\prime}=0$. Hence, the corresponding harmonic oscillator wave
functions will now be functions of $z_{\bot}$ and $z_{\bot}^{\prime}$.
Extending the profile of the wave function into the $z_{\bot}^{\prime}$
direction, we thus find that the wave function exhibits the profile of a
harmonic oscillator in directions transverse to the curve. Letting $z_{L}$
denote the local coordinate along the lepton curve, which is normal to the neutrino curve, it now follows that
$\Psi^{(I_{\bot})}$ will contain terms of the form:%
\begin{equation}
\Psi^{(I_{\bot})}\supset\left(  \frac{\overline{z_{L}}}{R_{\ast}}\right)
^{i}\exp\left(  -M_{\ast}^{2}z_{L}\overline{z_{L}}\right)  \label{PSIPERP}%
\end{equation}
for all $i\leq I_{\bot}$.

\subsubsection{Overlap Between Massive Modes and Zero
Modes\label{MassiveOverlap}}

Having estimated the profile of the massive mode wave functions, we now
evaluate the overlap integral:%
\begin{equation}
\widetilde{y}_{i,I}=\underset{\mathcal{U}_{B}}{\int}\widetilde{\Psi}_{H_{u}%
}\widetilde{\Psi}_{L}^{i}\widetilde{\Psi}_{N}^{(I)}\text{.}%
\end{equation}
Plugging in the rough form of the zero mode profiles for $\widetilde{\Psi
}_{H_{u}}$ and $\widetilde{\Psi}_{L}^{i}$, the Yukawa $\widetilde
{y}_{i,I}$ is then given by:%
\begin{equation}
\widetilde{y}_{i,I}\sim\int \rd^{2}z_{H}\rd^{2}z_{L}\rd^{2}z_{N}\left(  \frac{z_{L}%
}{R_{L}}\right)  ^{3-i}\cdot\widetilde{\Psi}_{N}^{(I)}\exp\left(
\mathcal{M}_{k\overline{l}}\cdot z_{k}\overline{z}_{l}\right)  \cdot
Gaussian\text{,} \label{YEST}%
\end{equation}
where we have used the local coordinates for the Higgs, lepton and neutrino
curve $z_{H}$, $z_{L}$ and $z_{N}$ to define the coordinates of the local
patch, and $\mathcal{M}$ denotes the contribution to the profile of the wave
functions from background fluxes. By inspection, the $z_H$ and $z_N$ integrals give order one answers, up to normalization of the wave functions. Thus, up to order one coefficients, the Yukawa $\widetilde{y}_{i,I}$
reduces to an integral over the $z_{L}$ coordinate:%
\begin{equation}
\widetilde{y}_{i,I}\sim\int \rd^{2}z_{L}\left(  \frac{z_{L}}{R_{L}}\right)
^{3-i}\cdot\widetilde{\Psi}_{N}^{(I)}\exp\left(  \mathcal{M}_{k\overline{l}%
}\cdot z_{k}\overline{z}_{l}\right)  \cdot Gaussian\text{.}%
\end{equation}

Recall that in directions normal to the right-handed neutrino curve, $\Psi^{(I)}$
behaves as a harmonic oscillator with terms of the form of line (\ref{PSIPERP}%
). It now follows that these $\overline
{z_{L}}$'s will saturate the overlap integral, and up to normalization of the wave functions, the Yukawa $\widetilde{y}_{i,I}$ is then given by:%
\begin{equation}
\widetilde{y}_{i,I}\sim\left(  \frac{1}{M_{\ast}R_{L}}\right)  ^{3-i}%
\cdot\theta_{3-i}\left(  I\right)
\end{equation}
where $\theta_{3-i}\left(  I\right)  $ is a step function which is $1$
for $I\geq 3-i$ and $0$ for $I< 3-i$. Finally, as in \cite{BHVII},
the small parameter $1/M_{\ast}R_{L}$ is related to the GUT\ fine structure
constant through the relation:%
\begin{equation}
\varepsilon\equiv\left(  \frac{1}{M_{\ast}R_{L}}\right)  ^{2}\sim\frac
{M_{GUT}^{2}}{M_{\ast}^{2}}\sim\alpha_{GUT}^{1/2}\text{.}%
\end{equation}
Writing $\widetilde{y}$ as a $3\times N$ matrix where $N\rightarrow\infty$ is
the number of massive modes participating in the Yukawa, we therefore have:%
\begin{equation}
\widetilde{y} \sim \left(
\begin{array}
[c]{ccccc}%
\varepsilon & \varepsilon & \varepsilon & \varepsilon & ...\\
\varepsilon^{1/2} & \varepsilon^{1/2} & \varepsilon^{1/2} & \varepsilon^{1/2}
& ...\\
1 & 1 & 1 & 1 & ...
\end{array}
\right)  \text{.} \label{YINF}%
\end{equation}

\subsection{Neutrino Yukawa Matrix \label{NEUTYUK}}

In this subsection we estimate the form of the neutrino Yukawas. In matrix
notation, this amounts to evaluating:
\begin{equation}
\frac{\lambda^{(\nu)}}{\Lambda_{\text{UV}}}=\widetilde{y}\cdot\frac{1}{M}%
\cdot\widetilde{y}^{T}\text{.}%
\end{equation}
To determine the rough structure of this matrix, consider equation
(\ref{YINF}) in the truncated case where $\widetilde{y}$ is given by a
$3\times N$ matrix with $N=4$. Summing over all of the massive excitations
which have characteristic scale $M_{\ast}$, it follows that:%
\begin{align}
\frac{\lambda^{(\nu)}\left(  N=4\right)  }{\Lambda_{\text{UV}}}  &
=\widetilde{y}_{N=4}\cdot\frac{1}{M}\cdot\widetilde{y}_{N=4}^{T}%
\label{MLINE}\\
&  \sim\frac{1}{M_{\ast}}\left(
\begin{array}
[c]{cccc}%
\varepsilon & \varepsilon & \varepsilon & \varepsilon\\
\varepsilon^{1/2} & \varepsilon^{1/2} & \varepsilon^{1/2} & \varepsilon
^{1/2}\\
1 & 1 & 1 & 1
\end{array}
\right)  \left(
\begin{array}
[c]{ccc}%
\varepsilon & \varepsilon^{1/2} & 1\\
\varepsilon & \varepsilon^{1/2} & 1\\
\varepsilon & \varepsilon^{1/2} & 1\\
\varepsilon & \varepsilon^{1/2} & 1
\end{array}
\right) \\
&  \sim\frac{1}{M_{\ast}}\left(
\begin{array}
[c]{ccc}%
\varepsilon^{2} & \varepsilon^{3/2} & \varepsilon\\
\varepsilon^{3/2} & \varepsilon & \varepsilon^{1/2}\\
\varepsilon & \varepsilon^{1/2} & 1
\end{array}
\right)  \text{,}%
\end{align}
where each entry of the matrix is multiplied by an order one entry. The
generalization to an infinite number of modes is now given by%
\begin{equation}
\frac{\lambda_{(\nu)}^{\text{Maj}}}{\Lambda_{\text{UV}}}=\widetilde{y}%
\cdot\frac{1}{M}\cdot\widetilde{y}^{T}\sim\frac{\Sigma}{M_{\ast}}\left(
\begin{array}
[c]{ccc}%
\varepsilon^{2} & \varepsilon^{3/2} & \varepsilon\\
\varepsilon^{3/2} & \varepsilon & \varepsilon^{1/2}\\
\varepsilon & \varepsilon^{1/2} & 1
\end{array}
\right)  \text{,} \label{ourLAMBDA}%
\end{equation}
where as before, each entry of the given matrix is multiplied by an order one
coefficent. Here, the overall coefficient $\Sigma$ reflects the normalization
due to the contribution of an infinite number of modes. In terms of bra-ket
notation, the overlap of wave functions leading to $\lambda^{(\nu)}%
/\Lambda_{\text{UV}}$ can be written as:%
\begin{equation}
\frac{\lambda_{ij}^{(\nu)}}{\Lambda_{\text{UV}}}=\underset{I}{\sum
}\left\langle \Psi_{H_{u}}\Psi_{L}^{i}\right\vert \frac{1}{\overline{\partial
}_{B_{3}}}\left\vert \Psi_{H_{u}}\Psi_{L}^{j}\right\rangle \text{.}%
\end{equation}
In other words, $\Sigma$ is specified by the Green's function associated with
the massive modes of the compactification.

Finally, we note that the presence of an infinite
sum over massive states addresses a potential subtlety in that although most
of the contributing modes in the infinite sum have mass of order $M_{\ast}$, a
subset of these modes correspond to massless six-dimensional fields. These
fields descend to massive four-dimensional modes, but with a slightly lower
Kaluza-Klein seesaw scale set by the radius of the corresponding matter curve.
Because of the lower seesaw scale, such modes might at first appear to provide
a dominant contribution to the seesaw. Note, however, that there is an
infinite number of massive six-dimensional fields of characteristic mass
$M_{\ast}$, which overwhelm the contributions from these massless
six-dimensional modes.

\subsection{Green's Functions and the Majorana Mass Scale\label{NORMALIZATION}%
}

In the previous section we obtained a rough estimate for the relative mass
ratios in the neutrino sector. In this subsection we discuss the overall
normalization of the neutrino masses set by the heaviest neutrino mass:%
\begin{equation}
m_{3}^{(\nu)}\sim\frac{\Sigma\cdot v_{u}^{2}}{M_{\ast}}\text{,}%
\end{equation}
where $\Sigma$ is a shorthand for the presence of a suitable regularization
scheme defined over the infinite modes of the theory. Insofar as $M_{\ast}$ is
near the GUT scale, the precise value of $\Sigma$ will determine whether a
given geometry will yield a viable mass scale for the light neutrinos on the
order of $0.05$ eV, or will end up being either too large or too small.

As mentioned previously, $\Sigma$ reflects the contribution from the Green's
functions associated with the massive transverse modes. Returning to the
discussion of subsection \ref{MassiveModes}, in a patch of the neutrino
interaction point, the Kaluza-Klein seesaw can be formulated in terms of
ten-dimensional fields as:%
\begin{equation}
\widetilde{W}\supset\int_{B_{3}}\mathcal{N}^{c}\overline{\partial
}_{\mathcal{A}}\mathcal{N+}\int_{B_{3}}\mathcal{H}_{u}\mathcal{LN+}\int
_{B_{3}}\mathcal{H}_{u}^{\prime}\mathcal{L}^{\prime}\mathcal{N}^{c}\text{,}%
\end{equation}
where the script fields correspond to ten-dimensional fields, and
$\overline{\partial}_{\mathcal{A}}$ denotes the Dolbeault operator with
respect to the ten-dimensional background gauge field $\mathcal{A}$. Upon
reduction to eight dimensions, this background corresponds to the background
$(2,0)$ form as well as the gauge field of the eight-dimensional
quasi-topological theory. Integrating out the right-handed neutrinos thus
yields:%
\begin{equation}
W\supset\int_{B_{3}}\mathcal{H}_{u}\mathcal{L}\frac{1}{\overline{\partial
}_{\mathcal{A}}}\mathcal{H}_{u}^{\prime}\mathcal{L}^{\prime}\text{.}%
\end{equation}
Since the neutrino interaction $\mathcal{H}_{u}\mathcal{LN}$ localizes near a
point of the threefold base, which we denote by $P$, while $\mathcal{H}%
_{u}^{\prime}\mathcal{L}^{\prime}\mathcal{N}^{c}$ localizes at $P^{\prime}$,
it follows that the resulting term can also be written as:%
\begin{align}
\widetilde{W}  &  \supset\int_{B_{3}}\mathcal{H}_{u}\mathcal{L}\delta
_{P}G_{\mathcal{A}}(z_{B},P^{\prime})\delta_{P^{\prime}}\mathcal{H}%
_{u}^{\prime}\mathcal{L}^{\prime}\\
&  =\mathcal{H}_{u}(P)\mathcal{L}(P)G_{\mathcal{A}}(P,P^{\prime}%
)\mathcal{H}_{u}^{\prime}(P^{\prime})\mathcal{L}^{\prime}(P^{\prime})\text{.}%
\end{align}
More generally, when multiple interaction terms participate in the
Kaluza-Klein seesaw, the net contribution is of the form:%
\begin{equation}
L_{eff}\supset\underset{P,P^{\prime}}{\sum}\mathcal{H}_{u}(P)\mathcal{L}%
(P)G_{\mathcal{A}}(P,P^{\prime})\mathcal{H}_{u}^{\prime}(P^{\prime
})\mathcal{L}^{\prime}(P^{\prime})\text{.}%
\end{equation}
We conclude that the overall normalization $\Sigma$
is%
\begin{equation}
\Sigma\sim\underset{P,P^{\prime}}{\sum}G_{\mathcal{A}}(P,P^{\prime})\text{.}%
\end{equation}
In particular, when $P\neq P^{\prime}$, we note that as $P\rightarrow
P^{\prime}$, $G_{\mathcal{A}}(P,P^{\prime})$ diverges, so that:%
\begin{equation}
P\neq P^{\prime}:\Sigma\gtrsim1\text{,} \label{SIGMABOUND}%
\end{equation}
lowering the effective seesaw mass scale.

We have also seen, however, that in some case the interaction points $P$ and
$P^{\prime}$ coincide in the covering theory, as in the $E_{8}$ enhancement
model discussed in subsection \ref{BIGBOY}. In this case, the geometry near
the interaction point $P$ is to be quotiented by the discrete group
$\mathfrak{S}$. Summing over all of the orbits, it follows that in this case,
$\Sigma$ is given as:%
\begin{equation}
\Sigma\sim\underset{P\rightarrow P^{\prime}}{\lim}\underset{\sigma
\in\mathfrak{S}}{\sum}G_{\mathcal{A}}(P,\sigma(P^{\prime}))\text{,}%
\end{equation}
where the limit procedure is defined by taking $\sigma(P^{\prime})$ to lie on
one of the matter curves in the orbit. Note that in this case, the singular
behavior of the Green's function will in general cancel out, so that in
principle, $\Sigma$ could be greater or less than one. It would be worth investigating this question further.

\section{Dirac Scenario\label{sec:DIRAC}}

Up to this point, we have focussed on Majorana neutrino scenarios. As we
now explain, the suggestive link between the neutrino, weak and GUT scales is \textit{also} present
in Dirac scenarios where the Dirac mass term is generated by the higher dimension operator:
\begin{equation}\label{ourOP}
\int \rd^{4}\theta \frac{H_{d}^{\dag} L N_{R}}{\Lambda_{\text{UV}}}.
\end{equation}
We will show later that this operator is
generated in an analogous fashion to the Giudice-Masiero operator $X^{\dag
}H_{u}H_{d}/\Lambda_{\text{UV}}$ obtained in \cite{HVGMSB} where
$\Lambda_{\text{UV}}$ is close to $M_{GUT}$.  {\it Moreover the scale of the neutrino
mass this leads to is automatically right}:
Indeed, the most important feature of the usual GUT scale seesaw is that:
\begin{equation}
m_{\nu}\sim\frac{M_{\text{weak}}^{2}}{\Lambda_{\text{UV}}}\sim\frac{v_{u}%
^{2}}{\Lambda_{\text{UV}}}\sim\frac{\overline{F_{H_{d}}}}{\Lambda_{\text{UV}}},
\end{equation}
where as usual, $v_{u}$ denotes the scale of the Higgs up vev, and $F_{H_{d}}$ denotes
the F-term component of the $H_{d}$ superfield.  Note that $F_{H_{d}}$ converts the D-term to a Dirac mass term
for the neutrinos:
\begin{equation}\label{ourfinalOP}
\int \rd^{4}\theta \frac{H_{d}^{\dag} L N_{R}}{\Lambda_{\text{UV}}}\rightarrow \int \rd^{2}\theta \frac{\mu \langle H_u\rangle L N_{R}}{\Lambda_{\text{UV}}}.
\end{equation}
 This last equality follows from the fact that the
MSSM superpotential contains the $\mu$-term:
\begin{equation}
W_{MSSM}\supset \mu H_{u} H_{d}
\end{equation}
so that the F-term equation of motion yields:
\begin{equation}
\overline{F_{H_{d}}}\sim\frac{\partial W_{MSSM}}{\partial H_{d}}\sim
\mu\left\langle H_{u}\right\rangle \sim10^{5}\text{ GeV}^{2}\text{,}%
\end{equation}
where we have used the fact that the $\mu$ parameter is typically between
$500-1000$ GeV in F-theory GUTs\cite{HVGMSB}. Similar Planck suppressed operators have been discussed for example in \cite{DvaliNirNeut}. Although the exact operator of line \eqref{ourOP} was not used, the idea of correlating supersymmetry breaking with the generation of viable Dirac masses has appeared for example in \cite{ArkaniHamed:2000bq}.

In this section we study minimal F-theory GUT scenarios which incorporate
Dirac masses through higher dimension operators of the effective theory. As
opposed to the case of the Kaluza-Klein seesaw, here, the right-handed
neutrinos correspond to four-dimensional zero modes of the compactification.
Moreover, the identification of $U(1)_{PQ}$ in $SO(10)\times U(1)_{PQ}\subset
E_{6}$ is compatible with Dirac neutrinos. To illustrate the main ideas, we
therefore restrict to three right-handed neutrino zero modes, and take
$U(1)_{PQ}$ to be embedded in $E_{6}$, as in \cite{HVGMSB}. Since the $U(1)_{PQ}$ charges of $L$ and $H_{d}$ are respectively $+1$ and $-2$,
it follows that $N_{R}$ has charge $-3$. Note that the PQ deformation of F-theory GUTs reviewed in section \ref{FREV} will then
induce a soft mass term for the right-handed sneutrinos on the order of $100-1000$ GeV.

Although the right-handed neutrinos correspond to zero modes, we will see that
Kaluza-Klein mode excitations of the higher dimensional theory still play a
prominent role in setting the overall mass scale of the neutrino sector.

The rest of this section is organized as follows. In the next subsection we
show that when the Higgs down, lepton doublet and right-handed neutrino curves
meet at a point, the required D-term is generated by integrating out massive
modes localized on the Higgs down curve. In this same subsection we also
show that all of the interaction terms of the MSSM and the neutrino sector can
geometrically unify at a point of $E_{8}$ enhancement. Next, we estimate the form of the
Yukawa matrix and find that the resulting mass hierarchy is quite similar to
the case of the Kaluza-Klein seesaw Majorana scenario. Additional discussion of Dirac scenarios in F-theory GUTs
is presented in Appendices A, B and C.

\subsection{Generating Higher Dimensional Operators}

\begin{figure}[ptb]
\begin{center}
\includegraphics[
height=4.9182in,
width=6.1237in
]%
{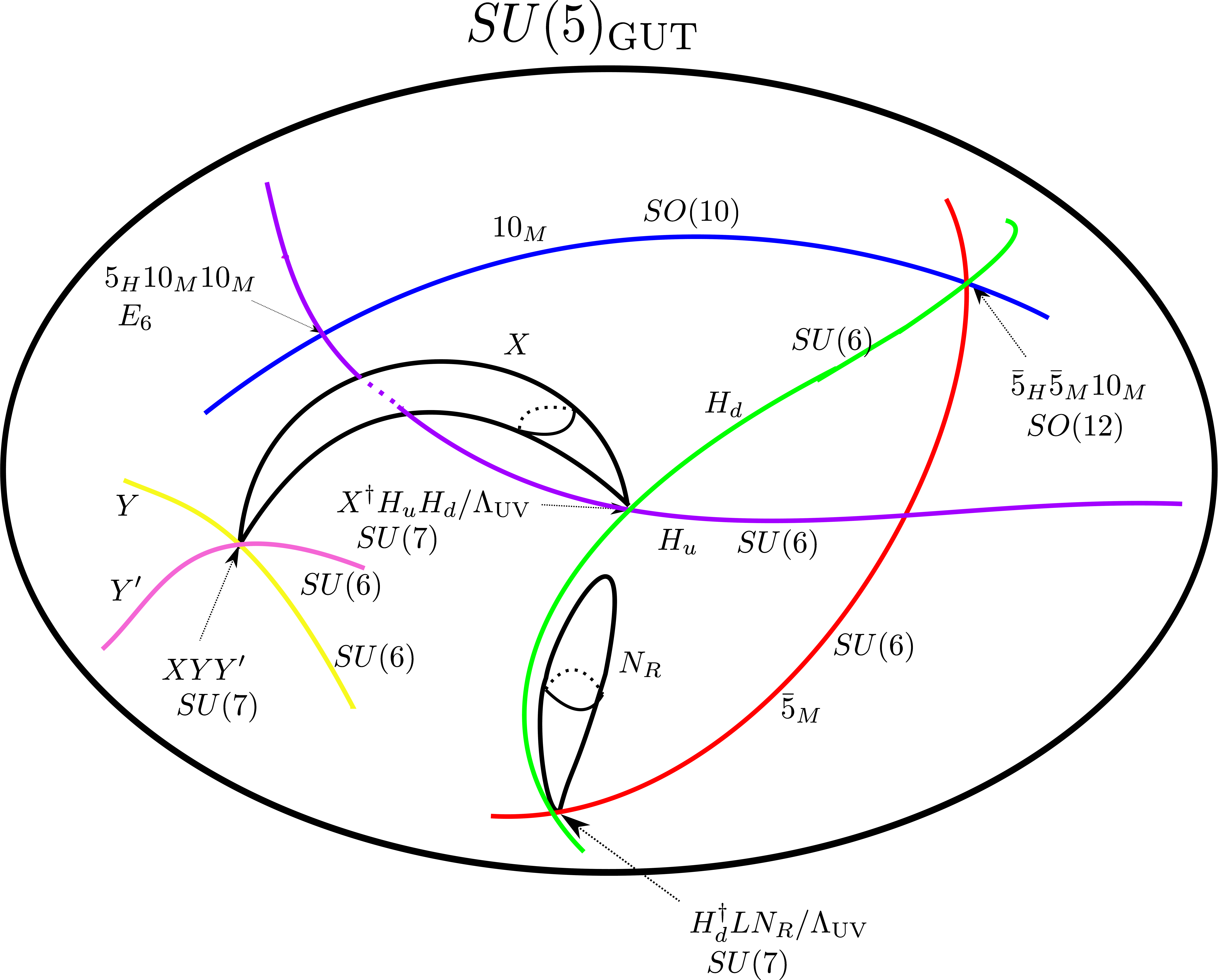}%
\caption{Depiction of a minimal F-theory GUT with a Dirac neutrino sector. In this case,
the Higgs down curve forms a triple intersection with the lepton doublet curve and the
right-handed neutrino curve. Integrating out massive modes localized on the Higgs down
curve then generates a higher dimension cubic D-term which induces a suitable Dirac
mass term in the low energy effective theory.}
\label{pic:DIRAC}
\end{center}
\end{figure}

We now demonstrate that the higher dimension operator:
\begin{equation}
\frac{\lambda_{ij}^{\text{Dirac}}}{\Lambda_{\text{UV}}}\int\mathrm{d}%
^{4}\theta H_{d}^{\dag}L^{i}N_{R}^{j} \label{cubic2}%
\end{equation}
is generated by integrating out massive modes localized on the Higgs down curve.
Here, the right-handed neutrinos localize on a curve which is normal to the
GUT seven-brane. See figure \ref{pic:DIRAC} for a depiction of a minimal F-theory
GUT which contains a Dirac neutrino sector.
As we now argue, this operator can originate from an $SU(7)$ point
of enhancement where the Higgs down, lepton doublet and right-handed neutrino
curve form a point of triple intersection.

We begin by writing the relevant terms of the higher-dimensional action in
terms of an infinite collection of $\mathcal{N}=1$ four-dimensional chiral
superfields labeled by points of the internal directions of the
compactification. As opposed to the Majorana scenario, the operator of line
(\ref{cubic2}) is obtained by integrating out massive modes localized on the
Higgs down curve. Treating the higher-dimensional fields as labelled by points
of the threefold base, the relevant interaction terms are given by:%
\begin{equation}
L\supset\int_{B_{3}}\rd^{4}\theta\mathcal{H}_{d}^{\dag}\mathcal{H}_{d}%
+\int_{B_{3}}\rd^{2}\theta\mathcal{H}_{d}^{c}\mathcal{LN}+\int_{B_{3}}%
\rd^{2}\theta\mathcal{H}_{d}^{c}\overline{\partial}_{\mathcal{A}}\mathcal{H}%
_{d}\text{.}\label{tendeff}%
\end{equation}
The F-term equation of motion for $\mathcal{H}_{d}^{c}$ yields:%
\begin{equation}
\overline{\partial}_{\mathcal{A}}\mathcal{H}_{d}+\mathcal{LN}=0\text{,}%
\end{equation}
or:%
\begin{equation}
\mathcal{H}_{d}=H_{d}-\frac{1}{\overline{\partial}_{\mathcal{A}}^{\prime}%
}\mathcal{LN}\text{,}\label{shifter}%
\end{equation}
where $H_{d}$ denotes the four-dimensional massless mode solution. Plugging
$\mathcal{H}_{d}$ back into the effective action of line (\ref{tendeff}), we
therefore obtain the effective operator:%
\begin{equation}
\frac{\lambda_{ij}^{\text{Dirac}}}{\Lambda_{\text{UV}}}\int\mathrm{d}%
^{4}\theta H_{d}^{\dag}L^{i}N_{R}^{j}=\int_{B_{3}}\rd^{4}\theta H_{d}^{\dag
}\frac{1}{\overline{\partial}_{\mathcal{A}}^{\prime}}L^{i}N_{R}^{j}\text{.}%
\end{equation}

In other words, the relevant Yukawa matrix is given by the overlap integral:%
\begin{equation}
\frac{\lambda_{ij}^{\text{Dirac}}}{\Lambda_{\text{UV}}}=\int_{B_{3}}%
\overline{\Psi}_{H_{d}}\frac{1}{\overline{\partial}_{\mathcal{A}}^{\prime}%
}\Psi_{L}^{i}\Psi_{N}^{j}\text{,}\label{OURDIRAC}%
\end{equation}
where the $\Psi$'s denote the zero mode wave functions. This can be rewritten in bra-ket notation by inserting a complete basis of states, so that the Dirac
Yukawa reduces to a sum over massive states $\left\vert \Psi_{\mathcal{H}%
}\right\rangle $:
\begin{equation}
\frac{\lambda_{ij}^{\text{Dirac}}}{\Lambda_{\text{UV}}}=\underset
{\Psi_{\mathcal{H}}}{\sum}\left\langle \Psi_{H_{d}}|\Psi_{\mathcal{H}%
}\right\rangle \frac{1}{M_{\Psi_{\mathcal{H}}}}\left\langle \Psi_{\mathcal{H}%
}|\Psi_{L}^{i}\Psi_{N}^{j}\right\rangle \text{.}\label{MASSIVESUM}%
\end{equation}
It follows
that to estimate the structure of $\lambda_{ij}^{\text{Dirac}}/\Lambda
_{\text{UV}}$, it is enough to compute the overlap of the massive mode wave
functions localized on the Higgs down curve with the lepton doublet and
neutrino zero mode wave functions:\footnote{We note that in general,
$\left\langle \Psi_{H_{d}}|\Psi_{\mathcal{H}}\right\rangle \neq0$. Indeed, this is essentially
the content of equation (\ref{shifter}).}%
\begin{equation}
\left\langle \Psi_{\mathcal{H}}|\Psi_{L}^{i}\Psi_{N}^{j}\right\rangle
=\int_{\mathcal{U}_{B}}\overline{\Psi}_{\mathcal{H}}\Psi_{L}^{i}\Psi_{N}%
^{j}\text{,}%
\end{equation}
where in the above, $\mathcal{U}_{B}$ denotes a patch in $B_{3}$ containing
the neutrino interaction point.

\subsubsection{Geometric $E_{8}$ Unification of All MSSM Interactions}

\begin{figure}
[ptb]
\begin{center}
\includegraphics[
height=4.9182in,
width=6.1237in
]%
{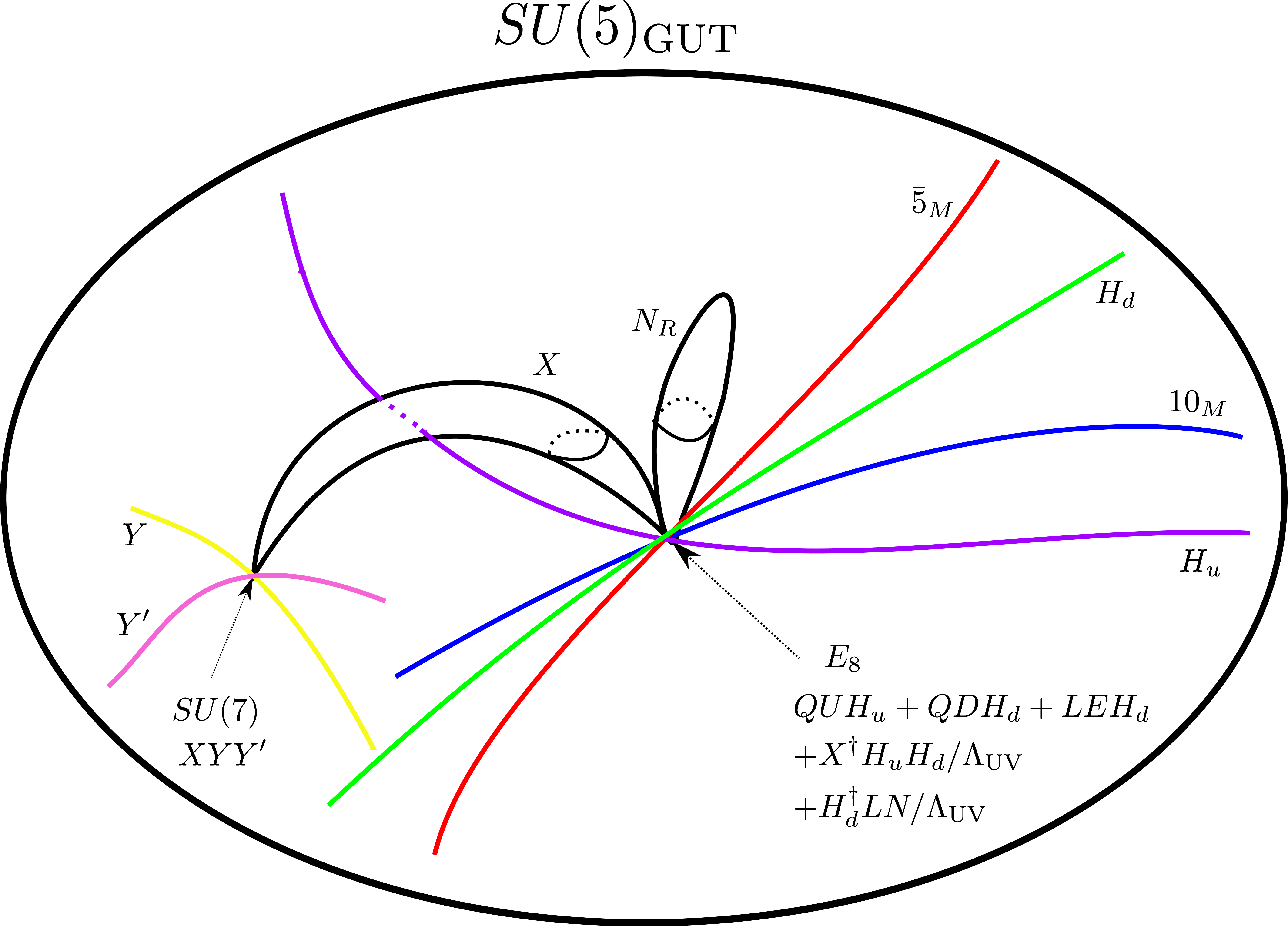}%
\caption{Depiction of a Dirac neutrino scenario in which all of the interaction terms geometrically unify at a single point of $E_{8}$ enhancement.}%
\label{fgutnugdirE8}%
\end{center}
\end{figure}

In the context of the Kaluza-Klein seesaw, we found in section \ref{KKMAJ}
that with an $E_{8}$ point of enhancement it is possible to unify all of the
interaction terms of the MSSM\ at a single point of the geometry. In this
subsection we show that a similar result also holds for the Dirac scenario. See
figure \ref{fgutnugdirE8} for a depiction of this model.
%TCIMACRO{\FRAME{ftbpFU}{4.7366in}{3.4091in}{0pt}{\Qcb{CAPTION}}%
%{\Qlb{fgutnugdirE8}}{fgutnugdirE8.pdf}{\special{ language "Scientific Word";
%type "GRAPHIC";  maintain-aspect-ratio TRUE;  display "USEDEF";
%valid_file "F";  width 4.7366in;  height 3.4091in;  depth 0pt;
%original-width 44.3078in;  original-height 31.8303in;  cropleft "0";
%filename 'fgutnugdirE8.pdf';file-properties "XNPEU";}} }%
%BeginExpansion
%EndExpansion%

With notation as in subsection \ref{BIGBOY}, we consider a specific discrete
subgroup $\mathfrak{S}\subset W\left(  SU(5)_{\bot}\right)  $ such that the
fields of the covering theory which are to be identified fall into
orbits of $\mathfrak{S}$. As an explicit example, we consider the order three
group generated by the cyclic permutation $(123)$ of the first three Cartan
parameters:%
\begin{equation}
(t_{1},t_{2},t_{3},t_{4},t_{5})\mapsto(t_{3},t_{1},t_{2},t_{4},t_{5})\text{.}%
\end{equation}
In this case, we consider fields in the covering theory which lie in the
following orbits:%
\begin{align}
\widetilde{\mathbf{5}}_{H} &  :\{-t_{1}-t_{2},-t_{2}-t_{3},-t_{3}%
-t_{1}\}\nonumber\\
\widetilde{\overline{\mathbf{5}}}_{H} &  :\{t_{1}+t_{4},t_{2}+t_{4}%
,t_{3}+t_{4}\}\\
\widetilde{\overline{\mathbf{5}}}_{M}: &  \{t_{1}+t_{5},t_{2}+t_{5}%
,t_{3}+t_{5}\}\\
\widetilde{\mathbf{10}}_{M} &  :\{t_{1},t_{2},t_{3}\},\\
\widetilde{N}_{R}: &  \{t_{4}-t_{5}\},\nonumber\\
\widetilde{X}: &  \{t_{4}-t_{1},t_{4}-t_{2},t_{4}-t_{3}\}\text{.}%
\end{align}
By inspection, this choice of matter curves allows the interaction terms
$\mathbf{5}_{H}\mathbf{\times10}_{M}\mathbf{\times10}_{M}$, $\overline
{\mathbf{5}}_{H}\mathbf{\times}\overline{\mathbf{5}}_{M}\mathbf{\times10}_{M}%
$, as well as the higher dimension operators $X^{\dag}H_{u}H_{d}/\Lambda_{\text{UV}}$ and
$H_{d}^{\dag}LN_{R}/\Lambda_{\text{UV}}$. In this case, the $U(1)_{PQ}$ symmetry associated with
the abelian factor of $SO(10)\times U(1)_{PQ}\subset E_{6}$ is given by the invariant
linear combination of $U(1)$'s:%
\begin{equation}
U(1)_{PQ}=U(1)_{1}+U(1)_{2}+U(1)_{3}-3U(1)_{4}\text{.}%
\end{equation}
One can check that with this identification of $U(1)_{PQ}$, we have the charge assignments
\begin{equation}%
\begin{tabular}
[c]{|c|c|c|c|c|c|c|}\hline
& $X$ & $H_{u}$ & $H_{d}$ & $\mathbf{10}_{M}$ & $\mathbf{5}_{M}$ & $N_R$ \\\hline
$U(1)_{PQ}$ & $-4$ & $-2$ & $-2$ & $+1$ & $+1$ & $-3$\\\hline
\end{tabular}
\text{ \ .}%
\end{equation}

\subsection{Neutrino Yukawa Matrix}

In this subsection, we show that the Yukawa matrix of the Dirac scenario
has a similar structure to that of the Majorana scenario. The zero mode wave
functions $\Psi_{L}^{i}$ and $\Psi_{N}^{j}$ can be organized according to
their order of vanishing, so that%
\begin{equation}
\Psi_{L}^{i}\sim\left(  \frac{z_{L}}{R_{L}}\right)  ^{3-i}\text{, }\qquad \Psi
_{N}^{j}\sim\left(  \frac{z_{N}}{R_{N}}\right)  ^{3-j}\text{,}%
\end{equation}
where $z_{L}$ (resp. $z_{N}$) denotes a local coordinate for the lepton
doublet (resp. neutrino) curve, and $R_{L}$ (resp. $R_{N}$) denotes the
characteristic length scale of this curve. As in the Majorana scenario, the
crucial point is that the massive modes will overlap with the zero mode wave
functions, inducing maximal violation of the corresponding $U(1)$ coordinate
rephasing symmetries in the directions transverse to the Higgs down
curve.\ Indeed, the massive mode wave function $\Psi_{\mathcal{H}}%
^{I_{L},I_{N}}$ will contain contributions of the form:%
\begin{equation}
\Psi_{\mathcal{H}}^{I_{L},I_{N}}\supset\left(  \frac{\overline{z_{L}}}%
{R_{\ast}}\right)  ^{i}\left(  \frac{\overline{z_{N}}}{R_{\ast}}\right)
^{j}\exp\left(  -\frac{z_{L}\overline{z_{L}}}{R_{\ast}^{2}}-\frac
{z_{N}\overline{z_{N}}}{R_{\ast}^{2}}\right)
\end{equation}
for all $i\leq I_{L}$, $j\leq I_{N}$. It now follows that the overlap is given
as:%
\begin{equation}
\langle\Psi_{\mathcal{H}}^{I_{L},I_{N}}|\Psi_{L}^{i}\Psi_{N}^{j}\rangle
\sim\sqrt{\varepsilon_{L}^{3-i}\varepsilon_{N}^{3-j}}\theta_{3-i}(I_{L}%
)\theta_{3-j}(I_{N})\text{,}%
\end{equation}
where here $\theta_{3-i}(I)$ denotes a step function which is $1$ for $I\geq 3-i$,
and $0$ for $I<3-i$, and we have introduced the small parameters:%
\begin{equation}
\varepsilon_{L}\equiv\left(  \frac{R_{\ast}}{R_{L}}\right)  ^{2}\text{,
}\qquad\varepsilon_{N}\equiv\left(  \frac{R_{\ast}}{R_{N}}\right)  ^{2}\text{.}%
\end{equation}
Summing over all of the massive mode contributions in equation
(\ref{MASSIVESUM}), it now follows that the Dirac matrix is given as:%
\begin{equation}
\frac{\lambda_{(\nu)}^{\text{Dirac}}}{\Lambda_{\text{UV}}}\sim\frac{\Sigma
}{M_{\ast}}%
\begin{pmatrix}
\varepsilon_{L}\varepsilon_{N} & \varepsilon_{L}^{1/2}\varepsilon_{N} &
\varepsilon_{N}\\
\varepsilon_{L}\varepsilon_{N}^{1/2} & \varepsilon_{L}^{1/2}\varepsilon
_{N}^{1/2} & \varepsilon_{N}^{1/2}\\
\varepsilon_{L} & \varepsilon_{L}^{1/2} & 1
\end{pmatrix}
\sim\frac{\Sigma}{M_{\ast}}%
\begin{pmatrix}
\varepsilon^{2} & \varepsilon^{3/2} & \varepsilon\\
\varepsilon^{3/2} & \varepsilon & \varepsilon^{1/2}\\
\varepsilon & \varepsilon^{1/2} & 1
\end{pmatrix}
\text{,}\label{ourLAMBDADIR}%
\end{equation}
where $\Sigma$ denotes the contribution from the convolution of the wave
functions by the Green's function, and in the final relation we have used the
approximation $\varepsilon_{L}\sim\varepsilon_{N}\sim\varepsilon$. Comparing
equations (\ref{ourLAMBDA}) and (\ref{ourLAMBDADIR}), we see that the two Yukawa matrices have the same hierarchical structure with respect to $\varepsilon$.
Note that since there is a single neutrino interaction point the convolution
of the Green's function becomes large near the interaction point. For this
reason, a similar argument to that given near equation (\ref{SIGMABOUND}) of
subsection \ref{NORMALIZATION} implies:%
\begin{equation}
\Sigma\gtrsim1\text{,}%
\end{equation}
which will again boost the value of the Dirac neutrino mass.

\section{Comparison with Experiment\label{s:obs}}

In previous sections we have seen that with the minimal geometric ingredients
required to accommodate neutrino physics, both the Majorana and Dirac
scenarios yield Yukawa couplings in the neutrino sector which are
qualitatively different from the case of the quarks and charged leptons. In
particular, the neutrino Yukawa matrix is given by:%
\begin{equation}
\lambda_{(\nu)}\sim%
\begin{pmatrix}
\varepsilon^{2} & \varepsilon^{3/2} & \varepsilon\\
\varepsilon^{3/2} & \varepsilon & \varepsilon^{1/2}\\
\varepsilon & \varepsilon^{1/2} & 1
\end{pmatrix}
\text{,} \label{lambdanunu}%
\end{equation}
where each entry of this matrix is understood to be multiplied by an order one
complex number. In this section we compare the expected form of these Yukawas
with experiment. Since we shall mainly be interested in order of magnitude
estimates, we will neglect the effects of running; this tends to be a
subdominant contribution on top of the theoretical uncertainties already present.

To make contact with experiment, we first extract the expected form of the
mixing matrix and masses. Depending on the actual geometry of the
compactification, the neutrino mixing matrix can either exhibit a hierarchy
which is milder than that of the CKM\ matrix, or can correspond to a unitary
matrix with little hierarchical structure. In both cases, the mixing angles
are expected to be large so that $\theta_{13}$ should be close to the current
experimental bound. Moreover, in the case where the mixing matrix exhibits a
hierarchical structure, we find the rough relation:%
\begin{equation}
\sin\theta_{13}\sim\sin\theta_{C}\sim\alpha_{GUT}^{1/2}\sim0.2\text{,}%
\end{equation}
where here, $\theta_{C}$ denotes the Cabibbo angle. Due to order one
ambiguities in the values of the underlying parameters, this should be viewed
primarily as an order of magnitude estimate.

After this analysis, we next turn
to the expected mass hierarchy in the neutrino sector. The neutrino masses exhibit
a ``normal'' hierarchy, with ratios:
\begin{equation}
m_{1}:m_{2}:m_{3} \sim \alpha_{GUT} : \alpha^{1/2}_{GUT} : 1.
\end{equation}
Again, we find that the milder hierarchy is in reasonable agreement with the observed mass splittings.
Moreover, using the structure of the Yukawa matrix, we extract the value of
the lightese neutrino mass $m_{1}$, and discuss the prospects for testing
these expected mass ranges, as well as the prospects for distinguishing
between the Majorana and Dirac scenarios.

\subsection{Neutrino Mixing Matrix}

As reviewed in section \ref{NEUTREV}, the neutrino mixing matrix is defined
by:%
\begin{equation}
U_{PMNS}=U_{L}^{(l)}\left(  U_{L}^{(\nu)}\right)  ^{\dag}\text{.}%
\end{equation}
The Yukawa matrices in the neutrino and charged lepton sectors both exhibit a
hierarchical structure. However, in terms of the geometry, there is \emph{a
priori} no reason for these hierarchies to be manifest in the same basis.
Indeed, recall that the hierarchy derives from the presence of a local $U(1)$
coordinate rephasing symmetry in a patch of a given interaction point. This
requires a particular choice of basis for holomorphic functions near this
point. Thus, when the neutrino and charged lepton interaction points $p_{\nu}$
and $p_{l}$ are far away, there is no reason to expect the basis of
holomorphic functions to be the same. On the other hand, when $p_{\nu}$ and
$p_{l}$ are close together, the two basis of holomorphic functions should be
approximately the same. In particular, if the two interactions occur at the
same point, as in the $E_{8}$ model, then the two Yukawa matrices should be in
the same basis. This leads to a mixing matrix with potentially more structure
in the parameter $\varepsilon$.

Thus, the neutrino mixing matrix depends on whether the two interaction points
are nearby or far away in the geometry. In the following subsections we
further discuss these two possibilities.

\subsubsection{Hierarchical Mixing}

As we alluded to previously, the form of the CKM\ matrix found in \cite{HVCKM}
strongly hints at the presence of a higher unification structure. As noted in
\cite{HVCKM}, a hierarchical structure in the CKM matrix requires the up and
down type interaction points $p_{u}$ and $p_{d}$ to roughly satisfy the
relation $\left\vert p_{u}-p_{d}\right\vert \lesssim0.1\times M_{GUT}^{-1}$.
Turning the discussion around, the hierarchy in this sector can then
be taken as evidence for the existence of a higher unification structure.
Unifying neutrinos with the remaining matter content of the MSSM, it is then
natural to perform the further identification $p_{u}=p_{d}=p_{l}=p_{\nu}$.
Indeed, for both the Majorana
and Dirac scenarios we presented models of this type, where all of the interaction terms unified in a single
point of enhancement to $E_{8}$.

When $p_{l}$ is close or equal to $p_{\nu}$, the rephasing symmetry of the local
coordinates will lead to a hierarchical structure in the neutrino and charged
lepton Yukawas with respect to the \textit{same} basis. These Yukawas are then
estimated to be:%
\begin{equation}
\lambda_{(\nu)}\sim%
\begin{pmatrix}
\varepsilon^{2} & \varepsilon^{3/2} & \varepsilon\\
\varepsilon^{3/2} & \varepsilon & \varepsilon^{1/2}\\
\varepsilon & \varepsilon^{1/2} & 1
\end{pmatrix}
\text{, }\qquad\lambda_{(l)}\sim%
\begin{pmatrix}
\varepsilon^{8} & \varepsilon^{6} & \varepsilon^{4}\\
\varepsilon^{6} & \varepsilon^{4} & \varepsilon^{2}\\
\varepsilon^{4} & \varepsilon^{2} & 1
\end{pmatrix}
\text{,}%
\end{equation}
where the form of $\lambda_{(l)}$ was found in \cite{HVCKM}. Introducing
matrices $U_{L}$ and $U_{R}$ such that $U_{L}^{(\nu)}\lambda_{(\nu)}\left(
U_{R}^{(\nu)}\right)  ^{\dag}$ and $U_{L}^{(l)}\lambda_{(l)}\left(
U_{R}^{(l)}\right)  ^{\dag}$ are diagonal, we note that since a matrix with
entries $\lambda_{ij}\sim\varepsilon^{a_{i}+a_{j}}$ has $\left(  U_{L}\right)
_{ij}\sim\left(  U_{R}\right)  _{ij}\sim\varepsilon^{\left\vert a_{i}%
-a_{j}\right\vert }$, we obtain:%
\begin{equation}
U_{L}^{(\nu)}\sim\left(
\begin{array}
[c]{ccc}%
1 & \varepsilon^{1/2} & \varepsilon\\
\varepsilon^{1/2} & 1 & \varepsilon^{1/2}\\
\varepsilon & \varepsilon^{1/2} & 1
\end{array}
\right)  \text{, }\qquad U_{L}^{(l)}\sim\left(
\begin{array}
[c]{ccc}%
1 & \varepsilon^{2} & \varepsilon^{4}\\
\varepsilon^{2} & 1 & \varepsilon^{2}\\
\varepsilon^{4} & \varepsilon^{2} & 1
\end{array}
\right)  \text{.}%
\end{equation}
The resulting form of the PMNS\ matrix is then dominated by the terms in
$U_{L}^{(\nu)}$ so that:%
\begin{equation}
U_{PMNS}^{F-th}=U_{L}^{(l)}\left(  U_{L}^{(\nu)}\right)  ^{\dag}\sim\left(
\begin{array}
[c]{ccc}%
1 & \varepsilon^{1/2} & \varepsilon\\
\varepsilon^{1/2} & 1 & \varepsilon^{1/2}\\
\varepsilon & \varepsilon^{1/2} & 1
\end{array}
\right)  \text{.}%
\end{equation}
Here, we have simply estimated each matrix element
to be an order one complex number multiplied by the appropriate power of
$\varepsilon$. The diagonal entries of $U_{PMNS}^{F-th}$ are expected to be
order one complex numbers, so that in the limit where $\varepsilon
\rightarrow0$, $U_{PMNS}^{F-th}$ tends to a diagonal unitary matrix. To be more precise, since the off-diagonal entries are small but not infinitesimally so, a more reliable estimate of the diagonal entries is obtained by imposing the constraint that $U_{PMNS}^{F-th}$ is unitary, which implies that:%
\begin{equation}
\underset{i=1}{\overset{3}{%
%TCIMACRO{\dsum }%
%BeginExpansion
{\displaystyle\sum}
%EndExpansion
}}\left\vert U_{li}\right\vert ^{2}=1\text{,} \label{UNconst}%
\end{equation}
for $l=e,\mu,\tau$.

In the context of
F-theory GUTs, the parameter $\varepsilon\sim M_{GUT}^{2}/M_{\ast}^{2}%
\sim\alpha_{GUT}^{1/2}$. Plugging in this value, we obtain the final expected
form for the neutrino mixing matrix:
\begin{equation}
U_{PMNS}^{F-th} \sim\left(
\begin{array}
[c]{ccc}%
U_{e1} & \alpha_{GUT}^{1/4} & \alpha_{GUT}^{1/2}\\
\alpha_{GUT}^{1/4} & U_{\mu2} & \alpha_{GUT}^{1/4}\\
\alpha_{GUT}^{1/2} & \alpha_{GUT}^{1/4} & U_{\tau3}%
\end{array}
\right)  \text{,} \label{OURU}%
\end{equation}
where the $U$'s along the diagonal are fixed by \eqref{UNconst}. More precisely, \eqref{OURU} provides an estimate for the magnitudes of the entries of the neutrino mixing matrix.

It is interesting to compare this form of the neutrino mixing matrix with that
of the CKM\ matrix obtained in \cite{HVCKM}:%
\begin{equation}
V_{CKM}^{F-th}\sim\left(
\begin{array}
[c]{ccc}%
1 & \varepsilon & \varepsilon^{3}\\
\varepsilon & 1 & \varepsilon^{2}\\
\varepsilon^{3} & \varepsilon^{2} & 1
\end{array}
\right)  \sim\left(
\begin{array}
[c]{ccc}%
1 & \alpha_{GUT}^{1/2} & \alpha_{GUT}^{3/2}\\
\alpha_{GUT}^{1/2} & 1 & \alpha_{GUT}\\
\alpha_{GUT}^{3/2} & \alpha_{GUT} & 1
\end{array}
\right)  \text{,} \label{CKM}%
\end{equation}
which is manifestly more hierarchical.

Let us now compare with experiments. We know that
$\alpha_{GUT}^{1/2}\sim0.2$. Plugging this value into our estimate for the
neutrino mixing matrix \eqref{OURU} and extracting the diagonal $U$'s using
the unitarity constraint (\ref{UNconst}), we obtain the rough
estimate for the magnitudes of the mixing matrix elements:
\begin{equation}
\left\vert U_{PMNS}^{F-th}\right\vert \sim%
\begin{pmatrix}
0.87 & 0.45 & 0.2\\
0.45 & 0.77 & 0.45\\
0.2 & 0.45 & 0.87
\end{pmatrix}
\text{.} \label{e:mixnumb}%
\end{equation}
This is to be compared with the experimental result which was quoted in
section \ref{NEUTREV}:
\begin{equation}
\left\vert U_{PMNS}^{3\sigma}\right\vert \sim\left(
\begin{array}
[c]{ccc}%
0.77-0.86 & 0.50-0.63 & 0.00-0.22\\
0.22-0.56 & 0.44-0.73 & 0.57-0.80\\
0.21-0.55 & 0.40-0.71 & 0.59-0.82
\end{array}
\right)  \text{.}%
\end{equation}
These two matrices look amazingly close! Given that we are working only up to
order one coefficients, this reveals a very interesting match between theory
and experiment. In fact, as in \cite{HVCKM} the mixing matrix seems to be
relatively insensitive to the various order one coefficients which appear in
the Yukawa matrices, since the results appear very close to the actual
experimental result. It would be interesting to study in more precise terms
whether these order one effects indeed tend to cancel out.

Given the rough numerical values of the mixing matrix in equation
(\ref{e:mixnumb}), we can also extract estimates for the values of the
neutrino mixing angles. Here it is important to stress that the theoretical
uncertainties present will mean that the numerical values of the angles thus
obtained should only be treated as crude approximations. To start with, from
the form of the matrix we see that $\theta_{12}$ and $\theta_{23}$ should take
similar values, while $\theta_{13}$ should be smaller. We roughly estimate
\begin{equation}
\theta_{13}\sim\alpha_{GUT}^{1/2}\sim\theta_{C}\sim0.2\text{,}%
\end{equation}
where $\theta_{C}$ is the Cabibbo angle, the value of which we have extracted
from \eqref{CKM}. Converting from radians to degrees, this yields the rough
expectation $\theta_{13}^{F-th}\sim 10^{\circ}$, where we have rounded to the first significant figure since order one uncertainties in the coefficients of the mixing matrix elements will also propagate to the mixing angles. Since the experimental upper bound on
$\theta_{13}$ is on the order of $13^{\circ}$, we conclude that from F-theory
we expect $\theta_{13}$ to be close to its upper bound.\footnote{After the
results of this paper had already been obtained, we learned from G. Feldman
that current results from the MINOS collaboration have indeed found evidence
that the mixing angle $\theta_{13}$ is non-zero, and is close to this upper
bound \cite{MINOS}.}

We can also extract values for the two other mixing angles. From the entries
$12$ and $23$ of the mixing matrix \eqref{e:mixnumb} we obtain
$\theta^{F-th}_{12}\sim\theta^{F-th}_{23}\sim 30^{\circ}$, where we have again
rounded to the nearest significant figure. These order of magnitude estimates are to be compared with the
experimental values extracted in \cite{GonzalezGarcia:2007ib} which at the $3\sigma$ level are: $\theta_{12}\sim30.5^{\circ}-39.3^{\circ}$ and $\theta_{23}\sim34.6^{\circ}-53.6^{\circ}$.

\subsubsection{Non-Hierarchical Mixing}

Although somewhat counter to the notion of unification, it is in principle
also possible to consider geometries where the neutrino and charged lepton
interaction points are not close together. In this class of geometries, the
matrices $U_{L}^{(l)}$ and $U_{L}^{(\nu)}$ are hierarchical, but \textit{in
different bases}. As a consequence, our expectation is that the mixing matrix
$U_{PMNS}$ should consist of a \textquotedblleft generic\textquotedblright%
\ unitary matrix with no particular structure. Although we do not have a precise
notion of genericity, as a substitute we can consider $U_{PMNS}$ to be a random
unitary matrix. To generate random unitary $N \times N$ matrices, one has to use the only probability measure which is invariant under $U(N)$ group multiplication, known as the Haar
measure. Perhaps surprisingly, even in this case where little structure is
present, we still obtain the qualitative expectation that the mixing angles
$\theta_{12}$ and $\theta_{23}$ should be comparable, while $\theta_{13}$
should be somewhat smaller.

\begin{figure}[ptb]
\begin{center}
\includegraphics[width=4in]
{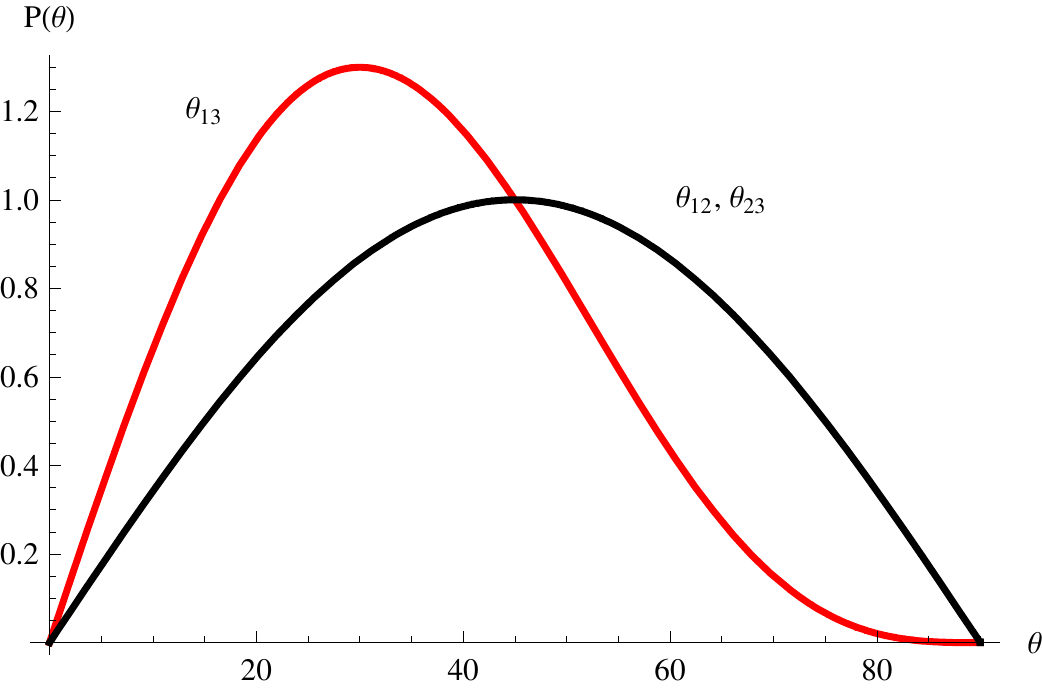}
\end{center}
\caption{Probability density functions for the three mixing angles
$\theta_{12}$, $\theta_{23}$ and $\theta_{13}$, for a random unitary neutrino
mixing matrix.}%
\label{distributions}%
\end{figure}

This directly follows from the parametrization of the neutrino matrix in
terms of the mixing angles $\theta_{ij}$. It is at first tempting to
think that generating uniformly distributed random mixing angles and CP
violating phases will generate random unitary matrices through the
parametrization \eqref{anglePARAM}. This is however too naive. In Appendix D we review the
parametrization of the Haar measure in terms of the three neutrino mixing
angles. With respect to this measure, the probability density functions for
the mixing angles are given by equation (\ref{PROBDIST}) of Appendix D:
\begin{gather}
P(\theta_{12})=2\sin(\theta_{12})\cos(\theta_{12}),\notag\\P(\theta
_{23})=2\sin(\theta_{23})\cos(\theta_{23}),\notag\\
\text{ }P(\theta_{13})=4\sin
(\theta_{13})\cos^{3}(\theta_{13}). \label{e:probdens}%
\end{gather}
This means that to generate random unitary matrices in terms of mixing angles,
we should \emph{not} consider uniformly distributed mixing angles, but rather
the probability densities of equation \eqref{e:probdens}. This may seem
surprising at first, but is again simply a consequence of the way that the
neutrino mixing angles parameterize a unitary matrix. See figure
\ref{distributions} for a plot of the probability densities for the three
neutrino mixing angles.

{}From the explicit form of these densities, we extract that the two mixing angles $\theta_{12}$
and $\theta_{23}$ behave similarly, while the mixing angle $\theta_{13}$ has a
drastically different probability density. In fact, from figure
\ref{distributions} one can see that both distributions of $\theta_{12}$ and
$\theta_{23}$ have mean value $45^{\circ}$, while the distribution of
$\theta_{13}$ has a lower mean value at $33.75^{\circ}$. Therefore, given a
random neutrino mixing matrix, we expect that $\theta_{12}$ and $\theta_{23}$
should be roughly of the same order, while $\theta_{13}$ should be smaller.
This fits relatively well with the current experimental data for the mixing
angles reviewed in section \ref{NEUTREV}. Amazingly, randomness itself
provides an explanation why $\theta_{13}$ should be smaller than the two other
mixing angles!

\begin{figure}[ptb]
\begin{center}
\includegraphics[width=4in]{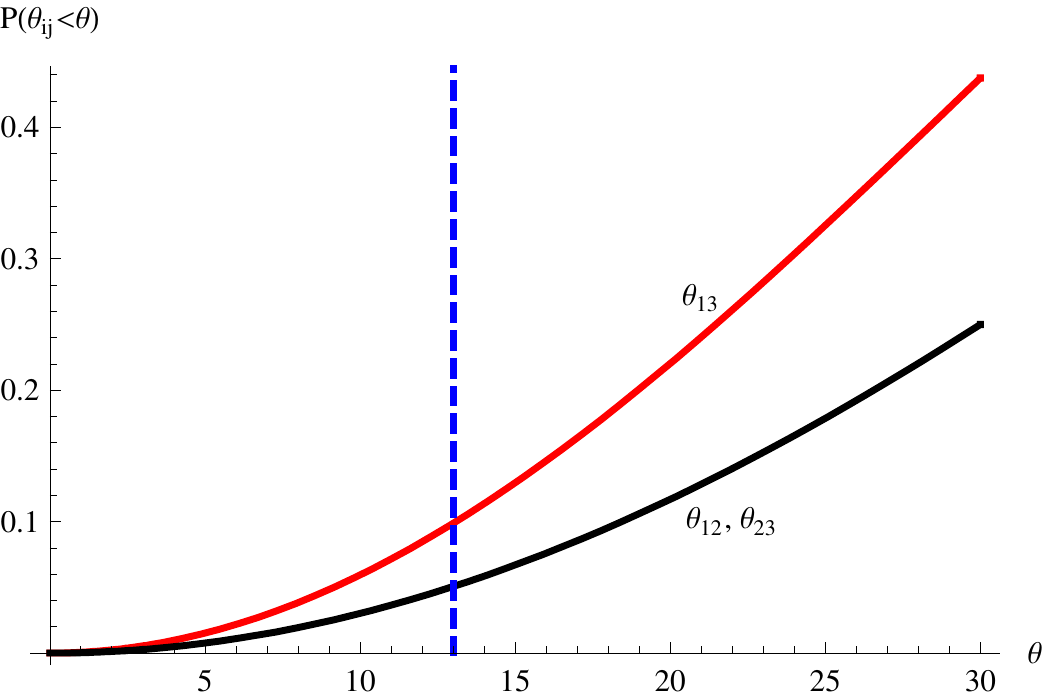}
\end{center}
\caption{Plot of the probability that a random unitary $3 \times3$ matrix has
a value of $\theta_{ij}$ less than a given cutoff $\theta$, where $0\leq
\theta\leq90^{\circ}$ (plotted up to $30^{\circ}$). The dashed vertical line
denotes the current experimental bound of roughly $13^{\circ}$.}%
\label{PROB}%
\end{figure}

However, the actual experimental values are somewhat smaller than the mean
values of the probability distributions. It is therefore worth asking what is
the probability that the angles have their measured values, using the
probability distributions relevant for random unitary matrices. From simple
integration of the probability densities shown in figure \ref{distributions},
we obtain the following probabilities:
\begin{equation}
P(\theta_{13}  < 13^{\circ}) = 9.9\%.
\end{equation}
We conclude that randomness of the neutrino mixing matrix is potentially
consistent with the experimental values, provided that $\theta_{13}$ is close
to its current upper bound. For instance, if the upper bound was lowered to
$1^{\circ}$, we would get the probability:
\begin{equation}
P\left(  \theta_{13}<1^{\circ}\right)  =0.06\%,
\end{equation}
which illustrates the general point that we expect $\theta_{13}$ to be as
close to the current experimental bound as possible. In figure \ref{PROB}, we
provide a plot of the probability that $\theta_{13}$ be lower than a given
angle (the cumulative distribution function), and compare with the same
probability for the other mixing angles.

We can also say something about the CP violating phases. As explained in
Appendix D, the phases of a random unitary matrix are uniformly distributed
between $0$ and $2\pi$. These correspond to the CP violating phases $\delta$,
$\alpha_{1}$ and $\alpha_{2}$ of the neutrino mixing matrix, where the latter two
are physical only in the Majorana case. Hence, for this class of geometries
in F-theory there is no reason why these phases should vanish; all values
are equally probable.

As a cautionary note this analysis should only be viewed as semi-quantitative.
This is because in a completely specified compactification, the neutrino
mixing matrix is not exactly random, since the actual overlap of all wave
functions can be computed. Thus, in a more precise computation randomness
should be supplanted by an appropriate notion of genericity. Even so,
this analysis provides a first estimate of what to expect.

\subsection{Neutrino Mass Hierarchy \label{neutrinomassestimate}}

We now turn to the mass hierarchy in F-theory GUTs. Returning to the form of
the Yukawa matrix, the neutrino masses are:%
\begin{equation}
m_{i}^{(\nu)}\sim\frac{\Sigma M_{\text{weak}}^{2}}{M_{GUT}}\cdot\varepsilon
^{3-i}\text{,}%
\end{equation}
where $\Sigma$ is the contribution from
the Green's function near the interaction point. Here, we have absorbed the
overall volume normalization from the Higgs up and lepton doublet wave
functions into the value of $\Sigma$. These normalization factors tend to decrease
the couplings by an additional factor of O($\alpha_{GUT})$.
 Let us now estimate the overall normalization of the heaviest neutrino.  Since the neutrinos exhibit
a normal mass hierarchy, we have that
\begin{equation}
\sqrt{\Delta m_{31}^{2}}    = \sqrt{m_{3}^{2}-m_{1}^{2}}\sim m_{3}\sim 50 \text{ meV}.%
\label{mmass}
\end{equation}
For the Majorana and Dirac scenarios we have the two slightly different estimates
\begin{align}
m^{\text{Maj}}_{3}  & \sim \frac{\Sigma_{\text{Maj}}v_{u}^{2}}{M_{GUT}}\\
m^{\text{Dirac}}_{3}  & \sim \frac{\Sigma_{\text{Dirac}}\mu v_{u}}{M_{GUT}}.
\end{align}
Using the values
$v_{u}\sim170$ GeV, $M_{GUT}\sim3\times10^{16}$ GeV, and
$\mu\sim 500-1000$ GeV \cite{HVGMSB},
 it follows that the overall
enhancement from the Green's function and volume dependence should be $\Sigma_{\text{Maj}}\sim 50$ and
$\Sigma_{\text{Dirac}}\sim 10$.
This seems to slightly favor the Dirac scenario.

Having discussed the overall mass scales, we now turn to the relative neutrino
mass ratios, which from lightest to heaviest are:
\begin{equation}
m_{1}:m_{2}:m_{3}\sim\varepsilon^{2}:\varepsilon:1\text{.} \label{neutRATIO}%
\end{equation}
Compare this with the parametric $\varepsilon$ dependence in the masses of the
charged leptons and quarks obtained in \cite{HVCKM}:
\begin{align}
m_{e}  &  :m_{\mu}:m_{\tau}\sim\varepsilon^{8}:\varepsilon^{4}%
:1,\label{elecRATIO}\\
m_{u}  &  :m_{c}:m_{t}\sim\varepsilon^{8}:\varepsilon^{4}:1,\label{upRATIO}\\
m_{d}  &  :m_{s}:m_{b}\sim\varepsilon^{5}:\varepsilon^{3}:1\text{.}
\label{downRATIO}%
\end{align}
It is clear that the neutrino hierarchy is much milder than that of the quark
and lepton sectors. In fact, the neutrino mass hierarchy turns out to be the
fourth root of the hierarchy in the charged lepton sector sector. In terms of
numerical values, recall that the parameter $\varepsilon$ is roughly estimated
to be
\begin{equation}
\varepsilon\sim\alpha_{GUT}^{1/2}\sim0.2. \label{pointTWO}%
\end{equation}
We note however that a more precise estimate of $\varepsilon$ will depend on
details of the geometry. In particular, as explained in \cite{HVCKM}, the
overlaps of distinct wave functions will lead to different possible values of
$\varepsilon$. In other words, in equations \eqref{neutRATIO}-\eqref{downRATIO},
the parameter $\varepsilon$ may in principle
take different values.

 The mass splittings are given by:
\begin{align}
\Delta m_{31}^{2}  &  =m_{3}^{2}-m_{1}^{2}\sim m_{3}^{2}(1-d_{31}%
\cdot\varepsilon^{4}),\\
\Delta m_{21}^{2}  &  =m_{2}^{2}-m_{1}^{2}\sim m_{2}^{2}(1-d_{21}%
\cdot\varepsilon^{2})\text{,}%
\end{align}
where the $d$'s are order one coefficients. The mass splittings then satisfy the relation:
\begin{equation}
\frac{\Delta m_{21}^{2}}{\Delta m_{31}^{2}}\sim\varepsilon^{2}\text{.}
\label{masssplitratio}%
\end{equation}
Using the values for the mass splittings reviewed for example in \cite{GonzalezGarcia:2007ib,GonzalezGarcia:2009ij}, the
mass splittings at the $3\sigma$ level are:%
\begin{equation}%
\begin{tabular}
[c]{|c|c|c|c|}\hline
& Min & Central & Max\\\hline
$\Delta m_{21}^{2}/(10^{-5}$ eV$^{2})$ & $7.06$ & $7.67$ & $8.34$\\\hline
$\Delta m_{31}^{2}/(10^{-3}$ eV$^{2})$ & $2.13$ & $2.49$ & $2.88$\\\hline
\end{tabular}
\text{ \ .}%
\end{equation}
The maximum and minimum values of equation (\ref{masssplitratio}) compatible
with this range of experimental values are then given as:%
\begin{equation}
\frac{\min\Delta m_{21}^{2}}{\max\Delta m_{31}^{2}}\leq\frac{\Delta m_{21}%
^{2}}{\Delta m_{31}^{2}}\leq\frac{\max\Delta m_{21}^{2}}{\min\Delta m_{31}%
^{2}}\text{.}%
\end{equation}
Extracting the minimal, maximal and central value of $\varepsilon$ consistent
with equation (\ref{masssplitratio}) we therefore obtain the following range
of experimental values for $\varepsilon$:%
\begin{equation}%
\begin{tabular}
[c]{|c|c|c|c|}\hline
& Min & Central & Max\\\hline
$\varepsilon$ & $0.16$ & $0.18$ & $0.20$\\\hline
\end{tabular}
\ \ \text{.} \label{neutrinoepsilon}%
\end{equation}
We note that this value derived from experimental results is consistent with the rough estimate $\varepsilon\sim\alpha_{GUT}^{1/2}\sim0.2$! Taking
into account that our estimates are only valid up to order one numbers, this
reveals a very interesting match with experiments.

We can now use the value just obtained to estimate the lightest
neutrino mass $m_{1}$. For simplicity, we use the central value
$\varepsilon \sim0.18$ obtained in (\ref{neutrinoepsilon}).  Given our prediction
that neutrino masses exhibit a
normal hierarchy we can take $m_{3}$
and $m_{2}$ given by the square root of the two mass splittings:%
\begin{align}
m_{3}^{obs}  &  \sim\sqrt{\Delta m_{31}^{2}}\sim 50 \pm 4\text{ meV}\\
m_{2}^{obs}  &  \sim\sqrt{\Delta m_{21}^{2}}\sim 8.7\pm 0.4\text{ meV.}%
\label{expms}
\end{align}
Using the relative mass ratios \eqref{neutRATIO}, we can then extract the value for $m_1$ predicted by F-theory. We obtain\footnote{It is interesting to compare this value to the landscape of $AdS_{3}$ Standard Model vacua, where it was found that in a Dirac scenario, an $AdS_{3}$ minimum requires a lightest neutrino of mass greater than $8.3$ meV, a metastable $dS_{3}$ minimum requires a mass of $7.1 - 8.3$ meV, and no minimum is present for lighter masses \cite{ArkaniHamed:2007gg}.}
\begin{equation}
m^{F-th}_1 \sim 1-3 \text{ meV}.
\label{ourms}
\end{equation}
Hence, we expect a relatively light value for $m_{1}$. As before, we note that this should
be interpreted only as a crude estimate on the value of $m_{1}$, since we are
working up to order one numbers.
The expected deviation in the value of $\varepsilon$ was estimated by comparing with
the variation present, for example, in the charged lepton sector. Fitting the masses of the electron, muon and tau to the hierarchy present
of equation \eqref{elecRATIO}, the central value of $\varepsilon_{L} \sim 0.36$. By contrast, using just the
mass ratios $m_{\mu}/m_{\tau} \sim \varepsilon_{L}^{4}$ and $m_{e}/m_{\mu} \sim \varepsilon_{L}^{4}$ respectively yield $\varepsilon \sim 0.49$ and $\varepsilon \sim 0.26$. Thus, we can expect an error of roughly $50 \%$ in extracting the value of $\varepsilon$ in the neutrino sector.

\subsection{Distinguishing Majorana and Dirac: Neutrinoless Double Beta Decay}

In this paper we have found that both the Majorana and Dirac neutrino
scenarios naturally fit within the F-theory GUT framework. Indeed, up to
multiplicative phases which cannot be removed in the Majorana case, the form
of the Yukawas are identical. It is in principle possible to distinguish
between these two scenarios through neutrinoless double beta decay
experiments. Here, the idea is that a nucleus with $Z$ nucleons and atomic
number $A$ can undergo two $\beta$ decays. The emitted neutrinos can then
annihilate each other provided a Majorana mass term couples these two states.
The associated decay rate for neutrinoless double $\beta$ decay scales with
the Majorana mass as:%
\begin{equation}
\Gamma\left(  (Z,A)\rightarrow(Z,A+2)e^{-}e^{-}\right)  \sim M\cdot
A_{\text{nuc}}^{2}\cdot\left\vert m_{\beta\beta}\right\vert ^{2}\text{,}%
\end{equation}
where $A_{\text{nuc}}$ is the contribution to the amplitude from nuclear
processes, $M$ is an overall phase space factor and the relevant Majorana mass
term is given by:%
\begin{equation}
\left\vert m_{\beta\beta}\right\vert ^{2}=\left\vert \underset{i=1}%
{\overset{3}{\sum}}m_{i}\left(  U_{ei}^{PMNS}\right)  ^{2}\right\vert
^{2}\text{.} \label{e:decay}%
\end{equation}
As reviewed in \cite{Law:2009vh}, experiments such as CUORE
\cite{Ardito:2005ar}, GERDA \cite{Abt:2004yk}\ and
\textit{Majorana} \cite{Aalseth:2004yt, Avignone:2007js}\ will likely be
sensitive to this decay rate at the level of $m_{\beta\beta}\sim 50$ meV. To
give a rough expectation for future experiments, as explained for example in
\cite{O'Sullivan:2008zz}, within ten years the EXO experiment is expected to
be sensitive down to $m_{\beta\beta}\sim4-40$ meV.

In neutrino models with a normal hierarchy, it is sometimes common to also
posit that the mixing angle $\theta_{13}$ is quite small. In such models, the
value of $\left\vert m_{\beta\beta}\right\vert ^{2}$ would instead be
controlled by $m_{1}$ and $m_{2}$, rendering this effect less observable. For
this reason, it is common to say that observing neutrinoless double beta decay
would appear to favor a Majorana scenario with an \textit{inverted} hierarchy.

But in the context of F-theory GUTs, we have seen that it is natural
to expect as large a value of $\theta_{13}$ as possible, consistent with
current experimental bounds. It is therefore of interest to study whether
upcoming neutrinoless double beta decay experiments will be sensitive
to the value of $m_{\beta\beta}$ expected in the class of models studied in this paper.
Including the effects of the CP\ violating phases, we obtain:
\begin{equation}
\left\vert m_{\beta\beta}\right\vert =\left\vert
m_{1}e^{i\alpha_{1}}\left\vert U_{e1}^{PMNS}\right\vert ^{2}+m_{2}%
e^{i\alpha_{2}}\left\vert U_{e2}^{PMNS}\right\vert ^{2}+m_{3}e^{-2i\delta
}\left\vert U_{e3}^{PMNS}\right\vert ^{2}\right\vert \text{.}%
\end{equation}
Depending on the relative phases of these contributions, the individual
summands can either add constructively, or destructively. For concreteness, we
take the rough numerical estimate for the magnitudes of the mixing matrix
elements obtained in equation (\ref{e:mixnumb}), with the values of the masses
given in equations \eqref{expms} and (\ref{ourms}), and range over the values of the CP violating phases. As
a function of $\alpha_{1}$, $\alpha_{2}$ and $\delta$, our rough estimate for
$m_{\beta\beta}$ is then:%
\begin{equation}
\left\vert m_{\beta\beta}^{F-th}\right\vert =\left\vert (1.4\pm 0.8)\cdot e^{i\alpha_{1}%
}+(2.8\pm 0.8)\cdot e^{i\alpha_{2}}+(1.3\pm 0.9)\cdot e^{-2i\delta}\right\vert \text{ meV.}%
\end{equation}
Thus, the maximal value of $m_{\beta\beta}$ expected is:%
\begin{equation}
m_{\beta\beta}^{\max}\sim 6\text{ meV,}%
\end{equation}
while the minimal value expected is consistent with zero. For generic
complex phases, we therefore roughly expect $m_{\beta\beta}$ on the order of
a few meV which is too small for observation in the current round of experiments,
but which is tantalizingly close to the limits of sensitivity expected
in the near future.

\subsection{Single Beta Decay}

Although the lightest neutrino mass $m_{1}$ we have found is likely to be too
small for direct detection, it is still of interest to consider constraints
from other experiments. Here we focus on constraints derived by precisely
measuring the masses of all of the visible decay products in single beta decays.
The effective mass of the electron neutrino, or $m_{\beta}$ is:%
\begin{equation}
\left\vert m_{\beta}\right\vert ^{2}=\underset{i=1}{\overset{3}{\sum}}%
m_{i}^{2}\left\vert U_{ei}^{PMNS}\right\vert ^{2}\text{.}%
\end{equation}
Results from the Troitsk experiment \cite{Lobashev:2001uu} and Mainz neutrino
mass experiment \cite{Kraus:2004zw} give only an upper bound of $2.5$ eV and
$2.3$ eV, respectively. The KATRIN experiment is expected to be sensitive to a
non-zero value of $m_{\beta}$ down to $0.2$ eV \cite{KATRIN}.

Again using the rough numerical estimate for the magnitudes of the mixing
matrix elements obtained in equation (\ref{e:mixnumb}), with the values of the
masses given in equation (\ref{ourms}), we obtain:%
\begin{equation}
\left\vert m_{\beta}^{F-th}\right\vert \sim 5-10\text{ meV.}%
\end{equation}
which is far too small to be observed by current direct detection experiments.

\section{Conclusions}

\label{CONC}

In this paper we have studied minimal realizations of F-theory GUTs which
contain a neutrino sector. We have found that small Majorana and Dirac neutrino masses can be accomodated naturally in minimal F-theory GUTs. In both scenarios, Kaluza-Klein modes play a prominent role. Owing to the fact that these massive Kaluza-Klein wave functions are not holomorphic, the neutrino Yukawa matrix exhibits a milder hierarchical structure than its quark and charged lepton counterparts.

For both the Majorana and Dirac scenarios, we have found that a normal hierarchy is expected, with relative mass ratios
$m_{1}:m_{2}:m_{3} \sim \alpha_{GUT}:\alpha^{1/2}_{GUT}:1$, which is
consistent with experimental values for the neutrino mass splittings. In analyzing
neutrino mixing, we have considered geometries where the neutrino and charged
leptons unify at a single point, as well as configurations where these
interactions do not unify. In the former case, we find a mild hierarchy
in the mixing matrix compatible with current observational constraints.
In particular, we find that the mixing angle $\theta_{13}$ and Cabibbo angle are related
through $\theta_{13}\sim\theta_{C} \sim \alpha^{1/2}_{GUT}\sim 0.2$. When the neutrino and charged lepton
interactions do not unify, we instead expect a generic neutrino mixing matrix. For this reason,
such models naturally realize large mixing angles. This in turn leads to the expectation that the mixing angle
$\theta_{13}$ is close to the current experimental bound. In the remainder of
this section we discuss further potential avenues of investigation.

In our implementation of the Majorana scenario in F-theory GUTs, it was necessary to
consider an alternative choice of Peccei-Quinn symmetry consistent with the
Majorana mass term. It would be interesting to investigate in more detail the
phenomenology associated with this new choice of Peccei-Quinn symmetry, and in
particular the expected form of LHC signals, much as in the analysis of
\cite{HKSV}.

In this paper we have also presented Majorana and Dirac scenarios where all
of the interactions of the MSSM geometrically unify at a single $E_8$ point
of enhancement. Our main purpose was to present examples in which the monodromy
group appropriately identifies curves in the quotient theory. Studying other subgroups
of the Weyl group of $SU(5)_{\bot}$ in the breaking pattern $SU(5)_{GUT}
\times SU(5)_{\bot} \subset E_{8}$ may provide further insight into realizations of F-theory neutrinos.

Finally, in extracting the neutrino Yukawa matrices, the overlap between
massive modes and zero modes enters in a crucial way in both the Majorana
and Dirac scenarios. It would be worth studying more precisely how these massive
modes fit into the eight-dimensional quasi-topological field theory framework,
and how the general equations of motion can be deformed to include massive excitations.

\section*{Acknowledgements}

We thank G. Feldman, L. Hall, D. Morrissey, L. Randall, M. Schwartz, D.
Simmons-Duffin and J. Thaler for helpful discussions. The work of the authors is supported
in part by NSF grant PHY-0244821. JJH thanks the University of Chicago
Particle Theory Group and the University of Texas at Austin Theory Group for
hospitality during part of this work.

\appendix

\section*{Appendices}

\section{Dirac Scenario Operator Analysis\label{Opanal}}

In this Appendix we discuss in more general terms cubic and quartic
operators which can potentially generate a viable Dirac neutrino
mass term. As in section \ref{sec:DIRAC}, we restrict attention to
$U(1)_{PQ}$ charge assignments compatible with the embedding
$SO(10) \times U(1)_{PQ} \subset E_{6}$.

At a minimal level, generating a Dirac mass for neutrinos requires the
presence of an operator which contains the product $L^{i}{N}_{R}^{j}$ for
$i,j=1,2,3$, where the ${N}_{R}^{j}$ are right-handed neutrinos, as well as
some additional fields which develop a suitable vev to generate a Dirac mass
term. Compatibility with $SU(2)_{L}\times U(1)_{Y}$ gauge invariance therefore
requires $L^{i}{N}_{R}^{j}$ to couple to either $H_{u}$, or $H_{d}^{\dag}$. In
keeping with the requirements of a minimal matter spectrum, the only fields
which develop a vev are $H_{u}$, $H_{d}$ and $X$, where this last field
develops a supersymmetry breaking vev%
\begin{equation}
\left\langle X\right\rangle =x+\theta^{2}F_{X}%
\end{equation}
with $x\sim10^{12}$ GeV and $F_{X}\sim10^{17}$ GeV$^{2}$. For this reason, we
shall restrict our attention to operators containing the fields $L$, $N_{R}$,
$H_{u}$, $H_{d}$ and $X$.

At the level of cubic terms in superfields, the possible invariant terms are:%
\begin{align}
O_{H_{u}LN_{R}}  &  =\int \rd^{2}\theta H_{u}L^{i}N_{R}^{j}\text{, }\\
O_{H_{d}^{\dag}LN_{R}}  &  =\int \rd^{4}\theta\frac{H_{d}^{\dag}L^{i}N_{R}^{j}%
}{\Lambda_{\text{UV}}}\text{.}%
\end{align}

Using the $U(1)_{PQ}$ charge assignments described in subsection \ref{e6PQ}
obtained from identifying $U(1)_{PQ}$ as the abelian factor of $SO(10)\times
U(1)\subset E_{6}$, it follows that $N_{R}$ must have charges:%
\begin{align}
\int \rd^{2}\theta H_{u}LN_{R}  &  \Longrightarrow q_{PQ}\left(  N_{R}\right)
=+1,\\
\int \rd^{4}\theta\frac{H_{d}^{\dag}L^{i}N_{R}^{J}}{\Lambda_{\text{UV}}}  &
\Longrightarrow q_{PQ}\left(  N_{R}\right)  =-3\text{.}%
\end{align}
In other words, in the first case $N_{R}$ comes from the $\mathbf{27}$ of
$E_{6}$ whereas in the second case $N_{R}$ comes from the $\mathbf{78}$.

Assuming that the overall coefficient of each operator is an order one number,
note that the resulting Dirac mass in each case is:%
\begin{align}
m_{H_{u}LN_{R}}  &  \sim v_{u}\sim170\text{ GeV ,}\\
m_{H^{\dag}LN_{R}}  &  \propto\frac{\overline{F_{H_{d}}}}{\Lambda_{\text{UV}}%
}\sim\frac{\mu v_{u}}{\Lambda_{\text{UV}}}\sim0.01\text{ eV,}%
\end{align}
where in the second line, we have set $\Lambda_{\text{UV}}\sim10^{16}$ GeV,
and used the value of the $\mu$ parameter obtained in \cite{HVGMSB} so that:%
\begin{equation}
\overline{F_{H_{d}}}\sim\frac{\partial W_{MSSM}}{\partial H_{d}}\sim
\mu\left\langle H_{u}\right\rangle \sim10^{5}\text{ GeV}^{2}\text{.}%
\end{equation}
Thus, $O_{H_{d}^{\dag}LN_{R}}$ generates a small Dirac mass term in a
potentially viable range, while $O_{H_{u}LN_{R}}$ generates a Dirac mass which
is too big.

It is also possible to consider operators which are quartic in the relevant
superfields. As before, we restrict attention to operators which contain a
factor of the form $L^{i}N_{R}^{j}$. There are precisely four possible quartic
operators involving $H_{u}$, $H_{d}$, $L^{i}$, $N_{R}^{j}$ and $X$:%
\begin{align}
\int \rd^{2}\theta\frac{XH_{u}L^{i}N_{R}^{j}}{\Lambda_{\text{UV}}}  &
\Longrightarrow q_{PQ}\left(  N_{R}\right)  =+5,\label{quartic1}\\
\int \rd^{4}\theta\frac{X^{\dag}H_{u}L^{i}N_{R}^{j}}{\Lambda_{\text{UV}}^{2}}
&  \Longrightarrow q_{PQ}\left(  N_{R}\right)  =-3,\label{quartic2}\\
\int \rd^{4}\theta\frac{XH_{d}^{\dag}L^{i}N_{R}^{j}}{\Lambda_{\text{UV}}^{2}}
&  \Longrightarrow q_{PQ}\left(  N_{R}\right)  =+1,\label{quartic3}\\
\int \rd^{4}\theta\frac{X^{\dag}H_{d}^{\dag}L^{i}N_{R}^{j}}{\Lambda_{\text{UV}%
}^{2}}  &  \Longrightarrow q_{PQ}\left(  N_{R}\right)  =-7\text{.}%
\end{align}
In particular, only the second and third possibilities are compatible with an
$E_{6}$ GUT\ structure because the decomposition of the $\mathbf{27}$,
$\overline{\mathbf{27}}$ and $\mathbf{78}$ only contain $U(1)_{PQ}$ of charges
magnitude between zero and four. The estimated size of the Dirac mass in these
two cases is:%
\begin{align}
\int \rd^{4}\theta\frac{X^{\dag}H_{u}L^{i}N_{R}^{j}}{\Lambda_{\text{UV}}^{2}}
&  \Longrightarrow m_{D}\sim v_{u}\cdot\frac{\overline{F_{X}}}{\Lambda
_{\text{UV}}^{2}}\sim\frac{\mu v_{u}}{\Lambda_{\text{UV}}}\sim0.01\text{ eV}\\
\int \rd^{4}\theta\frac{XH_{d}^{\dag}L^{i}N_{R}^{j}}{\Lambda_{\text{UV}}^{2}}
&  \Longrightarrow m_{D}\sim x\cdot\frac{\overline{F_{H_{d}}}}{\Lambda
_{\text{UV}}^{2}}\sim4\times10^{-5}\text{ eV,}%
\end{align}
where in the first line we have used the fact that in F-theory GUTs, the
Giudice-Masiero operator $X^{\dag}H_{u}H_{d}/\Lambda_{\text{UV}}$ generates
the $\mu$-term in the effective theory. Both of these values are close to the
required values for the neutrinos, although the first possibility is somewhat
closer to the required mass scale necessary for matching to the observed mass splittings.

Although it is in principle possible to consider operators with a larger
number of fields, note that the largest field vev is set at the scale
$x\sim10^{12}$ GeV. As a consequence, each successive operator will be
suppressed by a factor of roughly $x/M_{GUT}\sim10^{-4}$, so that only quartic
or lower operators are relevant for the present discussion.

\section{Quartic Operator Dirac Scenario}

In this Appendix we discuss geometric configurations in F-theory GUTs which
realize the quartic operators of lines \eqref{quartic2} and \eqref{quartic3}
in Appendix A:%
\begin{equation}
L_{eff}\supset\int \rd^{4}\theta\frac{X^{\dag}H_{u}L^{i}N_{R}^{j}}%
{\Lambda_{\text{UV}}^{2}}\text{, }\qquad \int \rd^{4}\theta\frac{XH_{d}^{\dag}%
L^{i}N_{R}^{j}}{\Lambda_{\text{UV}}^{2}}\text{.} \label{otherguy}%
\end{equation}
As a piece of notation, we shall denote the triple intersection of three
matter curves $\Sigma_{A}$, $\Sigma_{B}$ and $\Sigma_{C}$ by $\Sigma_{A}%
\Sigma_{B}\Sigma_{C}$. Further, we shall often be interested in two
configurations of triple intersections which share a common curve. For
example, if the points joining $\Sigma_{A}\Sigma_{B}\Sigma_{C}$ and
$\Sigma_{C}\Sigma_{D}\Sigma_{E}$ both lie on the curve $\Sigma_{C}$, we shall
sometimes denote such a configuration as $\Sigma_{A}\Sigma_{B}\Sigma_{C}%
\oplus_{\Sigma_{C}}\Sigma_{C}\Sigma_{D}\Sigma_{E}$. We now turn to an analysis
of various matter curve configurations which can generate the appropriate
higher dimension operators.

To see how the quartic operators are generated, consider a configuration where
the $X$, $H_{u}$ and $H_{d}$ curves form a triple intersection such that
$X^{\dag}H_{u}H_{d}$ is gauge invariant such that $H_{d}$ also forms a triple
intersection with the $L$ and $N_{R}$ curves. In terms of the same abstract
ten-dimensional formulation provided earlier, the relevant interaction terms
are given as:%
\begin{align}
L_{eff}  &  \supset\int_{B_{3}}\rd^{4}\theta\mathcal{X}^{\dag}\mathcal{X}%
+\int_{B_{3}}\rd^{2}\theta\mathcal{X}^{c}\overline{\partial}_{\mathcal{A}%
}\mathcal{X}+\int_{B_{3}}\rd^{2}\theta\mathcal{H}_{d}^{c}\overline{\partial
}_{\mathcal{A}}\mathcal{H}_{d}\\
&  +\int_{B_{3}}\rd^{2}\theta\mathcal{X}^{c}\mathcal{H}_{u}\mathcal{H}_{d}%
+\int_{B_{3}}\rd^{2}\theta\mathcal{H}_{d}^{c}\mathcal{LN}_{R}+h.c.
\end{align}
The first two F-terms originate from the covariant derivative on the
appropriate curve. The second two F-terms originate from the triple overlap of
matter curves. The $\mathcal{X}^{c}$ and $\mathcal{H}_{d}^{c}$ F-term
equations of motion therefore contain the terms:%
\begin{align}
&  \frac{\partial}{\partial\mathcal{H}_{d}^{c}}\Longrightarrow\mathcal{H}%
_{d}=H_{d}-\frac{1}{\overline{\partial}_{\mathcal{A}}^{\prime}}\left(
\mathcal{LN}_{R}\right), \\
&  \frac{\partial}{\partial\mathcal{X}^{c}}\Longrightarrow\mathcal{X}%
=X-\frac{1}{\overline{\partial}_{\mathcal{A}}^{\prime}}\left(  \mathcal{H}%
_{u}\mathcal{H}_{d}\right)  \text{,}%
\end{align}
so that:%
\begin{equation}
\mathcal{X}=X-\frac{1}{\overline{\partial}_{\mathcal{A}}^{\prime}}%
\mathcal{H}_{u}\frac{1}{\overline{\partial}_{\mathcal{A}}^{\prime}}\left(
\mathcal{LN}_{R}\right)  +...
\end{equation}
where we have dropped terms which will not figure in our discussion. Plugging
this into the resulting D-term $\mathcal{X}^{\dag}\mathcal{X}$, we therefore
obtain:%
\begin{equation}
L_{eff}\supset\int \rd^{4}\theta X^{\dag}H_{u}L^{i}N_{R}^{J}\cdot\int_{B_{3}%
}\overline{\Psi}_{X}\frac{1}{\overline{\partial}_{\mathcal{A}}^{\prime}}%
\Psi_{H_{u}}\frac{1}{\overline{\partial}_{\mathcal{A}}^{\prime}}\Psi_{L}%
^{i}\Psi_{N}^{J}\text{.}%
\end{equation}

In principle, this quartic operator can also be generated in configurations
where $L$ and $N$ do not even meet at a common point. In this case, we can
consider a configuration where $X$, $N$ and some additional curve meet at some
point in the threefold base. Assuming that this curve also forms a triple
intersection with $H_{u}$ and $L$, it follows that by integrating out the
massive modes $\mathcal{S}\mathbb{\oplus}\mathcal{S}^{c}$ localized on the
singlet curve, an analogous expression will again be generated. To be
explicit, in this case, we consider the configuration of matter curves
$\Sigma_{X}\Sigma_{N}\Sigma_{S}\oplus_{\Sigma_{S}}\Sigma_{S}\Sigma_{L}%
\Sigma_{H_{u}}$. We can write down the superpotential terms as before, solve
the F-term equations of motion for $\mathcal{X}^{c}$ and $\mathcal{S}^{c}$,
and plug the result back into the $\mathcal{X}^{\dagger}\mathcal{X}$ D-term. A
similar analysis then yields the coupling:
\begin{equation}
L_{eff}\supset\int \rd^{4}\theta X^{\dag}H_{u}L^{i}N_{R}^{J}\cdot\int_{B_{3}%
}\overline{\Psi}_{X}\frac{1}{\overline{\partial}_{\mathcal{A}}^{\prime}}%
\frac{1}{\overline{\partial}_{\mathcal{A}}^{\prime}}\Psi_{H_{u}}\Psi_{L}%
^{i}\Psi_{N}^{J}\text{.}%
\end{equation}
Finally, although we will not present explicit geometric configurations here,
we note that the second quartic operator of line \eqref{otherguy} can also
be generated by integrating out massive modes.

\section{Other Neutrino Scenarios}

In this section we collect some other possible neutrino scenarios which it
would be interesting to develop further. Our aim here is not so much to
provide an exhaustive list of alternative scenarios, but rather, to present
some other potential avenues of investigation. To this end, we first discuss
some additional Dirac mass scenarios where right-handed neutrinos localize in
the bulk, and also discuss the numerology of instanton induced Dirac mass
terms. After this, we briefly mention another seesaw of potential
interest based on massive string excitations.

\subsection{Miscellaneous Dirac Scenarios}

In section \ref{sec:DIRAC} and Appendices A and B, we presented an analysis of Dirac
mass terms where the right-handed neutrinos localize on curves normal to the
GUT\ seven-brane. In that context, higher dimension operators generated the
necessary suppression in the mass scale below the scale of electroweak
symmetry breaking. Here, we discuss scenarios where the right-handed neutrinos
propagate in the bulk of the GUT seven-brane, and models where instanton
effects can potentially generate a viable mass term.

\subsubsection{$N_{R}$ From the Bulk}

So far in this paper we focused on the case where $N_{R}$ lives on a matter
curve. In this subsection we briefly note that in the Dirac scenario, it is
also possible to consider models where $N_{R}$ propagates in the
bulk.\footnote{Although in different settings, there are neutrino models with
right handed neutrino as bulk neutrinos. The main advantage is that the
geometry gives the desired small mass. See \cite{Mohapatra:1999af}.} This is
especially natural in configurations where the bulk gauge group is of the form
$E_{6}$, so that $N_{R}$ descends from a spinor of $SO(10)$ with PQ\ charge
$-3$. On the other hand, as noted in \cite{BHVII,HVGMSB}, such models
typically contain extraneous zero mode states beyond those present in the MSSM.

Putting aside this potential issue, we now consider a geometry where there is
a local enhancement from $E_{6}$ to $E_{7}$ along curves, and $E_{6}$ to
$E_{8}$ so that the corresponding modes trapped on the curves can form the
$\mathbf{27}^{3}$ interaction term. We consider configurations where $X$
descends from the $\mathbf{\overline{27}}$, while $H_{u}$, $L$ descend from
the $\mathbf{27}$ and $N_{R}$ from the $\mathbf{78}$. Note that the operator
$X^{\dag}H_{u}LN_{R}$ is indeed invariant. Since $X^{\dag}$ is in the
$\mathbf{27}$ we have the fusion rule\cite{Slansky}:%
\begin{equation}
\mathbf{78}\times\mathbf{27}=\mathbf{27}+\mathbf{351}+\mathbf{1728},
\end{equation}
the interaction term $\mathbf{27}\times\mathbf{78}\times\mathbf{27}%
\times\mathbf{27}$ contains a $\mathbf{27}^{3}$ term, and thus a singlet as well.

In addition to this zero mode content, we will also keep track of the
$\mathbf{\overline{16}}$ Kaluza-Klein mode excitations on the $X$-curve, which
we denote by $\mathcal{S}^{c}$. In this case, the relevant interaction terms
are:%
\begin{align}
L_{eff}\supset&\int_{B_{3}}\rd^{4}\theta\mathcal{X}^{\dag}\mathcal{X}+\int
_{B_{3}}\rd^{2}\theta\mathcal{S}^{c}\overline{\partial}_{X}\mathcal{S}%
\mathbb{+}\int_{B_{3}}\rd^{2}\theta\mathcal{X}^{c}\overline{\partial
}_{\mathcal{A}}\mathcal{X}\notag\\
&\mathbb{+}\int_{B_{3}}\rd^{2}\theta\mathcal{SN}%
_{R}\mathcal{X}^{c}+\int_{B_{3}}\rd^{2}\theta\mathcal{S}^{c}\mathcal{H}%
_{u}\mathcal{L}+h.c.
\end{align}
where in the above, the first two F-terms originate from the associated
kinetic terms on the X-curve. The third F-term originates from a coupling
between a bulk gauge field and two chiral fields localized on the same curve
(an $S\Sigma\Sigma$ coupling, in the terminology of \cite{BHVII}), and the
last originates from the triple intersection of three matter curves.

We now proceed to integrate out the relevant Kaluza-Klein modes for
$\mathcal{S}$ and $\mathcal{X}$. We
solve the F-term equations of motion for $\mathcal{X}^{c}$ and $\mathcal{S}%
^{c}$, and plug the result back in the D-term $\mathcal{X}^{\dag}\mathcal{X}$
to obtain the operator:
\begin{equation}
L_{eff}\supset\int \rd^{4}\theta X^{\dag}H_{u}L^{i}N_{R}^{J}\cdot\int_{B_{3}%
}\overline{\Psi}_{X}\frac{1}{\overline{\partial}_{\mathcal{A}}^{\prime}}%
\frac{1}{\overline{\partial}_{\mathcal{A}}^{\prime}}\Psi_{H_{u}}\Psi_{L}%
^{i}\Psi_{N}^{J}\text{.}%
\end{equation}
We caution however that for $N_{R}$ to live in the bulk, we must have an
$E_{6}$ bulk gauge group. In this case, it is not clear whether it is possible
to obtain a low energy spectrum completely free of exotic fields. It would be
interesting to study the consequences of such a scenario in greater detail,
and in particular to establish whether potentially problematic exotics can
indeed be removed from such models.

\subsubsection{Instanton Induced Dirac Masses}

Instanton generated Dirac mass terms have been considered in intersecting
brane models, for example in \cite{Cvetic:2008hi}. In the context of F-theory
GUTs, the characteristic size of instanton effects is determined by the
requirement that the instanton induced Polonyi term:%
\begin{equation}
\int \rd^{2}\theta F_{X}X=M_{PQ}^{2}\int \rd^{2}\theta q_{(4)}X
\end{equation}
responsible for supersymmetry breaking generates a value of $F_{X}=M_{PQ}%
^{2}q_{(4)}$ of order:%
\begin{equation}
M_{PQ}^{2}q_{(4)}\sim F_{X}\sim10^{17}\text{ GeV}^{2}\text{.}%
\end{equation}
Here, $M_{PQ}$ denotes the characteristic mass scale of the Peccei-Quinn
seven-brane, which we shall take to be roughly the GUT scale, and $q_{(4)}$
denotes the suppression factor associated with a D3-instanton wrapping the
same surface as the Peccei-Quinn seven-brane. Note that since $X$ has $-4$
units of PQ charge, $q_{(4)}$ will have $+4$ units of PQ charge. Returning to
the operator:%
\begin{equation}
\int \rd^{2}\theta H_{u}LN_{R}\text{,}%
\end{equation}
when $H_{u}$, $L$ and $N_{R}$ have respective PQ\ charges $-2,+1,+1$, this
operator is invariant under $U(1)_{PQ}$, and so will not be generated by
instanton effects. On the other hand, when $N_{R}$ has PQ\ charge $-3$, so
that it descends from the $\mathbf{78}$ of $E_{6}$, the resulting operator
$H_{u}LN_{R}$ will have PQ\ charge $-4$ and so can in principle be generated
by instanton effects.\footnote{Although it is tempting to consider instanton
effects which directly generate the operator $(H_{u}L)^{2}/\Lambda_{\text{UV}%
}$ in a Majorana scenario, note that instantons will generate such operators
with unviably small coefficients.} The resulting coefficient is then given as:%
\begin{equation}
q_{(4)}\sim\frac{F_{X}}{M_{PQ}^{2}}\sim10^{-15}-10^{-17}\text{.}%
\end{equation}
As a consequence, the resulting Dirac mass term will be of order:%
\begin{equation}
m_{\text{Dirac}}^{(inst)}\sim q_{(4)}\cdot v_{u}\sim10^{-4}-10^{-6}\text{ eV,}%
\end{equation}
which is slightly too small. In principle, however, such effects could be
present and may generate additional subleading corrections. It would be
interesting to evaluate the expected flavor hierarchy derived from estimating
the overlap of instanton zero modes.

\subsection{Symmetric Representation Seesaw}

As a final possibility, we consider another Majorana scenario derived from
fields transforming in two index symmetric representations (the $\mathbf{15}$
or $\overline{\mathbf{15}}$) of $SU(5)$. In terms of representations of the
$SU(2)_{L}$ factor of the Standard Model gauge group, these fields transform in the triplet of $SU(2)$. This scenario then realizes
the triplet seesaw mechanism.

We consider a configuration of matter curves where the Higgs up
self-intersects, and the lepton doublet curve self-intersects, such that both
self-intersections form a triple intersection with a curve $\Sigma_{15}$ where
six-dimensional fields transforming in the $\mathbf{15\oplus}\overline
{\mathbf{15}}$ of $SU(5)$ localize. In terms of the notation introduced in
Appendix B, this can be described as the matter curve configuration
$\Sigma_{H_{u}}\Sigma_{H_{u}}\Sigma_{15}\oplus_{\Sigma_{15}}\Sigma_{15}%
\Sigma_{L}\Sigma_{L}$. Letting $N_{15}\oplus N_{15}^{c}$ denote a vector-like
pair of matter fields localized on $\Sigma_{15}$, the superpotential will
contain the terms:%
\begin{equation}
W\supset H_{u}H_{u}N_{15}^{c}+LLN_{15}+MN_{15}N_{15}^{c}\text{,}%
\end{equation}
which would realize a variant of the Kaluza-Klein seesaw. Note that fields
transforming in the two index anti-symmetric representation of $SU(5)$ would
not couple to $H_{u}$.

The resulting light neutrino masses are either difficult to accomodate within
a GUT framework, or tend to be too small. The essential problem stems from the
fact that the $\mathbf{15}$ of $SU(5)$ is a two index symmetric representation
of $SU(5)$, and so as a massless six-dimensional field necessarily localizes
on a curve where $SU(5)$ enhances to $USp(10)$. This can be arrived at by
noting that the $\mathbf{10}$ of $SU(5)$ localizes on a curve of $SO(10)$
enhancement. Unfortunately, $USp(10)$ does not embed in $E_{8}$; such a
configuration is thus somewhat counter to the notion of E-type structures, which
have figured prominently in F-theory GUTs. On the other hand, as is well known
in the context of perturbative orientifold constructions, when the massless
sector consists of fields transforming in the $\mathbf{10}$ of $SU(5)$, the
first excited string state will transform in the $\mathbf{15}$ of $SU(5)$.
Although the analogue of the perturbative string states are not known in the
present context, it is likely that some massive modes localized on a curve
where $SU(5)$ enhances to $SO(10)$ will indeed transform in the $\mathbf{15}$
of $SU(5)$. Since these fields correspond to the analogue of massive string
modes, they are quite heavy, and as such, will tend to have a seesaw
suppression scale which is far too high. We therefore conclude that resulting
light neutrino mass scale again tends to be slightly too small. Even so, it
could nevertheless be of potential interest to study such a scenario in more detail.

\section{Haar Measure and Mixing Angles}

In this section we review the parameterization of the Haar measure for $3
\times3$ unitary matrices in terms of Euler angles. The Haar measure must be
used to generate random unitary matrices. Using the parameterization in terms
of Euler angles, we extract the probability distributions for the three
neutrino mixing angles for random unitary matrices.

There are various algorithms to generate random unitary matrices using the
Haar measure. A particularly simple one, using parameterizations of unitary
matrices in terms of Euler angles, is explained in \cite{RUM}. Any $3\times3$
unitary matrix $U$ can be written as
\begin{equation}
U = \mathrm{e}^{i \alpha} E^{(2,3)}(\theta_{23},\psi_{23},\eta_{23})
E^{(1,3)}(\theta_{13},\psi_{13},0) E^{(1,2)}(\theta_{12},\psi_{12},\eta_{12}),
\end{equation}
where the $E^{(i,j)}$ are $3\times3$ unitary matrices with entries
\begin{equation}
E_{k,l}^{(i,j)}(\theta_{ij},\psi_{ij},\eta_{ij}) =
\begin{cases}
\delta_{k l} & \text{for $k,l=1,2,3$ and $k,l \neq i,j$}\\
\cos\theta_{ij} \mathrm{e}^{i \psi_{ij} } & \text{for $k=l=i$}\\
\cos\theta_{ij} \mathrm{e}^{-i \psi_{ij} } & \text{for $k=l=j$}\\
\sin\theta_{ij} \mathrm{e}^{i \eta_{ij} } & \text{for $k=i$ and $l=j$}\\
- \sin\theta_{ij} \mathrm{e}^{- i \eta_{ij} } & \text{for $k=j$ and $l=i$}.
\end{cases}
\end{equation}
The three angles and six phases take values in the intervals
\begin{equation}
0 \leq\theta_{ij} \leq\frac{\pi}{2}, \qquad0 \leq\psi_{ij}, \eta_{ij},
\alpha\leq2 \pi.
\end{equation}
To make contact with our parameterization of the neutrino mixing matrix given
in \eqref{anglePARAM}, we note that the three angles $\theta_{ij}$ correspond
to the three mixing angles. Out of the six phases, three are physically
irrelevant, and the three other ones correspond to the CP violating phases
$\delta$, $\alpha_{1}$ and $\alpha_{2}$, where the latter two are only physical in the Majorana scenario.

In this parameterization the Haar measure can be written down explicitly.
Following \cite{RUM}, it reads:\footnote{The minor difference between our
expression for the Haar measure and the one presented in \cite{RUM} can be
traced back to a different ordering in the product of the matrices $E^{(i,j)}$
above. We use the ordering that makes contact with the standard
parametrization of the neutrino mixing matrix \eqref{anglePARAM}.}
\begin{equation}
P_{U}(\mathrm{d}U)=C\mathrm{d}\alpha\prod_{1<j\leq3}\mathrm{d}\eta_{j-1,j}%
\prod_{1\leq i<j\leq3}\mathrm{d}\psi_{ij}\mathrm{d}\left(  \cos^{2(j-i)}%
\theta_{ij}\right)  ,
\end{equation}
where $C$ is some normalization constant. From this explicit expression for
the Haar measure we can generate random unitary matrices as follows. First, we
generate random phases $\alpha$, $\eta_{ij}$ and $\psi_{ij}$ uniformly
distributed between $0$ and $2\pi$. However, we must not generate uniformly
distributed random angles $\theta_{ij}$. Rather, we first generate random
parameters $\xi_{ij}$ uniformly distributed between $0$ and $1$. Then, the
angles are given by
\begin{equation}
\theta_{12}=\arccos{\left(  \xi_{12}^{1/2}\right)  },\qquad\theta_{23}%
=\arccos{\left(  \xi_{23}^{1/2}\right)  },\qquad\theta_{13}=\arccos{\left(
\xi_{13}^{1/4}\right)  }.
\end{equation}
In other words, the probability density functions for the theta angles that
must be used to generate random unitary matrices are given by the functions:
\begin{gather}
P(\theta_{12})=2\sin(\theta_{12})\cos(\theta_{12}),\notag\\
P(\theta_{23})=2\sin(\theta_{23})\cos(\theta_{23}),\notag\\
P(\theta_{13})=4\sin(\theta
_{13})\cos^{3}(\theta_{13}). \label{PROBDIST}%
\end{gather}
As a further check, we
also generated 100,000 random unitary matrices using the numerical algorithm
presented in \cite{mezzadri-2007-54}. We then extracted the mixing angles from
these matrices, and indeed obtained the probability densities \eqref{PROBDIST}.

\newpage
\bibliographystyle{ssg}
\bibliography{fgutsnus}

\begingroup\raggedright\begin{thebibliography}{10}

\bibitem{homestake98}
B.~T. Cleveland {\em et.~al.}, ``{Measurement of the solar electron neutrino
  flux with the Homestake chlorine detector},'' {\em Astrophys. J.} {\bf 496}
  (1998) 505--526.

\bibitem{superkamiokande98}
{\bf Super-Kamiokande} Collaboration, Y.~Fukuda {\em et.~al.}, ``{Evidence for
  oscillation of atmospheric neutrinos},'' {\em Phys. Rev. Lett.} {\bf 81}
  (1998) 1562--1567, \href{http://xxx.lanl.gov/abs/hep-ex/9807003}{{\tt
  hep-ex/9807003}}.

\bibitem{Mohapatra:1986uf}
R.~N. Mohapatra, {\em Unification and Supersymmetry. The Frontiers of Quark -
  Lepton Physics, 3rd ed.}
\newblock Springer, Berlin, Germany, 2003.

\bibitem{BHVI}
C.~Beasley, J.~J. Heckman, and C.~Vafa, ``GUTs and Exceptional Branes in
  F-theory - I,'' {\em JHEP} {\bf 01} (2009) 058,
  \href{http://xxx.lanl.gov/abs/arXiv:0802.3391 [hep-th]}{{\tt arXiv:0802.3391
  [hep-th]}}.

\bibitem{BHVII}
C.~Beasley, J.~J. Heckman, and C.~Vafa, ``GUTs and Exceptional Branes in
  F-theory - II: Experimental Predictions,'' {\em JHEP} {\bf 01} (2009) 059,
  \href{http://xxx.lanl.gov/abs/arXiv:0806.0102 [hep-th]}{{\tt arXiv:0806.0102
  [hep-th]}}.

\bibitem{DonagiWijnholt}
R.~Donagi and M.~Wijnholt, ``Model Building with F-theory,''
  \href{http://xxx.lanl.gov/abs/arXiv:0802.2969 [hep-th]}{{\tt arXiv:0802.2969
  [hep-th]}}.

\bibitem{WatariTATARHETF}
H.~Hayashi, R.~Tatar, Y.~Toda, T.~Watari, and M.~Yamazaki, ``New Aspects of
  Heterotic--F Theory Duality,'' \href{http://xxx.lanl.gov/abs/arXiv:0805.1057
  [hep-th]}{{\tt arXiv:0805.1057 [hep-th]}}.

\bibitem{DonagiWijnholtBreak}
R.~Donagi and M.~Wijnholt, ``Breaking GUT Groups in F-Theory,''
  \href{http://xxx.lanl.gov/abs/arXiv:0808.2223 [hep-th]}{{\tt arXiv:0808.2223
  [hep-th]}}.

\bibitem{HVGMSB}
J.~J. Heckman and C.~Vafa, ``F-theory, GUTs, and the Weak Scale,''
  \href{http://xxx.lanl.gov/abs/arXiv:0809.1098 [hep-th]}{{\tt arXiv:0809.1098
  [hep-th]}}.

\bibitem{HVLHC}
J.~J. Heckman and C.~Vafa, ``From F-theory GUTs to the LHC,''
  \href{http://xxx.lanl.gov/abs/arXiv:0809.3452 [hep-ph]}{{\tt arXiv:0809.3452
  [hep-ph]}}.

\bibitem{Font:2008id}
A.~Font and L.~E. Ib\'{a}\~{n}ez, ``{Yukawa Structure from U(1) Fluxes in
  F-theory Grand Unification},'' {\em JHEP} {\bf 02} (2009) 016,
  \href{http://xxx.lanl.gov/abs/arXiv:0811.2157 [hep-th]}{{\tt arXiv:0811.2157
  [hep-th]}}.

\bibitem{HVCKM}
J.~J. Heckman and C.~Vafa, ``{Flavor Hierarchy From F-theory},''
  \href{http://xxx.lanl.gov/abs/arXiv:0811.2417 [hep-th]}{{\tt arXiv:0811.2417
  [hep-th]}}.

\bibitem{Blumenhagen:2008zz}
R.~Blumenhagen, V.~Braun, T.~W. Grimm, and T.~Weigand, ``GUTs in Type IIB
  Orientifold Compactifications,''
  \href{http://xxx.lanl.gov/abs/arXiv:0811.2936 [hep-th]}{{\tt arXiv:0811.2936
  [hep-th]}}.

\bibitem{Blumenhagen:2008aw}
R.~Blumenhagen, ``{Gauge Coupling Unification in F-Theory Grand Unified
  Theories},'' {\em Phys. Rev. Lett.} {\bf 102} (2009) 071601,
  \href{http://xxx.lanl.gov/abs/arXiv:0812.0248 [hep-th]}{{\tt arXiv:0812.0248
  [hep-th]}}.

\bibitem{FGUTSCosmo}
J.~J. Heckman, A.~Tavanfar, and C.~Vafa, ``Cosmology of F-theory GUTs,''
  \href{http://xxx.lanl.gov/abs/arXiv:0812.3155 [hep-th]}{{\tt arXiv:0812.3155
  [hep-th]}}.

\bibitem{Bourjaily:2009vf}
J.~L. Bourjaily, ``{Local Models in F-Theory and M-Theory with Three
  Generations},'' \href{http://xxx.lanl.gov/abs/arXiv:0901.3785 [hep-th]}{{\tt
  arXiv:0901.3785 [hep-th]}}.

\bibitem{Hayashi:2009ge}
H.~Hayashi, T.~Kawano, R.~Tatar, and T.~Watari, ``Codimension-3 Singularities
  and Yukawa Couplings in F-theory,''
  \href{http://xxx.lanl.gov/abs/arXiv:0901.4941 [hep-th]}{{\tt arXiv:0901.4941
  [hep-th]}}.

\bibitem{Andreas:2009uf}
B.~Andreas and G.~Curio, ``From Local to Global in F-Theory Model Building,''
  \href{http://xxx.lanl.gov/abs/arXiv:0902.4143 [hep-th]}{{\tt arXiv:0902.4143
  [hep-th]}}.

\bibitem{Chen:2009me}
C.-M. Chen and Y.-C. Chung, ``{A Note on Local GUT Models in F-Theory},''
  \href{http://xxx.lanl.gov/abs/arXiv:0903.3009 [hep-th]}{{\tt arXiv:0903.3009
  [hep-th]}}.

\bibitem{HKSV}
J.~J. Heckman, G.~L. Kane, J.~Shao, and C.~Vafa, ``The Footprint of F-theory at
  the LHC,'' \href{http://xxx.lanl.gov/abs/arXiv:0903.3609 [hep-ph]}{{\tt
  arXiv:0903.3609 [hep-ph]}}.

\bibitem{DonagiWijnholtIII}
R.~Donagi and M.~Wijnholt, ``Higgs Bundles and UV Completion in F-Theory,''
  \href{http://xxx.lanl.gov/abs/arXiv:0904.1218 [hep-th]}{{\tt arXiv:0904.1218
  [hep-th]}}.

\bibitem{RandallSimmonsDuffin}
L.~Randall and D.~Simmons-Duffin, ``{Quark and Lepton Flavor Physics from
  F-Theory},'' \href{http://xxx.lanl.gov/abs/arXiv:0904.1584 [hep-ph]}{{\tt
  arXiv:0904.1584 [hep-ph]}}.

\bibitem{Pontecorvo:1957cp}
B.~Pontecorvo, ``{Mesonium and antimesonium},'' {\em Sov. Phys. JETP} {\bf 6}
  (1957) 429.

\bibitem{Maki:1962mu}
Z.~Maki, M.~Nakagawa, and S.~Sakata, ``{Remarks on the unified model of
  elementary particles},'' {\em Prog. Theor. Phys.} {\bf 28} (1962) 870.

\bibitem{GonzalezGarcia:2007ib}
M.~C. Gonzalez-Garcia and M.~Maltoni, ``{Phenomenology with Massive
  Neutrinos},'' {\em Phys. Rept.} {\bf 460} (2008) 1--129,
  \href{http://xxx.lanl.gov/abs/arXiv:0704.1800 [hep-ph]}{{\tt arXiv:0704.1800
  [hep-ph]}}.

\bibitem{GonzalezGarcia:2009ij}
M.~C. Gonzalez-Garcia, ``{Neutrino Physics},''
  \href{http://xxx.lanl.gov/abs/arXiv:0901.2505 [hep-ph]}{{\tt arXiv:0901.2505
  [hep-ph]}}.

\bibitem{Chooz}
{\bf CHOOZ} Collaboration, M.~Apollonio {\em et.~al.}, ``{Search for neutrino
  oscillations on a long base-line at the CHOOZ nuclear power station},'' {\em
  Eur. Phys. J.} {\bf C27} (2003) 331--374,
  \href{http://xxx.lanl.gov/abs/hep-ex/0301017}{{\tt hep-ex/0301017}}.

\bibitem{MINOS0901}
{\bf MINOS} Collaboration, J.~M. Paley, ``{Recent Results and Future Prospects
  from MINOS},'' \href{http://xxx.lanl.gov/abs/arXiv:0901.2131 [hep-ex]}{{\tt
  arXiv:0901.2131 [hep-ex]}}.

\bibitem{Fogli:2008jx}
G.~L. Fogli, E.~Lisi, A.~Marrone, A.~Palazzo, and A.~M. Rotunno, ``{Hints of
  $\theta_{13}>0$ from global neutrino data analysis},'' {\em Phys. Rev. Lett.}
  {\bf 101} (2008) 141801, \href{http://xxx.lanl.gov/abs/arXiv:0806.2649
  [hep-ph]}{{\tt arXiv:0806.2649 [hep-ph]}}.

\bibitem{MINOS}
M.~Diwan, ``Talk at XIII International Workshop on Neutrino Telescopes, Venice,
  March 10-13, 2009.,''.

\bibitem{Mohapatra:2005wg}
R.~N. Mohapatra {\em et.~al.}, ``{Theory of neutrinos: A white paper},'' {\em
  Rept. Prog. Phys.} {\bf 70} (2007) 1757--1867,
  \href{http://xxx.lanl.gov/abs/hep-ph/0510213}{{\tt hep-ph/0510213}}.

\bibitem{Buchmuller:2007zd}
W.~Buchmuller, K.~Hamaguchi, O.~Lebedev, S.~Ramos-Sanchez, and M.~Ratz,
  ``{Seesaw Neutrinos from the Heterotic String},'' {\em Phys. Rev. Lett.} {\bf
  99} (2007) 021601, \href{http://xxx.lanl.gov/abs/hep-ph/0703078}{{\tt
  hep-ph/0703078}}.

\bibitem{BlumenhagenWeigandINST}
R.~Blumenhagen, M.~Cveti\v{c}, and T.~Weigand, ``Spacetime Instanton
  Corrections in 4D String Vacua ( - The Seesaw Mechanism for D-Brane Models -
  ),'' {\em Nucl. Phys.} {\bf B771} (2007) 113--142,
  \href{http://xxx.lanl.gov/abs/hep-th/0609191}{{\tt hep-th/0609191}}.

\bibitem{IbanezUrangaMajorana}
L.~E. Ib\'{a}\~{n}ez and A.~M. Uranga, ``Neutrino Majorana Masses From String
  Theory Instanton Effects,'' {\em JHEP} {\bf 03} (2007) 052,
  \href{http://xxx.lanl.gov/abs/hep-th/0609213}{{\tt hep-th/0609213}}.

\bibitem{Cvetic:2007ku}
M.~Cveti\v{c}, R.~Richter, and T.~Weigand, ``{Computation of D-brane instanton
  induced superpotential couplings - Majorana masses from string theory},''
  {\em Phys. Rev.} {\bf D76} (2007) 086002,
  \href{http://xxx.lanl.gov/abs/hep-th/0703028}{{\tt hep-th/0703028}}.

\bibitem{Jiang:2008yf}
J.~Jiang, T.~Li, D.~V. Nanopoulos, and D.~Xie, ``{$\mathcal{F}-SU(5)$},''
  \href{http://xxx.lanl.gov/abs/arXiv:0811.2807 [hep-th]}{{\tt arXiv:0811.2807
  [hep-th]}}.

\bibitem{ConlonNeut}
J.~P. Conlon and D.~Cremades, ``{The neutrino suppression scale from large
  volumes},'' {\em Phys. Rev. Lett.} {\bf 99} (2007) 041803,
  \href{http://xxx.lanl.gov/abs/hep-ph/0611144}{{\tt hep-ph/0611144}}.

\bibitem{Antoniadis:2002qm}
I.~Antoniadis, E.~Kiritsis, J.~Rizos, and T.~N. Tomaras, ``{D-branes and the
  standard model},'' {\em Nucl. Phys.} {\bf B660} (2003) 81--115,
  \href{http://xxx.lanl.gov/abs/hep-th/0210263}{{\tt hep-th/0210263}}.

\bibitem{KatzMorrison}
S.~Katz and D.~R. Morrison, ``Gorenstein Threefold Singularities with Small
  Resolutions via Invariant Theory for Weyl Groups,'' {\em J.Alg.Geom.} {\bf 1}
  (1992) 449, \href{http://xxx.lanl.gov/abs/alg-geom/9202002}{{\tt
  alg-geom/9202002}}.

\bibitem{KatzVafa}
S.~H. Katz and C.~Vafa, ``{Matter from geometry},'' {\em Nucl. Phys.} {\bf
  B497} (1997) 146--154, \href{http://xxx.lanl.gov/abs/hep-th/9606086}{{\tt
  hep-th/9606086}}.

\bibitem{FGUTSNC}
S.~Cecotti, M.~C.~N. Cheng, J.~J. Heckman, and C.~Vafa, ``{Yukawa Couplings in
  F-theory and Non-Commutative Geometry},''
  \href{http://xxx.lanl.gov/abs/arXiv:0910.0477 [hep-th]}{{\tt arXiv:0910.0477
  [hep-th]}}.

\bibitem{DvaliNirNeut}
G.~R. Dvali and Y.~Nir, ``{Naturally light sterile neutrinos in gauge mediated
  supersymmetry breaking},'' {\em JHEP} {\bf 10} (1998) 014,
  \href{http://xxx.lanl.gov/abs/hep-ph/9810257}{{\tt hep-ph/9810257}}.

\bibitem{ArkaniHamed:2000bq}
N.~Arkani-Hamed, L.~J. Hall, H.~Murayama, D.~Tucker-Smith, and N.~Weiner,
  ``{Small Neutrino Masses from Supersymmetry Breaking},'' {\em Phys. Rev.}
  {\bf D64} (2001) 115011, \href{http://xxx.lanl.gov/abs/hep-ph/0006312}{{\tt
  hep-ph/0006312}}.

\bibitem{ArkaniHamed:2007gg}
N.~Arkani-Hamed, S.~Dubovsky, A.~Nicolis, and G.~Villadoro, ``{Quantum horizons
  of the standard model landscape},'' {\em JHEP} {\bf 06} (2007) 078,
  \href{http://xxx.lanl.gov/abs/hep-th/0703067}{{\tt hep-th/0703067}}.

\bibitem{Law:2009vh}
S.~S.~C. Law, ``{Neutrino Models and Leptogenesis},''
  \href{http://xxx.lanl.gov/abs/arXiv:0901.1232 [hep-ph]}{{\tt arXiv:0901.1232
  [hep-ph]}}.

\bibitem{Ardito:2005ar}
R.~Ardito {\em et.~al.}, ``{CUORE: A cryogenic underground observatory for rare
  events},'' \href{http://xxx.lanl.gov/abs/hep-ex/0501010}{{\tt
  hep-ex/0501010}}.

\bibitem{Abt:2004yk}
I.~Abt {\em et.~al.}, ``{A new $^{76}$Ge Double Beta Decay Experiment at
  LNGS},'' \href{http://xxx.lanl.gov/abs/hep-ex/0404039}{{\tt hep-ex/0404039}}.

\bibitem{Aalseth:2004yt}
{\bf Majorana} Collaboration, C.~E. Aalseth {\em et.~al.}, ``{The Majorana
  neutrinoless double-beta decay experiment},'' {\em Phys. Atom. Nucl.} {\bf
  67} (2004) 2002--2010, \href{http://xxx.lanl.gov/abs/hep-ex/0405008}{{\tt
  hep-ex/0405008}}.

\bibitem{Avignone:2007js}
{\bf Majorana} Collaboration, I.~Avignone, Frank~T., ``{The MAJORANA $^{76}$Ge
  neutrino less double-beta decay project: A brief update},'' {\em J. Phys.
  Conf. Ser.} {\bf 120} (2008) 052059,
  \href{http://xxx.lanl.gov/abs/arXiv:0711.4808 [nucl-ex]}{{\tt arXiv:0711.4808
  [nucl-ex]}}.

\bibitem{O'Sullivan:2008zz}
{\bf EXO} Collaboration, K.~O'Sullivan, ``{The Enriched Xenon Observatory},''
  {\em J. Phys. Conf. Ser.} {\bf 120} (2008) 052056.

\bibitem{Lobashev:2001uu}
V.~M. Lobashev {\em et.~al.}, ``{Direct search for neutrino mass and anomaly in
  the tritium beta-spectrum: Status of 'Troitsk neutrino mass' experiment},''
  {\em Nucl. Phys. Proc. Suppl.} {\bf 91} (2001) 280--286.

\bibitem{Kraus:2004zw}
C.~Kraus {\em et.~al.}, ``{Final Results from phase II of the Mainz Neutrino
  Mass Search in Tritium $\beta$ Decay},'' {\em Eur. Phys. J.} {\bf C40} (2005)
  447--468, \href{http://xxx.lanl.gov/abs/hep-ex/0412056}{{\tt
  hep-ex/0412056}}.

\bibitem{KATRIN}
{\bf KATRIN} Collaboration, A.~Osipowicz {\em et.~al.}, ``{KATRIN: A next
  generation tritium beta decay experiment with sub-eV sensitivity for the
  electron neutrino mass},'' \href{http://xxx.lanl.gov/abs/hep-ex/0109033}{{\tt
  hep-ex/0109033}}.

\bibitem{Mohapatra:1999af}
R.~N. Mohapatra and A.~Perez-Lorenzana, ``{Sterile neutrino as a bulk
  neutrino},'' {\em Nucl. Phys.} {\bf B576} (2000) 466--478,
  \href{http://xxx.lanl.gov/abs/hep-ph/9910474}{{\tt hep-ph/9910474}}.

\bibitem{Slansky}
R.~Slansky, ``Group Theory for Unified Model Building,'' {\em Phys. Rept.} {\bf
  79} (1981) 1--128.

\bibitem{Cvetic:2008hi}
M.~Cveti\v{c} and P.~Langacker, ``{D-Instanton Generated Dirac Neutrino
  Masses},'' {\em Phys. Rev.} {\bf D78} (2008) 066012,
  \href{http://xxx.lanl.gov/abs/arXiv:0803.2876 [hep-th]}{{\tt arXiv:0803.2876
  [hep-th]}}.

\bibitem{RUM}
K.~Zyczkowski and M.~Kus, ``Random unitary matrices,'' {\em J. Phys. A: Math.
  Gen.} {\bf 27} (1994) 4235--4245.

\bibitem{mezzadri-2007-54}
F.~Mezzadri, ``How to generate random matrices from the classical compact
  groups,'' {\em Notices of the AMS} {\bf 54} (2007) 592.

\end{thebibliography}\endgroup

\end{document}